\documentclass[acmlarge,screen]{acmart}

\AtBeginDocument{%
  }

\setcopyright{acmlicensed}
\copyrightyear{2018}
\acmYear{2018}
\acmDOI{XXXXXXX.XXXXXXX}

\acmConference[Conference acronym 'XX]{Make sure to enter the correct
  conference title from your rights confirmation emai}{June 03--05,
  2018}{Woodstock, NY}
\acmISBN{978-1-4503-XXXX-X/18/06}

\usepackage[caption=false,font=footnotesize,labelfont=sf,textfont=sf]{subfig}
\usepackage{textcomp}
\usepackage{stfloats}
\usepackage{url}
\usepackage{verbatim}
\usepackage{graphicx}
\usepackage{booktabs}

\usepackage{booktabs}
\usepackage{multirow}
\usepackage{color}
\usepackage[countmax]{subfloat}
\usepackage{caption}

\setcopyright{none}
\renewcommand\footnotetextcopyrightpermission[1]{}
\begin{document}

\title{Mitigating Propensity Bias of Large Language Models for Recommender Systems}

\author{Guixian Zhang}
\orcid{0000-0002-7632-8411}
\author{Guan Yuan$^{\ast}$}
\orcid{0000-0003-3148-9817}
\affiliation{%
	\institution{School of Computer Science and Technology, Mine Digitisation Engineering Research Center of the Ministry of Education, China University of Mining and Technology}
	\city{Xuzhou}
	\state{Jiangsu}
	\country{China}
}

\author{Debo Cheng$^{\ast}$}
\orcid{0000-0002-0383-1462}
\author{Lin Liu}
\orcid{0000-0003-2843-5738}
\author{Jiuyong Li}
\orcid{0000-0002-9023-1878}
\affiliation{\institution{UniSA STEM, University of South Australia}
	\city{Adelaide}
        \state{SA}
	\country{Australia}}

\author{Shichao Zhang}
\orcid{0000-0003-4052-1823}
\affiliation{%
	\institution{Guangxi Key Lab of Multi-source Information Mining \& Security,
Guangxi Normal University}
	\city{Guilin}
	\state{Guangxi}
	\country{China}
}
\thanks{*Corresponding authors}


\authorsaddresses{%
Authors’ address: Guixian Zhang, Guan Yuan (corresponding author), School of Computer Science and Technology, Engineering Research Center of Mine Digitalisation, Artificial Intelligence Research Institute, China University of Mining and Technology, Xuzhou 221116, China; emails: \{guixian, yuanguan\}@cumt.edu.cn; Debo Cheng (corresponding author), Lin Liu, Jiuyong Li, UniSA STEM, University of South Australia, Adelaide 5095, Australia; emails:\{debo.cheng, lin.liu, jiuyong.li\}@unisa.edu.au; Shichao Zhang, Guangxi Key Lab of Multi-source Information Mining \& Security, 
Guangxi Normal University, Guilin 541004, China; email: zhangsc@mailbox.gxnu.edu.cn.
}
\renewcommand{\shortauthors}{Zhang et al.}

\begin{abstract}
The rapid development of Large Language Models (LLMs) creates new opportunities for recommender systems, especially by exploiting the side information (e.g., descriptions and analyses of items) generated by these models. However, aligning this side information with collaborative information from historical interactions poses significant challenges. The inherent biases within LLMs can skew recommendations, resulting in distorted and potentially unfair user experiences. On the other hand, propensity bias causes side information to be aligned in such a way that it often tends to represent all inputs in a low-dimensional subspace, leading to a phenomenon known as dimensional collapse, which severely restricts the recommender system's ability to capture user preferences and behaviours. To address these issues, we introduce a novel framework named \textbf{C}ounterfactual \textbf{LLM} \textbf{R}ecommendation (CLLMR). Specifically, we propose a spectrum-based side information encoder that implicitly embeds structural information from historical interactions into the side information representation, thereby circumventing the risk of dimension collapse. Furthermore, our CLLMR approach explores the causal relationships inherent in LLM-based recommender systems. By leveraging counterfactual inference, we counteract the biases introduced by LLMs. Extensive experiments demonstrate that our CLLMR approach consistently enhances the performance of various recommender models.
\end{abstract}



\received{20 February 2007}
\received[revised]{12 March 2009}
\received[accepted]{5 June 2009}

\maketitle

\section{Introduction}
Recommender systems are essential tools used on many platforms to help users navigate the vast amount of available information and find items that match their preferences~\cite{gao2023enhanced,wang2018modeling,wu2023faster}. Due to the information overload problem~\cite{zhang2017learning,he2020lightgcn,deng2022multi}, collaborative filtering-based recommender systems have become widely used. The main research direction in this field focuses on improving the performance of these models by increasing the effectiveness of user and item representations~\cite{zhang2019deep}. To achieve this, many models have been proposed to mine useful information to enhance these representations. Graph neural networks play an important role in this endeavor~\cite{wei2022causal,sun2022graph,wu2024graph,zhang2024bayesian}. Recently, graph neural network-based recommender systems modeling user preferences based on historical interactions have achieved promising results~\cite{wu2022graph,liu2022federated,wei2022heterogeneous,bao2025spatial,SSGCC,Epi_Survey_KDD24}. Nevertheless, collaborative filtering models still face the problem of data sparsity~\cite{cailightgcl,yu2023xsimgcl}.

Researchers have argued that relying solely on interaction information may ignore other valuable data~\cite{gao2023enhanced,wang2020ckan}, such as rich information related to users and items in the real world. The lack of additional information can lead to a less informative representation learning. Previous research has demonstrated the importance of side information, also known as user and item descriptions, for collaborative filtering recommender systems to perform effective representation learning~\cite{wei2022causal, zhang2023knowledge, ren2024representation}. However, these approaches tend to be based on domain-specific knowledge and are not sufficiently generalised for recommendation tasks that they do not know much about~\cite{peng2023knowledge}.

\begin{figure}[ht]
  \centering
\includegraphics[width=0.46\textwidth]{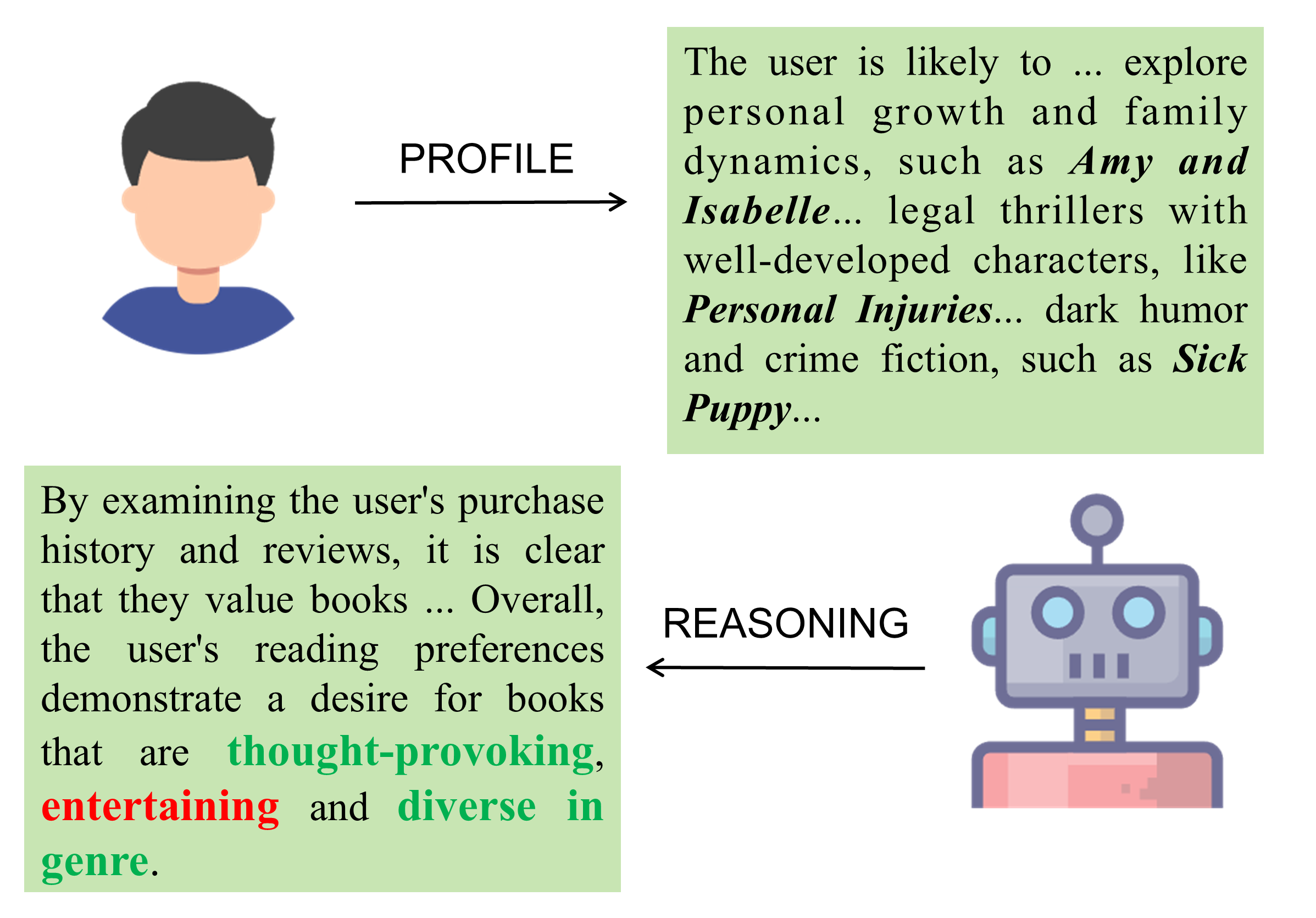}
  \caption{An example illustrating the introduction of propensity bias when using LLMs to generate profiles and reason about user 5596 in the Amazon dataset.}
  \label{example}
\end{figure}


Recently, Large Language Models (LLMs) have garnered attention due to their rich knowledge and powerful reasoning capabilities, with numerous studies~\cite{lu2024miko,wei2024llmrec,rong2024llm} showcasing their superior performance. These LLMs, equipped with the ability to process vast amounts of text data, offer promising opportunities for enhancing traditional recommender systems. However, directly deploying off-the-shelf LLMs for recommendation tasks presents several challenges~\cite{wei2024llmrec,ren2024representation,zhang2024text,lu2024aligning}. A key issue lies in the inherent mismatch between the generalisation capabilities of LLMs and the specific demands of personalised recommendation, particularly when it comes to capturing the nuanced preferences and behaviors of individual users. To bridge this gap, researchers have attempted to leverage LLMs to model human-like descriptions of users and items, thereby obtaining richer side information. The enriched data can then be aligned with collaborative representations learned from users' historical interactions via techniques such as contrastive learning~\cite{ren2024representation}, representing a promising direction for integrating the reasoning capabilities of LLMs with the precision required for personalised recommendations.

However, existing research has pointed out that the integration of LLM may bring new issues and challenges~\cite{ma2024causal,wu2024causality,Liu2024LargeLM}. Specifically, the inherent biases in LLM training datasets~\cite{jiang2024item} inevitably introduce propensity bias, which may generate inaccurate or irrelevant content. Specifically, LLM may exhibit biases toward certain cultures and viewpoints~\cite{tao2023auditing,NEURIPS2023_ae9500c4,sivaprasad2024exploring}, but not all propensity biases are harmful. Real-world knowledge exists to favour certain items possibly because they possess good intrinsic properties, and blindly removing this part of the propensity bias (named \textit{user/item propensity bias} in this paper) would make the model lose such real-world knowledge. However, the propensity bias which is not related to users/items (named \textit{user/item unrelated propensity bias} in this paper) can result in overly generalized representation, ignoring user and item specific details. As shown in Figure~\ref{example}, a user with ID 5596 in the Amazon dataset~\cite{he2016ups} has previously read a range of thrilling and exciting books, yet the LLM overlooks this in its reasoning and generalises the user's preference as  simply ``entertaining''.

\begin{figure*}[t]
  \centering
  \subfloat[Amazon]{\includegraphics[width=0.35\textwidth]{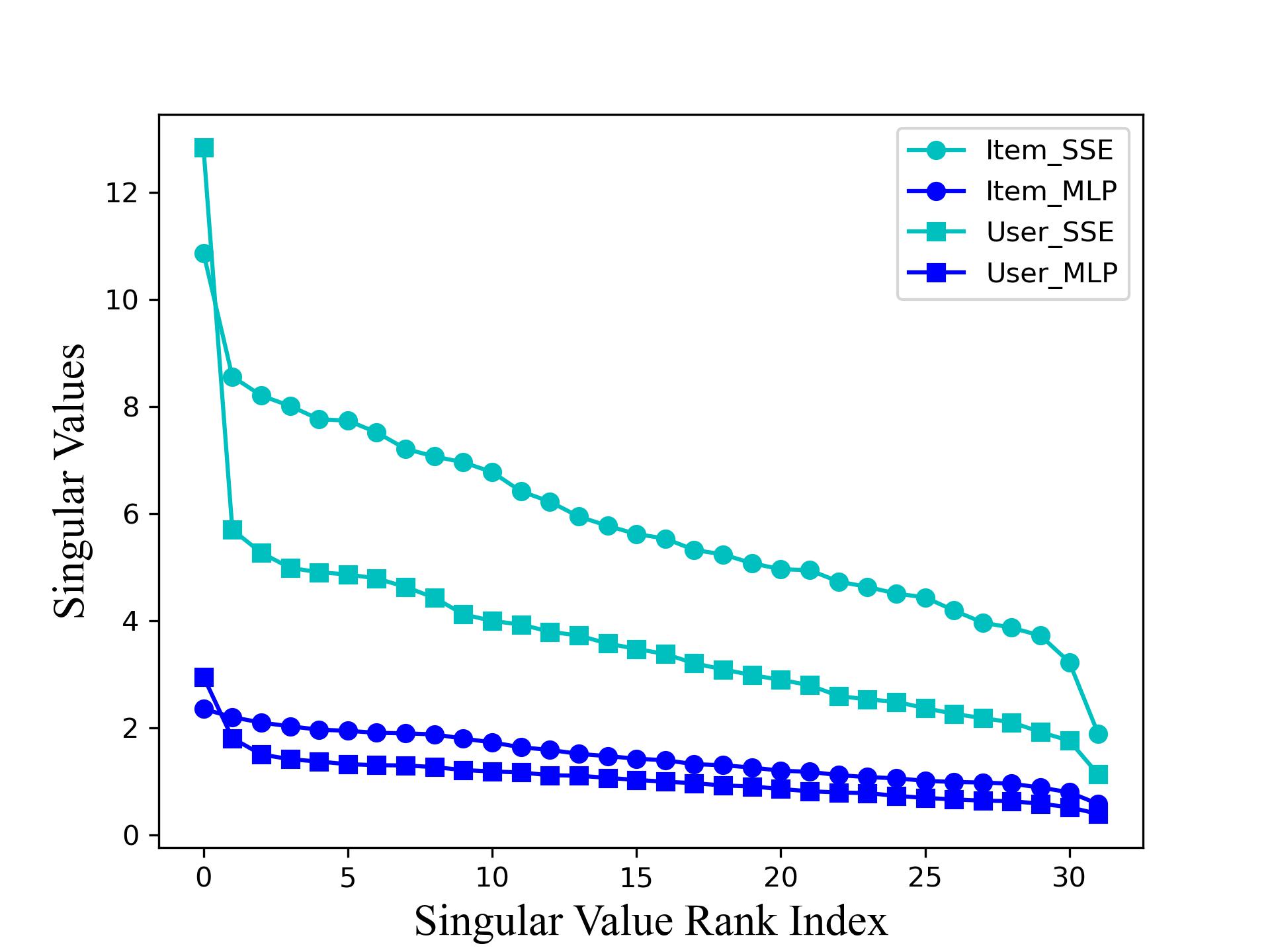}}
\subfloat[Yelp]{\includegraphics[width=0.35\textwidth]{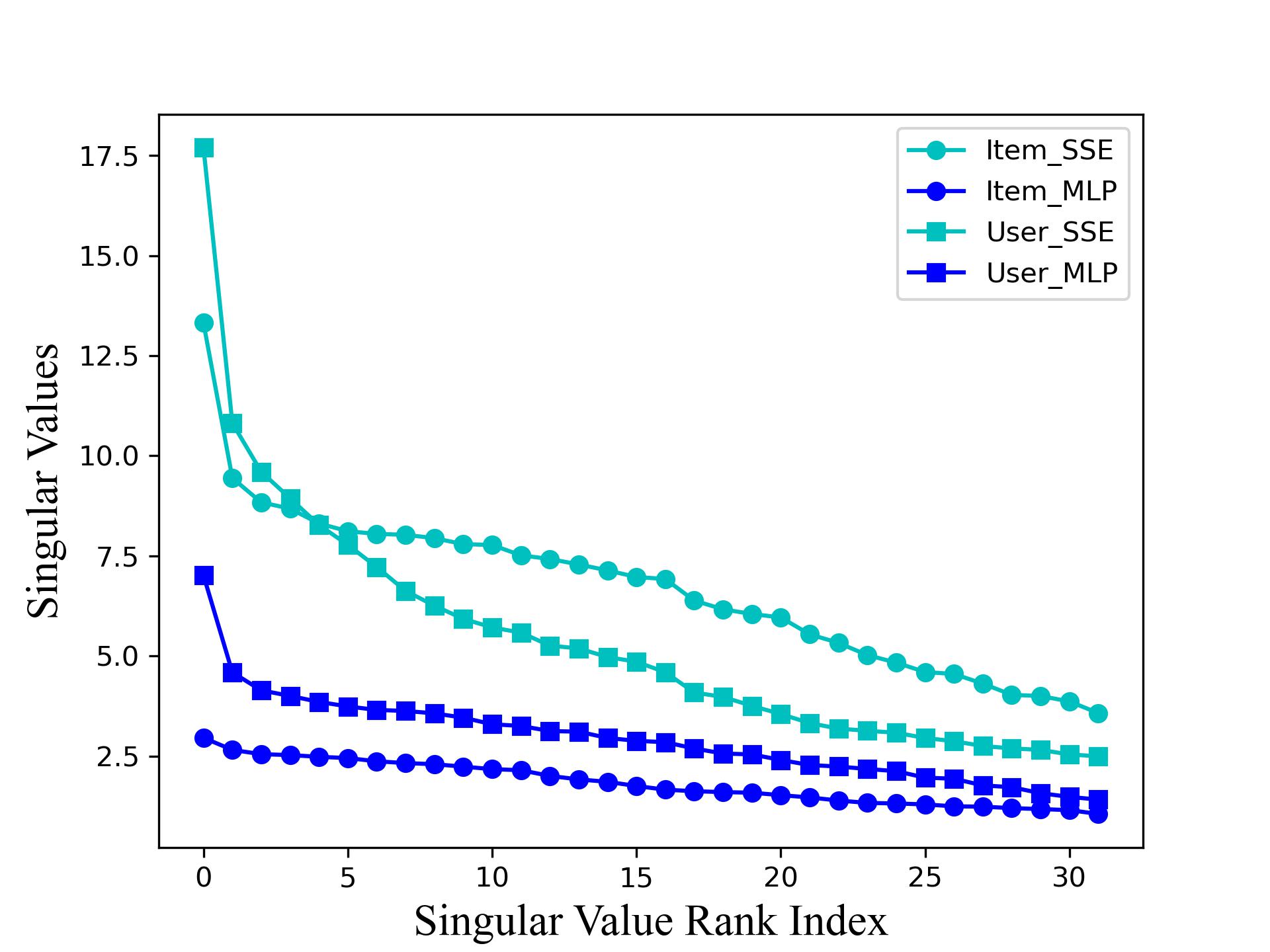}}
  \subfloat[Steam]{\includegraphics[width=0.35\textwidth]{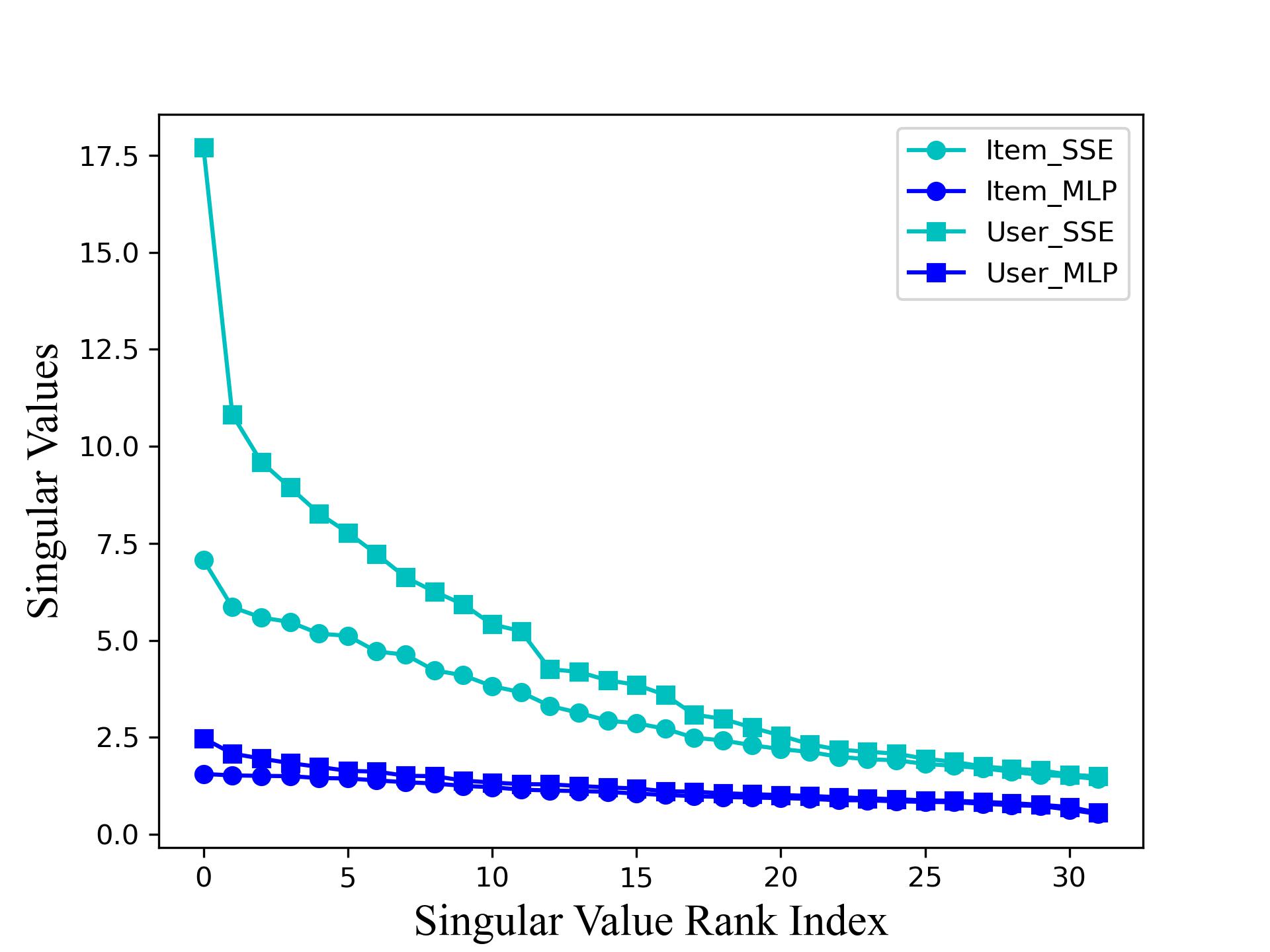}}
  \caption{The singular values of the trained representations show that directly aligning the collaborative information and the side information of LLMs often leads to dimensional collapse, whereby the singular values of each dimension are very low and similar, losing the distinction between different elements. However, our proposed SSE method effectively mitigates this issue and guarantees the quality of the representation.}
  \label{svdexample}
\end{figure*}

Furthermore, the unrelated propensity bias not only impacts recommendation accuracy but also contributes to the homogenisation of LLM-generated representations~\cite{barbero2024transformers}, which increases the risk of dimensional collapse. When using LLMs~\cite{guoembedding} or contrastive learning~\cite{jing2022understanding}, side representations are more likely to be restricted to a certain portion of the low-dimensional subspace, futher limiting their representational capability~\cite{zhang2023mitigating}. As shown in Figure.~\ref{svdexample}, when using LightGCN as the backbone method to align LLMs for recommendation through contrastive learning, the model exhibits dimensional collapse when side information is encoded using MLP. This occurs because the side information repsentation generated by the LLM lacks structural information, making it difficult to align with collaborative information derived from historical interactions, thereby exacerbating the risk of dimensional collapse.

To address these issues, we propose a novel LLM-based recommender system, named \textbf{C}ounterfactual \textbf{L}arge \textbf{L}anguage \textbf{M}odel \textbf{R}ecommendation (\textbf{CLLMR}). Specifically, we propose a \textbf{S}pectrum-based \textbf{S}ide information \textbf{E}ncoder (\textbf{SSE}) that incorporates structural information of historical interactions while capturing the rich information provided by LLMs. SSE effectively learns the various factors within the data, preventing dimensional collapsing during the alignment process. Additionally, we inject small amounts of noises into the spectrum to enhance the generalisation ability of the SSE and increase the diversity of side information. To mitigate the propensity bias introduced by large models, we employ causal inference techniques to evaluate the effect of LLM propensities on node representations and eliminate this bias through counterfactual inference. The contributions of the paper can be summarised as follows:

\begin{itemize}
    \item We propose SSE that implicitly captures structural information from the rich knowledge generated by LLMs while preventing dimensional collapse during the alignment process.
    \item We explore a causal perspective on propensity bias in LLM-based recommender systems and develop a counterfactual LLM framework (CLLMR) to train recommender systems, enabling counterfactual inference based on the designed causal graph to mitigate the propensity bias caused by LLMs during the recommendation inference stage.
    \item Comprehensive experiments on three real-world datasets demonstrate the effectiveness of the proposed CLLMR framework compared to state-of-the-art LLM recommendation alignment methods.
\end{itemize}

The rest of the paper is structured as follows: In Section II we present related work. Section III describes the preliminary work. Section IV provides a detailed description of the proposed method. Extensive experiments are presented in Section V. Finally,  Section VI concludes the paper.

\section{Related Work}

In this section, we discuss the related work, including the research on bias in artificial intelligence and recommender systems with side information.

\subsection{Bias in Artificial Intelligence}

With the widespread use of artificial intelligence in real life~\cite{deng2016efficient,deng2021pulse,deng2023tts,zhang2022hyper,zhang2023knn,zhang2022challenges}, bias in artificial intelligence has attracted a great deal of attention~\cite{zhang2023fpgnn, ling2024fair}. In a causal inference perspective~\cite{wang2023sound, li2024distribution}, bias is often due to spurious correlations caused by confounders~\cite{zhang2025disentangled}. For example, Buolamwini and Gebru~\cite{pmlr-v81-buolamwini18a} showed that business gender classification systems used in various fields such as marketing, entertainment and healthcare have different representations of users' skin colour and gender. Zhang et al.~\cite{zhang2024learning} demonstrated that the interaction structure in social networks can also contain preference information. It follows that artificial intelligence systems may display biases against specific groups embedded in the data or even amplify human biases. Natural Language Processing (NLP) systems are susceptible to bias as a result of the training corpus. Bolukbasi et al.~\cite{bolukbasi2016man} and Caliskan et al.~\cite{caliskan2017semantics} first found the presence of human-like preferences in word embedding models. Zhao et al.~\cite{zhao2018gender} found a correlation between gender pronouns and stereotypes. De-Arteaga et al.~\cite{de2019bias}'s study of semantic representation bias can lead to potential harm. 

Bias is particularly evident in generative models, where LLMs tends to amplify biases against specific groups. For example, Abid et al.~\cite{abid2021persistent} found that LLMs commonly associated Muslims with violence. Lucy and Bamman~\cite{lucy2021gender} found that female roles tended to be associated with family and emotions, and male roles tended to be associated with politics, sports, and crime. Lee et al.~\cite{lee2024large} found that ChatGPT also portrays women as more homogeneous than men. Psychologically, people tend to perceive their out-group members as more homogeneous than their in-group members, and LLMs inevitably outputs homogeneous representations of the same group~\cite{lee2024large}. Cheng et al.~\cite{cheng2023compost} found that LLMs tends to essentialise the characters of marginalised groups and produce positive and homogeneous narratives. In this paper, we collectively refer to these bias as LLMs' propensity bias, as there are not only human attributes in the recommendation scene, but also factors such as region and culture.

\subsection{Recommender Systems with Side Information}
The fundamental concept of recommender systems is to leverage user-item interactions and associated side information to compute matching scores between users and items~\cite{zhang2023knowledge, gao2023enhanced}. More specifically, various recommender models are designed based on collaborative behavior between users and items, which are then used to learn user and item representations~\cite{ma2022mixed}. Collaborative information modeling plays a crucial role in personalised recommendation~\cite{qin2024learning}, and deep neural networks have been widely adopted to advance recommender systems due to their superior representation learning capabilities across various domains~\cite{zhang2019deep}. Considering users' online behavior as graph-structured data, graph neural networks have been applied to recommender systems as an advanced representation learning technique for capturing user and item representations~\cite{wu2022graph}. At the same time, researchers have focused on enhancing recommender system effectiveness by addressing factors like popularity bias~\cite{yu2023xsimgcl, cailightgcl} and exposure bias~\cite{wang2022towards} in historical data.

Despite the significant success of these approaches, ID-based methods still have inherent limitations. The reason is that pure ID indexes of users and items are naturally discrete and lack the semantic information needed to effectively capture user and item representations, especially in cases of sparse user-item interactions~\cite{yuan2023go}. In recent years, the impressive natural language processing capabilities of LLMs have inspired researchers to explore their application in recommendation tasks. Early implementations utilised language models as feature extractors to create knowledge-aware recommendation embeddings~\cite{wu2022userbert}, providing valuable insights into user preferences. 

The adaptation of LLMs to recommendation scenarios has primarily relied on techniques like fine-tuning and in-context learning. Wang et al.~\cite{wang2023rethinking} utilised ChatGPT for session-based recommendations via zero-shot prompting. Agent4Rec~\cite{zhang2024generative} employs LLM-enabled generative agents to simulate users using user side information specifically tailored for recommender systems. Geng et al. introduced a unified framework integrating five recommendation tasks by fine-tuning LLMs. TALLRec~\cite{bao2023tallrec} fine-tunes LLMs for natural language generation tasks via instruction fine-tuning. However, the direct use of LLMs in recommendation tasks faces challenges such as high computational costs and slow inference times. Ren et al.~\cite{ren2024representation} used mutual information maximisation to combine LLM knowledge with collaborative relationship modeling. Lu et al.~\cite{lu2024aligning} employed reinforcement learning to enhance LLMs' ability to respond to user intent and mitigate formatting errors. Zhang et al.~\cite{zhang2024text} converted collaborative information from external models into binary sequences for alignment with LLMs. 

While these approaches improve the performance of LLM-based recommender systems, they often overlook the propensity bias inherent in these models. Different from these previous works, our proposed CLLMR framework specifically addresses the pro bias of LLM  on recommender systems and aims to enhance the overall performance of LLM-based recommender systems.

\section{Preliminary}
In this section, we present our problem setting and introduce the basic concepts of graphical causal models, along with key definitions in causal inference.

\subsection{Problem Setting}

With an existing recommendation method $\mathcal{R}$,  we can obtain the user collaborative representation $\mathbf{\hat{e}}_{u}$ and item collaborative representation $\mathbf{\hat{e}}_{i}$ based on the historical interactions between the user $\mathcal{U}$ and the item $\mathcal{I}$, and then compute the probability that $\mathcal{U}$ and $\mathcal{I}$ will interact with each other in the future. Concurrently, we leverage an LLM to generate user side information $\mathcal{P}_{u}$ and item side information $\mathcal{P}_{i}$, which are then projecting into the feature space via a text embedding model $\mathcal{T}$. An encoder is subsequently utilised to learn the user side representation $\mathbf{\hat{s}}_{u}$ and item side representation $\mathbf{\hat{s}}_{i}$. Finally, these side representations $\mathbf{\hat{s}}_{u}$ and $\mathbf{\hat{s}}_{i}$ are aligned with the collaborative representations $\mathbf{\hat{e}}_{u}$ and $\mathbf{\hat{e}}_{i}$ through contrastive learning, enhancing the overall recommendation performance of the model. However, LLM produces side information often with propensity bias, which has an impact on recommender systems. The aim of this paper is to improve the effectiveness of LLM for recommender systems by mitigating the unrelated propensity bias. A summary of the notations used in this paper is shown in Table~\ref{notation}.

\begin{table}[ht]
    \centering
    \caption{A summary of the notations used in this paper.}
    \begin{tabular}{c p{0.8\textwidth}}
    \hline
    Notations & Description \\
    \hline
    $\mathcal{P}_{u}$, $\mathcal{P}_{i} $     & the LLM-generated user side information and item side information \\
    $\mathbf{\hat{s}}_{u}$ $\mathbf{\hat{s}}_{i}$   & the user side text embedding and the item side text embedding\\

    $\mathcal{T}$ & the text embedding model\\
 
    $\mathbf{s}_{u}$ $\mathbf{s}_{i}$ & the user side representation and the item side representation learned by SSE\\
    $\hat{M}$ & the spectrum of the historical interaction information\\
    $M$ & the spectrum information added with random noise\\
    $f_{P}$  & the estimated model for the propensity bias\\
    $\sigma$ &  the sigmoid function\\
    $\mathbf{\hat{e}}_{u}$, $\mathbf{\hat{e}}_{i}$ & the user collaborative representation and the item collaborative representation obtained by traditional recommendation methods \\
    $\mathbf{e}_{u}$, $\mathbf{e}_{i}$ & the user collaborative representation and the item collaborative representation affected by propensity bias in the alignment process with side information brought by LLM \\
    \hline
    \end{tabular}
    \label{notation}
\end{table}

\subsection{Graphical Causal Model}
\begin{figure}[ht]
    \centering
    \includegraphics[width=0.45\linewidth]{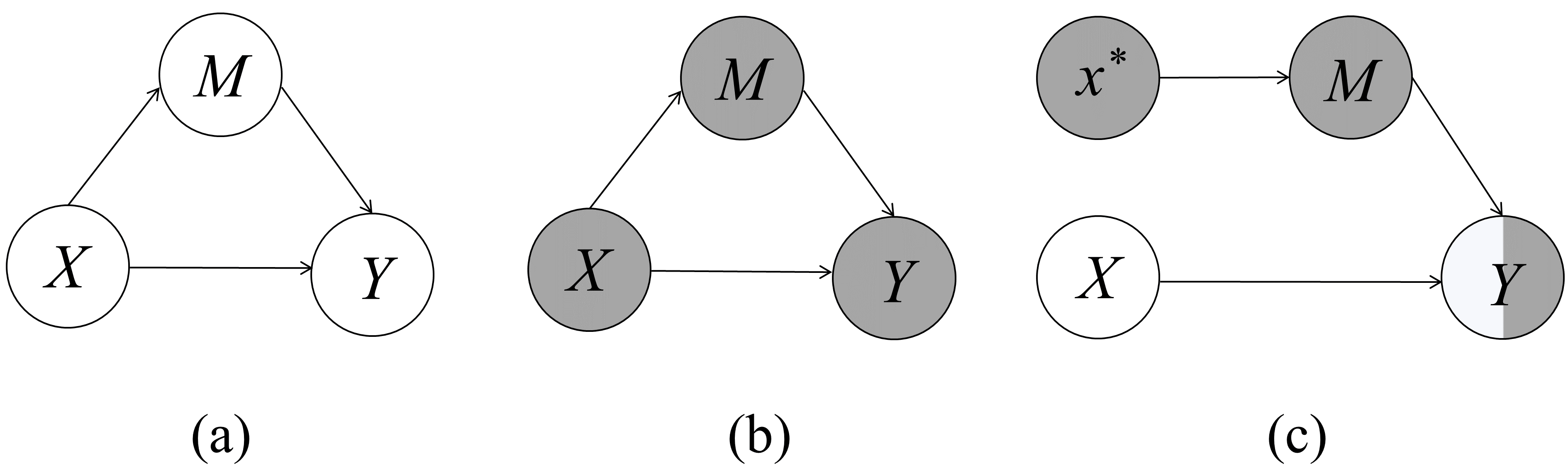}
    \caption{Example of causal DAG where $X$, $Y$, and $M$ denote exposure, outcome and mediator variables, respectively. Gray nodes indicate that the variables are set to reference values (e.g., $X$ = $x^{*}$), the half shaded node represents the result affected by the reference value. In this figure, (a) represents the causal graph for the factual scenario, (b) shows the causal graph where X is set to the reference state, and (c) illustrates the causal graph for the counterfactual scenario, where X is set to different values.}
    \label{causal1}
\end{figure}

In a graphical causal model, causal relationships between variables are typically represented using a directed acyclic graph (DAG)~\cite{pearl2009, cheng2024data}. A DAG is defined as $G= (V, E)$, where  $V$  denotes the set of variables and $E$ is the set of edges and they represents the cause-effect relations among these variables. We use capital letters (e.g. $X$) represent variables, while lowercase letters (e.g. $x$) denote the observed values of those variables. A directed edge from node $X$ to node $Y$, i.e., $X \rightarrow Y$, indicates that $X$ is a direct cause of $Y$. Indirect causal effects can also be represented using paths in a DAG. For example, the path $X \rightarrow M \rightarrow Y$ in Figure.~\ref{causal1} (a) indicates that $X$ has an indirect effect on $Y$, with $M$ acting as a mediator variable. Analyses of mediator variables require assumptions based on the temporal priorities of exposure, mediator, and outcome~\cite{rijnhart2021mediation}, which means that changes in exposure are assumed to precede changes in the mediator, and changes in the mediator are assumed to precede changes in the outcome.

\subsection{Causal Effect and Counterfactual Inference}
\label{Causal Effect}
The causal effect of the exposure, e.g. $X$ on the outcome, e.g. $Y$ is defined as the difference between two counterfactual outcomes that would be observed if an individual were exposed to different levels of the exposure of interest~\cite{hicks2011causal, imai2010general, cheng2022cau, xu2023disentangled}. In formal terms, if we denote the outcome as $ Y $, and the exposure values of interest as $ x $ and $ x^* $, the causal effect for an individual $ j $ is $ Y_j(x) - Y_j(x^*) $, where $ Y_j(x) $ is the counterfactual outcome for individual $ j $ under exposure level $ x $, and $ Y_j(x^*) $ is the counterfactual outcome under exposure level $ x^* $. 

To ensure a valid causal interpretation of the Pure Natural Direct Effect (PNDE) and Total Natural Indirect Effect (TNIE)~\cite{pearl2009, rijnhart2021mediation} at the population-average level, the following four assumptions must hold~\cite{pearl2009,rijnhart2021mediation,cheng2022cau,imbens2015causal, xu2023disentangled}:

\begin{itemize}
    \item \textbf{No unmeasured confounding of the exposure-outcome effect}: All confounders that affect both the exposure \( X \) and the outcome \( Y \) must be observed and appropriately adjusted for. This ensures that the direct effect of \( X \) on \( Y \) is not confounded by unmeasured variables.
    \item \textbf{No unmeasured confounding of the mediator-outcome effect}: All variables influencing the mediator \( M \) and the outcome \( Y \) must be measured. This ensures that the indirect effect of \( X \) on \( Y \) via \( M \) is not biased by hidden confounders. 
    \item \textbf{No unmeasured confounding of the exposure-mediator effect}: Every variable that confounds the relationship between the exposure \( X \) and the mediator \( M \) must be observed and accounted for. This ensures that the causal effect of \( X \) on \( M \) is not confounded by unmeasured variables. 
    \item \textbf{No confounders of the mediator-outcome effect that are affected by the exposure}: This assumption requires that no variables confounding the mediator-outcome relationship \( M \rightarrow Y \) are themselves influenced by the exposure \( X \). 
\end{itemize}

The total effect (TE)~\cite{pearl2009, rijnhart2021mediation} is the overall change in the outcome that would be observed if an individual's exposure level were changed $x$, taking into account causal paths from the exposure to the outcome, including both direct path and indirect paths via mediators. Mathematically, it is represented as:

\begin{equation}
TE = Y_j(x, M_j(x)) - Y_j(x^*, M_j(x^*)),    
\end{equation}
where $Y_j(x, M_j(x))$ is the counterfactual outcome for individual $j$ under exposure level $x$ and the mediator level that would be observed under the same exposure level, and $Y_j(x^*, M_j(x^*))$ is the counterfactual outcome under exposure level $x^*$ and the mediator level that would be observed under exposure level $x^*$.

The TE is crucial as it provides a complete picture of the impact of an exposure on an outcome, incorporating both the direct effects and the indirect effects mediated through a mediator. The counterfactual results in the mediator model depend not only on the exposure value but also on the mediator value. Counterfactual inference involves hypothetically intervening to change the value of a single variable and imagining the consequences of this change on other variables within the system. As illustrated in Figure~\ref{causal1} (b), the grey nodes indicate that the corresponding variable has been set to a reference value independently of the facts. As shown in Figure~\ref{causal1} (c), the introduction of reference states helps to divide the causal graph into different parts, i.e. to realize a counterfactual scenario. This scenario is termed a `counterfactual scenario' because it entails the simultaneous consideration of both a fact state, denoted by $X=x$, and a reference state, represented by $X=x^{*}$. 

The PNDE is the effect of the exposure on the outcome that occurs without any influence from the mediator:

\begin{equation}
    PNDE = Y_j(x, M_j(x^*)) - Y_j(x^*, M_j(x^*)).
\end{equation}

When we want to explore the effects through the mediator variable path after removing the direct effect, we can subtract the PNDE from the TE to obtain the TNIE:

\begin{equation}
TNIE = TE - PNDE = Y_j(x, M_j(x)) - Y_j(x, M_j(x^*)),
\end{equation}
where $Y_j(x, M_j(x))$ is the potential outcome for individual $j$ under exposure level $x$ and the mediator level that would be observed under exposure level $x$, and $Y_j(x, M_j(x^*))$ is the counterfactual potential outcome under exposure level $x$ but with the mediator level that would be observed under exposure level $x^*$.

\section{The Proposed CLLMR Framework}
\begin{figure*}[t]
\centering
\includegraphics[width=0.99\textwidth]{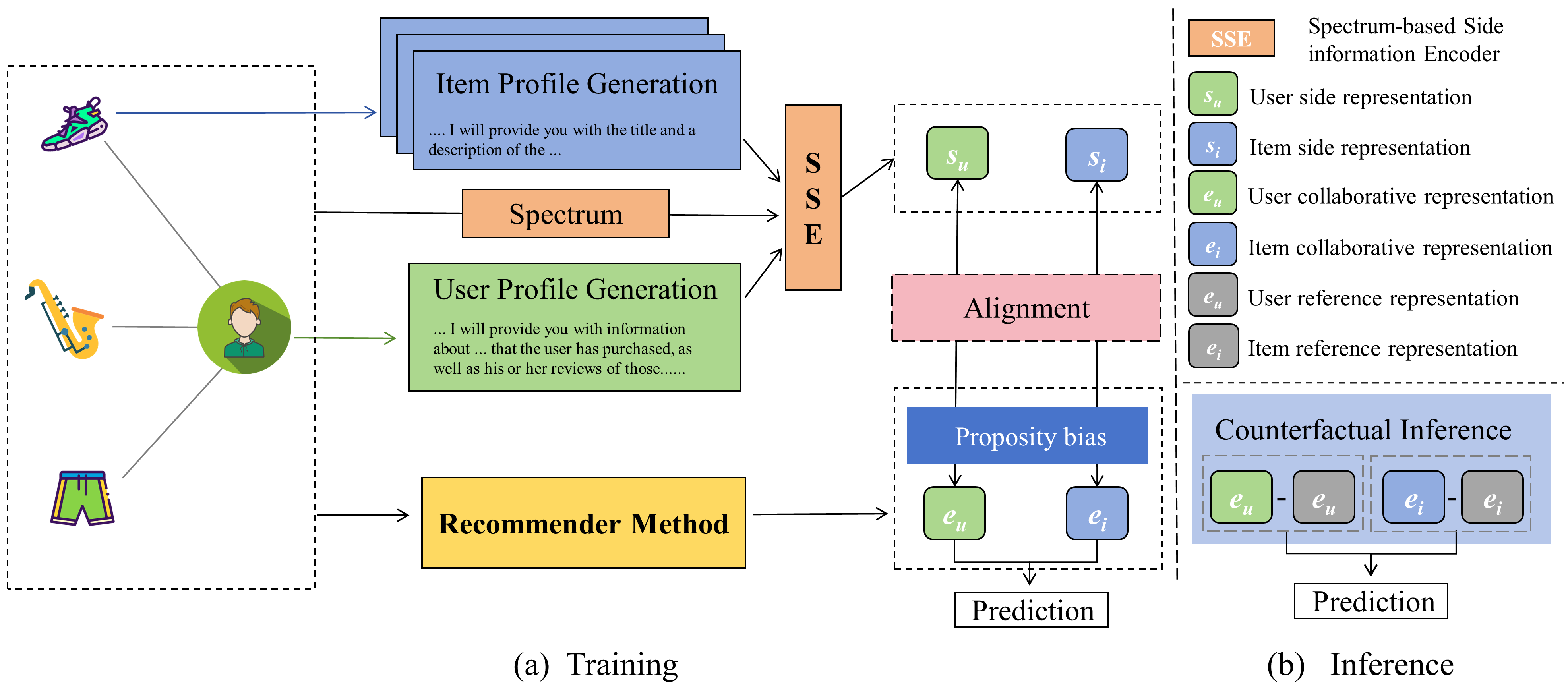} 
\caption{An overview of our proposed CLLMR framework for training LLM-based recommender systems and performing counterfactual inference using causal graphs to eliminate LLM propensity bias during the recommendation inference stage.}
\label{framework}
\end{figure*}

We provide an overview of our CLLMR framework in Figure.~\ref{framework}. The CLLMR framework is divided into two stages: training and inference. In the training stage, we focus on real-world causality and account for the influence of propensity bias on the model. Specifically, we first use an LLM to construct side information for both users and items, then propose SSE to introduce structural information into the representation of this side information. Finally, we align the side information with collaborative information affected by propensity bias. In the inference stage, we perform model inference based on causal relationships in the counterfactual world, thereby mitigating propensity bias. In this section, we describe each component of the framework in detail.

\subsection{Constructing Side Information}
Previous research has demonstrated the importance of side information~\cite{wei2022causal, zhang2023knowledge, ren2024representation}, also known as descriptions of users and items, for collaborative filtering recommender systems to perform effective representation learning. Specifically, user and item edge information that effectively characterises preferences and attractiveness is crucial for modeling latent relationships between users and items.

In this work, we utilise an LLM to infer about collaborative signals and generate useful side information, following~\cite{ren2024representation}. Items often have more freely available descriptive information compared to privacy-sensitive user profiles. Thus, our method first applies a uniform prompt to a pretrained LLM to produce an initial description summarizing each item based on its profile data. These item descriptions are then used to infer side information for both users and items via collaborative reasoning. Specifically, the LLM analyzes the descriptions of all items with which a given user has interacted with to formulate a description $\mathcal{P}_{u}$ characterizing that user's preferences. Similarly, the LLM analyses which types of users interacted with a given item to formulate a description $\mathcal{P}_{i} $ characterizing the item's attractiveness to different audiences. 

Compared to simply aggregating sparse profile fields, our approach allows the LLM to leverage its powerful language understanding and commonsense knowledge to generate more informative and coherent side information. After obtaining the textual description of the side information given by the LLM, it is necessary to transform the text into embeddings using the following formulas:

\begin{equation}
\mathbf{\hat{s}}_{u}=\mathcal{T}\left(\mathcal{P}_{u}\right), \quad \mathbf{\hat{s}}_{i}=\mathcal{T}\left(\mathcal{P}_{i}\right),
\end{equation}
where $\mathbf{\hat{s}}_{u}$ is the user side text embedding, $\mathbf{\hat{s}}_{i}$ is the item side text embedding, and $\mathcal{T}$ stands for the text embedding model~\cite{su2023one}.

\subsection{Spectrum-based Side Information Encoder}
When dealing with LLMs, side information often lacks structural information of historical interactions and may be affected by biases inherent in the corpus. These biases can lead to propensity bias~\cite{de2019bias,NEURIPS2023_ae9500c4,zhang2023mitigating}, which in turn trigger dimensional collapse during the alignment of side information with the collaborative information of the recommendation algorithm. To address this issue, we propose a \textbf{S}pectrum-based \textbf{S}ide information \textbf{E}ncoder (SSE) to reduce the text embedding of side information into a low-dimensional representation. Our SSE method leverages the identifiable variational autoencoder (iVAE)~\cite{khemakhem2020ivae} to enhance the identifiability of side information and implicitly introduce structural information from historical interactions.
 
The set of parameters for SSE is defined as $\boldsymbol{\theta}=(f, T, \lambda)$.  In SSE, we introduce a conditional prior distribution $p_{T, \lambda}(Z \mid M)$, where $Z$ denotes the latent variables and $M$ represents additional observed variables. In our SSE framework, the spectrum of historical interaction information is derived through singular value decomposition (SVD). However, users may occasionally click by mistake when performing an action, and overfitting historical interaction data may result in overfitting, which may not accurately capture user preferences. To avoid overfitting and increase the generalisation capabilities of the model, we add a small amount of noise to the spectrum:

\begin{equation}
   M = \hat{M} +  \Delta, \quad   \Delta=\omega \odot \operatorname{sign}\left(\hat{M}\right),
\end{equation}
where $\omega \in \mathbb{R}^{d} \sim U(0,1)$ and $\hat{M}$ is the original spectrum, thus ensuring that $\hat{M}$ and $\Delta$ have the same tendency.

This conditional prior enables the model to learn the distribution of the latent variable $Z$ given $M$, thereby facilitating the downscaling of the side information $X$. Specifically, we consider the following conditional generative model:
\begin{equation}
p_{\boldsymbol{\theta}}(X, Z \mid M)=p_{f}(X \mid Z) p_{T, \lambda}(Z \mid M).
\end{equation}

To elaborate, the model's structure comprises two components: the likelihood $p_{f}(X \mid Z) $ and the conditional prior $p_{T, \lambda}(Z \mid M)$. The value of  $X$  can be decomposed as  $X=f(Z)+\boldsymbol{\varepsilon}$  where  $\boldsymbol{\varepsilon}$  is an independent noise variable. In this work, we assume that the function  $f: \mathbb{R}^{n} \rightarrow \mathbb{R}^{d}$ is injective.

Let the prior over the latent variables $p_{\boldsymbol{\theta}}(Z \mid M)$ be conditionally factorial, such that each element $z_{i} \in Z$ within the vector is governed by a univariate exponential family distribution conditioned on $M$. The conditioning relies on an arbitrary function $\lambda(M)$, which could be a neural network, to yield the parameters $\lambda_{i, j}$ of the exponential family for each element.

The probability density function is articulated as follows:
\begin{equation}
p_{T, \lambda}(Z \mid M)=\prod_{i} \frac{Q_{i}\left(z_{i}\right)}{N_{i}(m_i)} \exp \left[\sum_{j=1}^{k} T_{i, j}\left(z_{i}\right) \lambda_{i, j}(m_i)\right],
\end{equation}
where $Q_{i}\left(z_{i}\right)$ serves as the base measure for the exponential family distribution of each $z_{i}$. $N_{i}(m)$ is the normalizing constant, ensuring that the probabilities across the distribution sum to unity. $T(Z)=\left(T_{1, 1}(z_1), \ldots, T_{n, k}(z_n)\right)$ represents the sufficient statistics that encapsulate the relevant information from the data. $\lambda(M)=\left(\lambda_{1, 1}(M), \ldots, \lambda_{n, k}(M)\right)$ are the parameters derived from the conditioning variable $M$, which are pivotal in determining the characteristics of the distribution. In this paper, the exponential family distribution adopted here is the Gaussian distribution.

The true posterior, denoted as $p_{\boldsymbol{\theta}}(Z \mid X, M)$, is often computationally intractable; hence, we utilise an approximation $q_{\boldsymbol{\phi}}(Z \mid X, M) $ to facilitate our modeling process.

The conditional marginal distribution of the observations is defined by integrating over the latent variables, as expressed by: 
\begin{align}
p_{\boldsymbol{\theta}}(X \mid M)=   \int p_{\boldsymbol{\theta}}(X, Z \mid M) \mathrm{d} Z,
\end{align}
where $p_{\boldsymbol{\theta}}(X \mid M)$ captures the likelihood of the observations given the conditioning variable $M$, while the empirical data distribution derived from the dataset $\mathcal{D}$ is represented by $q_{\mathcal{D}}(X)$. The parameters $(\boldsymbol{\theta}, \boldsymbol{\phi})$ of the SSE are estimated by maximizing the Evidence Lower Bound (ELBO) $\mathcal{L}(\boldsymbol{\theta}, \boldsymbol{\phi})$, which serves as a surrogate for the data's log-likelihood. The ELBO is formulated as follows:
\begin{align}
\mathbb{E}_{q_{\mathcal{D}}}\left[\log p_{\boldsymbol{\theta}}(X \mid M)\right] \geq \mathcal{L}(\boldsymbol{\theta}, \boldsymbol{\phi}):=    
\mathbb{E}_{q_{\mathcal{D}}}\left[\mathbb{E}_{q_{\boldsymbol{\phi}}(Z \mid X, M)}\left[\log p_{\boldsymbol{\theta}}(X, Z \mid M)-\log q_{\boldsymbol{\phi}}(Z \mid X, M)\right]\right].
\end{align}

This formulation encapsulates the expected log-likelihood of the observations, adjusted by the Kullback-Leibler divergence between the true posterior and its variational approximation.

To address the challenge of sampling from the variational distribution $q_{\boldsymbol{\phi}}(Z \mid X)$, we employ the reparameterisation trick as done in~\cite{kingma2013auto}. This technique allows for the generation of samples that are used to construct a low-variance gradient estimator for the ELBO with respect to the variational parameters $\phi$. The training process aligns with the standard VAE approach, leveraging these gradients to optimise the model parameters. Eventually, we can get user side representation $\mathbf{s}_{u}$ and item side representation $\mathbf{s}_{i}$:

\begin{equation}
\mathbf{s}_{u} = \operatorname{SSE}\left( \mathbf{\hat{s}}_{u} \right), \quad \mathbf{s}_{i} = \operatorname{SSE}\left( \mathbf{\hat{s}}_{i} \right),
\end{equation}
where $\mathbf{\hat{s}}_{u}$ and $\mathbf{\hat{s}}_{i}$ are the text embedding of the side information.

\subsection{Counterfactual Debiased Recommendation}

\begin{figure}[ht]
    \centering
    \includegraphics[width=0.6\linewidth]{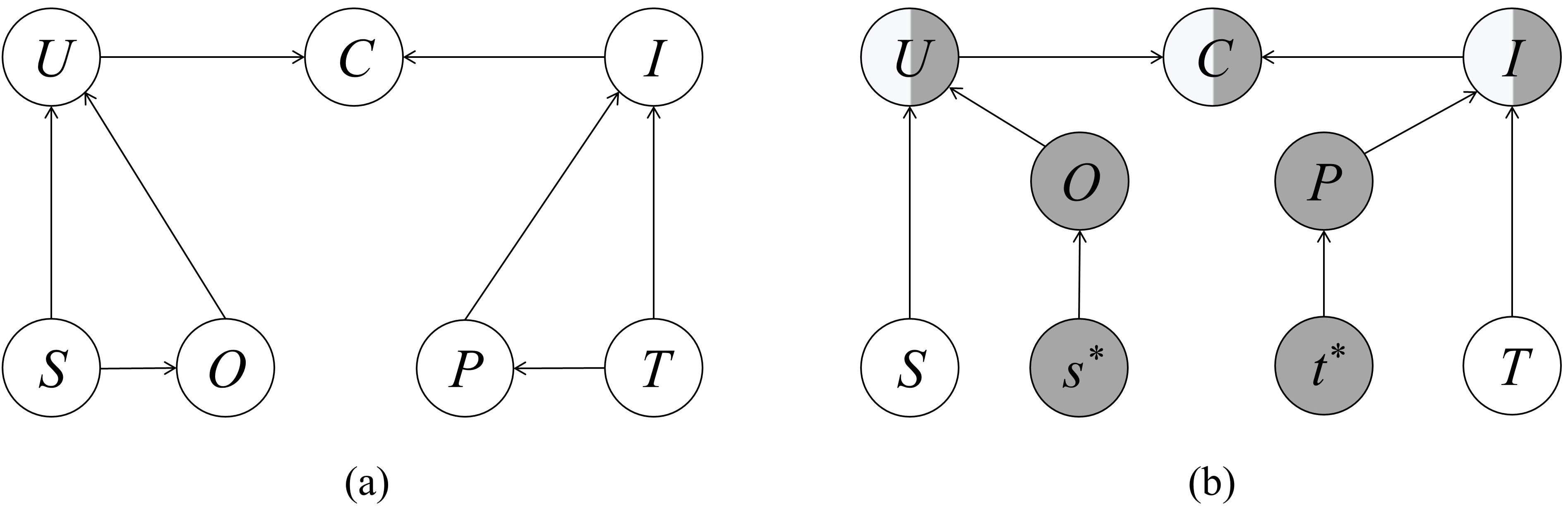}
    \caption{Figure (a) represents the causal map of the factual world, and Figure (b) represents the causal map of the counterfactual world. Gray nodes represent reference states, and half-shaded nodes represent that they are influenced by the reference state.}
    \label{causal2}
\end{figure}

In this work, we consider addressing the propensity bias from the influence of LLMs. In order to analyse the causality of the LLM-based recommender system, we build the causal graph as shown in Figure.~\ref{causal2} (a) to model the problem.  The causal graph explains the causes of propensity bias in LLM-based recommender system. The nodes in the causal graph $G$ are explained as follows:

\begin{itemize}
 \item $U$  represents user collaborative representation.
\item $I$ represents the item collaborative representation.
\item $C$ represents the choice of user and item matching.
\item $T$ represents the item side representation of the LLM to the recommender system.
\item $S$ represents the user side representation of the LLM to the recommender system.
\item $P$ represents the item propensity bias. 
\item $O$ represents the user propensity bias. 
\end{itemize}

The relationships between nodes in the causal graph $G$ are represented by edges, which are defined as follows:
\begin{itemize}
   \item $U \rightarrow C$ and $I \rightarrow C$ denote that both the user and the item collaborative representation have direct edges with $C$. This indicates that the user makes a choice of $C$ when the user's preference $U$ matches the attribute $I$ of the exposed item.

   \item $T \rightarrow P$ and $S \rightarrow O$ denote that part of the side representation is affected by the user/item propensity bias.

   \item $T \rightarrow I$ and $S \rightarrow U$ denote that the side representation is affected by both the user/item unrelated and user/item propensity biases, which directly affects the user/item collaborative representation.

   \item $P \rightarrow I$ and $O \rightarrow U$ denote that user/item side representations  affected by the user/item propensity bias are aligned with user/item collaborative representations to improve recommendation performance.
\end{itemize}

The aim of this paper is to mitigate the effect of user/item unrelated propensity bias on recommendations. Under the designed causal DAG, the problem of mitigating the unrelated propensity bias of the LLM becomes how to estimate the indirect effects of $O$ and $P$ on $U$ and $I$. Following the causal effect introduced in Section~\ref{Causal Effect}, we first construct the counterfactual-world causal graph shown in Figure~\ref{causal2} (b). Gray nodes represent reference states, and half-shaded nodes represent that they are influenced by the reference state. As we discussed in Section 3.3, to ensure a valid causal interpretation of the PNDE and TNIE in the causal graph in Figure~\ref{causal2}, the following assumptions must hold:

\begin{itemize}
    \item \textbf{No unmeasured confounding of the exposure-outcome effect}: All confounders that affect both the exposure \( S \) or \( T \) and the outcome \( U \) or \( I \) must be observed and appropriately adjusted for. This ensures that the direct effects are not confounded by unmeasured variables. 
    \item \textbf{No unmeasured confounding of the mediator-outcome effect}: All variables influencing the mediator \( O \) or \( P \) and the outcome \( U \) or \( I \) must be measured. This ensures that the indirect effects are not biased by hidden confounders. 
    \item \textbf{No unmeasured confounding of the exposure-mediator effect}: Every variable that confounds the relationship between the exposure \( S \) or \( T \) and the mediator \( O \) or \( P \) must be observed and accounted for. This ensures that the causal effects are not confounded by unmeasured variables. 
    \item \textbf{No confounders of the mediator-outcome effect that are affected by the exposure}: This assumption requires that no variables confounding the mediator-outcome relationship \( O \rightarrow U \) or \( P \rightarrow I \) are themselves influenced by the exposure \( S \) or \( T \). 
\end{itemize}

These assumptions ensure that the causal effects are identified without bias from unmeasured variables or violations of the causal structure implied by the graph. The total effect (TE) of individual $j$ ($j$ may be a user or a item) can be calculated as:

\begin{equation}
TE(U_j)=U_j(s, O_j(s)) - U_j(s^*, O_j(s^*)),
\end{equation}
\begin{equation}
TE(I_j)=I_j(t, P_j(t)) - I_j(t^*, P_j(t^*)).
\end{equation}

The indirect paths (i.e.,  $S \rightarrow O \rightarrow U$  and  $T \rightarrow P \rightarrow I$  ) are blocked by setting  $O$  and  $P$  to reference states, i.e.,  $Y=Y(S=   \left.s^{*}\right)$  and  $P=P(T=   \left.t^{*}\right)$. After that, we can calculate the PNDE of individual $j$ as:

\begin{equation}
PNDE(U_j)=U_j(s, O_j(s^*)) - U_j(s^*, O_j(s^*)),
\end{equation}

\begin{equation}
PNDE(I_j)=I_j(t, P_j(t^*)) - I_j(t^*, P_j(t^*)).
\end{equation}

According to Eq. (3), we can eliminate the effect of unrelated propensity bias on recommendation results by reducing PNDE in TE:
\begin{equation}
TE(U_j) - PNDE(U_j) = U_j(s, O_j(s)) - U_j(s, O_j(s^*)),
\end{equation}

\begin{equation}
TE(I_j) - PNDE(I_j) = I_j(t, P_j(t)) - I_j(t, P_j(t^*)).
\end{equation}

\subsection{Training and Inference}

With an existing collaborative filtering recommendation method $R$, we can obtain a user collaborative representation $\mathbf{\hat{e}}_{u}$ and an item collaborative representation $\mathbf{\hat{e}}_{i}$. However, our aim during training is to match the model predictions to the distribution of the training set, not to make recommendations for users. Therefore we use different approaches to estimate user-item interaction probabilities in the training and inference phases. Specifically, since the training set is created from a causal graph in the factual world as shown in Figure~\ref{causal2} (a), where all causal effects are not moderated, we need to take into account the effect of propensity bias on the collaborative information when estimating predictions:

\begin{equation}
 \mathbf{e}_{u} = \sigma \left (f_{O} \left ( \mathbf{s}_{u} \right )\right ) * \mathbf{\hat{e}}_{u}, 
\end{equation}
\begin{equation}
 \mathbf{e}_{i} = \sigma \left (f_{P} \left ( \mathbf{s}_{i} \right )\right ) * \mathbf{\hat{e}}_{i}, 
\end{equation}
where $f_{P}$ is the estimated model for the propensity bias and $\sigma$ is the sigmoid function.

We align the collaborative information with the side information to capture the information from user and item, which is then fed into the traditional recommendation method. The loss during the alignment process is as follows:
\begin{equation}
\begin{aligned}
\mathcal{L}_{\text {a}}  =  -\mathbb{E} \log \left[\frac{f\left ( \mathbf{e}_{u}, \mathbf{s}_{u} \right ) }{\sum_{u \in \mathbf{U}} f\left ( \mathbf{e}_{u}, \mathbf{s}_{u} \right ) }\right] -\mathbb{E} \log \left[\frac{f\left ( \mathbf{e}_{i}, \mathbf{s}_{i} \right)}{\sum_{i \in \mathbf{I}} f\left ( \mathbf{e}_{i}, \mathbf{s}_{i} \right)}\right],
\end{aligned}
\end{equation}
where $f(a,b) = exp(sim(a,b))$, and $sim(\cdot )$ represents the cosine similarity.

Subsequently, we can calculate the interaction probability $\hat{y_{ui}}$ and compute the loss of the recommendation method as follows:

\begin{equation}
L_{R}=\sum_{\left(u, i^{+}, i^{-}\right) \in O}-\ln \sigma\left(\hat{y}_{u, i^{+}}-\hat{y}_{u, i^{-}}\right),
\end{equation}
where  $O=\left\{\left(u, i^{+}, i^{-}\right) \mid\left(u, i^{+}\right) \in \mathcal{R},\left(u, i^{-}\right) \in \mathcal{R}^{-}\right\}$ denotes the training set, and  $\left(u, i^{+}, i^{-}\right)$ is a training sample. Specifically, $i^{+}$ denotes an item that has interacted with $u$ and $i^{-}$ denotes an item that has not interacted with $u$. Here,  $\mathcal{R}$  is the set of observed user-item interactions, and $i^{+}$ is a positive sample that user $u$ has interacted with.

In the inference stage, instead of relying on predictions in the factual world, we use Equation. 15 and Equation. 16 to mitigate the PNDE to obtain counterfactual world predictions:

\begin{equation}
\mathbf{\tilde{e}}_{u} = \mathbf{e}_{u}  -   \alpha * \sigma \left (f_O \left ( \mathbf{s}_{u} \right )\right ), 
\end{equation}
\begin{equation}
\mathbf{\tilde{e}}_{i} = \mathbf{e}_{u}  -   \alpha * \sigma \left (f_P \left ( \mathbf{s}_{i} \right )\right ),
\end{equation}
where $\alpha$ is a hyperparameter that denotes the reference state.

\section{Experimental evaluation}

In this section, we presents the experimental evaluations of the performance of our proposed CLLMR using multiple datasets to address the following research questions (RQ):
\begin{itemize}
\item \textbf{RQ1}: Does our proposed CLLMR improve existing state-of-the-art recommendation methods in various experimental settings? 
\item \textbf{RQ2}: Does SSE mitigate dimensional collapse? 
\item \textbf{RQ3}: Does the SSE approach and counterfactual framework of CLLMR contribute to improved recommendation performance? 
\item \textbf{RQ4}: Is each module in SSE function necessary?
\item \textbf{RQ5}: How does the spectrum of different ranks affect the model's performance? 
\end{itemize}

\subsection{Experiment Settings}

In this section, we describe the experimental setting, including the dataset, evaluation metrics, implementation details, and comparison algorithms.

\subsubsection{Datasets}

\begin{table}[ht]
\caption{Statistics of the experimental datasets.}
\resizebox{0.5\linewidth}{!}{
\begin{tabular}{ccccc}
\hline
Dataset     & \#Users & \#Items & \#Interactions & Density \\
\hline
Amazon & 11,000  & 9,332   & 120,464        & 1.2e-3 \\
Yelp        & 11,091  & 11,010  & 166,620        & 1.4e-3 \\
Steam       & 23,310  & 5,237   & 316,190        & 2.6e-3 \\
\hline
\end{tabular}}
\label{dataset}
\end{table}

\begin{table}[hp]
\caption{Instructions for obtaining side information using LLM on the three datasets following~\cite{ren2024representation}. ``You ... summarise'' is short for ``You will serve as an assistant to help me summarise''.  ``You ... determine'' is short for ``You will serve as an assistant to help me determine''.}
\label{instruction}
\centering
\begin{tabular}{c c p{0.85\textwidth}}
\hline
\multicolumn{2}{c}{Dataset}              & \multicolumn{1}{c}{Instruction}  \\
\hline  
\multirow{2}{*}[-15ex]{Amazon}      & \raisebox{-5ex}{Item}      & You ... summarise which types of users would enjoy a specific book. I will provide you with the title and a description of the book. Here are the instructions:
1. I will provide you with information in the form of a JSON string that describes the book:
{``title'': ``the title of the book'',  ``description'': ``a description of the book'', }    \\
 \cline{2-3}
                             & \raisebox{-8ex}{User}      & You ... determine which types of books a specific user is likely to enjoy.
I will provide you with information about books that the user has purchased, as well as his or her reviews of those books. Here are the instructions: 1. Each purchased book will be described in JSON format, with the following attributes: {``title'': ``the title of the book'', 
    ``description'': ``a description of what types of users will like this book'',
    "review'': ``the user's review on the book''} 2. The information I will give you: a list of JSON strings describing the items that the user has purchased.  \\
                             \hline  
\multirow{2}{*}[-15ex]{Yelp}        & \raisebox{-8ex}{Item}      & You ... summarise which types of users would enjoy a specific video game.
I will provide you with the basic information (name, publisher, genres and tags) of that game and also some feedback of users for it.
Here are the instructions:
1. The basic information will be described in JSON format, with the following attributes:
{
    ``name'': ``the name of the video game'',
    ``publisher'': ``the publisher of the game'', 
    ``genres'': ``the genres of the game'', 
    ``tags'': ``several tags describing the game''
}
2. Feedback from users will be managed in the following List format: [``the first feedback'', ``the second feedback'', ....]
3. The information I will give you: a JSON string describing the basic information about the game; a List object containing some feedback from users about the game.                                                                          \\
 \cline{2-3}
                             & \raisebox{-8ex}{User}      & You ... determine which types of game a specific user is likely to enjoy.
I will provide you with information about games that the user has interacted, as well as his or her reviews of those games. Here are the instructions: 1. Each interacted game will be described in JSON format, with the following attributes:
{ ``title'': ``the name/title of the game'', ``description'': ``a description of what types of users will like this game'',  ``review'': ``the user's review on the game''}
2. The information I will give you: a list of JSON strings describing the games that the user has played.                                                                \\
                             \hline  
\multirow{2}{*}[-20ex]{Steam}       & \raisebox{-10ex}{Item}      & You ... summarise which types of users would enjoy a specific business.
I will provide you with the basic information (name, city and category) of that business and also some feedback of users for it.
Here are the instructions:
1. The basic information will be described in JSON format, with the following attributes:
{
    ``name'': ``the name of the business'',
    ``city'': ``city where the company is located'', (if there is no city, I will set this value to ``None'')
    ``categories'': ``several tags describing the business'' (if there is no categories, I will set this value to ``None'')
}
2. Feedback from users will be managed in the following List format:
[
    ``the first feedback'',
    ``the second feedback'',
    ....
]
3. The information I will give you: a JSON string describing the basic information about the business. a list object containing some feedback from users about the business.
               \\
               \cline{2-3}
                             & \raisebox{-10ex}{User}      & You ... determine which types of business a specific user is likely to enjoy.
I will provide you with information about businesses that the user has interacted, as well as his or her reviews of those businesses.
Here are the instructions:
1. Each interacted business will be described in JSON format, with the following attributes:
{
    ``title'': ``the name of the business'', 
    ``description'': ``a description of what types of users will like this business'',
    ``review'': ``the user's review on the business'' 
}
2. The information I will give you: a list of JSON strings describing the businesses that the user has interacted.
           \\
\hline  
\end{tabular}
\end{table}

\begin{table*}[ht]
\caption{Comparison of recommendation performance of all methods on different datasets. The best value is \textbf{bolded}, the runner-up value is \underline{underlined}, and the improvement over the runner-up by the best performer is abbreviated as ``Imprv.''.}
\LARGE
\resizebox{1.0\linewidth}{!}{
\renewcommand{\arraystretch}{1.5}
\begin{tabular}{cc|cccccc|cccccc|cccccc}
\\
\hline
\multicolumn{2}{c}{Dataset}                               & \multicolumn{6}{c}{Amazon}                                 & \multicolumn{6}{c}{Yelp}                                   & \multicolumn{6}{c}{Steam}                                \\
\hline
\multicolumn{1}{c|}{Backbone}                  & Variants & R@10    & R@30    & R@50    & N@10     & N@30    & N@50    & R@10     & R@30    & R@50    & N@10    & N@30    & N@50    & R@10     & R@30    & R@50    & N@10    & N@30    & N@50  \\
\hline
\multicolumn{1}{c|}{\multirow{5}{*}{LRGCCF}}     & Base     & 0.08759 & 0.17070 & 0.22783 & 0.06568  & 0.09161 & 0.10711 & 0.06400  & 0.14130 & 0.20260 & 0.05212 & 0.07790 & 0.09534 & 0.08265  & 0.17022 & 0.23122 & 0.06595 & 0.09452 & 0.11156 \\
\multicolumn{1}{c|}{}                          & RLMGen   & \underline{0.08777} & \underline{0.17404} & 0.22836 & \underline{0.06730}  & \underline{0.09425} & \underline{0.10897} & 0.06689  & 0.14768 & 0.20882 & 0.05480 & 0.08154 & 0.09904 & 0.08741  & 0.17862 & 0.24034 & 0.07003 & 0.09978 & 0.11720 \\
\multicolumn{1}{c|}{}                          & RLMCon   & 0.08527 & 0.17164 & \underline{0.22925} & 0.06429  & 0.09124 & 0.10685 & \underline{0.069366} & \underline{0.15301} & \underline{0.21383} & \underline{0.05721} & \underline{0.08475} & \underline{0.10221} & \underline{0.08793}  & \underline{0.17905} & \underline{0.24109} & \underline{0.07130} & \underline{0.10093} & \underline{0.11843} \\
\multicolumn{1}{c|}{}                          & CLLMR    & \textbf{0.09252} & \textbf{0.18098} & \textbf{0.23817} & \textbf{0.06906}  & \textbf{0.09685} & \textbf{0.11233} & \textbf{0.07189}  & \textbf{0.15758} & \textbf{0.22073} & \textbf{0.05896} & \textbf{0.07417} & \textbf{0.10513} & \textbf{0.08814}  & \textbf{0.17912} & \textbf{0.24279} & \textbf{0.07214} & \textbf{0.10882} & \textbf{0.12093} \\
\multicolumn{1}{c|}{}                          & \textbf{Imprv.}   & 1.15\%  & 0.66\%  & 0.58\%  & 0.31\%   & 0.08\%  & 0.35\%  & 0.87\%   & 1.02\%  & 0.98\%  & 1.59\%  & 1.22\%  & 0.92\%  & 0.24\%   & 0.04\%  & 0.71\%  & 1.18\%  & 7.82\%  & 2.11\%  \\
\hline
\multicolumn{1}{c|}{\multirow{5}{*}{LightGCN}} & Base     & 0.08714 & 0.17301 & 0.22771 & 0.066824 & 0.09374 & 0.10853 & 0.06905  & 0.15162 & 0.21356 & 0.05699 & 0.08419 & 0.10201 & 0.08456  & 0.17293 & 0.23586 & 0.06808 & 0.09683 & 0.11448 \\
\multicolumn{1}{c|}{}                          & RLMGen   & 0.09247 & 0.17997 & 0.23733 & 0.07159  & 0.09894 & 0.11432 & \underline{0.07617}  & \underline{0.16411} & \underline{0.22410} & \underline{0.06220} & \underline{0.09123} & \underline{0.10880} & \underline{0.09040}  & \underline{0.18356} & \underline{0.24678} & \underline{0.07251} & \underline{0.10297} & \underline{0.12076} \\
\multicolumn{1}{c|}{}                          & RLMCon   & \underline{0.09665} & \underline{0.18534} & \underline{0.24209} & \underline{0.07313}  & \underline{0.10112} & \underline{0.11636} & 0.07411  & 0.16305 & 0.22448 & 0.06088 & 0.09024 & 0.10792 & 0.08942  & 0.18312 & 0.24675 & 0.07243 & 0.10281 & 0.12071 \\
\multicolumn{1}{c|}{}                          & CLLMR    & \textbf{0.09813} & \textbf{0.18611} & \textbf{0.24551} & \textbf{0.07357}  &\textbf{ 0.10128} & \textbf{0.11723} & \textbf{0.07738}  & \textbf{0.16453} & \textbf{0.23119} & \textbf{0.06331} & \textbf{0.09205} & \textbf{0.11119} & \textbf{0.09179}  & \textbf{0.18582} & \textbf{0.25055} & \textbf{0.07396} & \textbf{0.10397} & \textbf{0.12134} \\
\multicolumn{1}{c|}{}                          & \textbf{Imprv.}   & 1.53\%  & 0.42\%  & 1.41\%  & 0.60\%   & 0.16\%  & 0.75\%  & 1.59\%   & 0.26\%  & 2.99\%  & 1.78\%  & 0.90\%  & 2.20\%  & 1.54\%   & 1.23\%  & 1.53\%  & 2.00\%  & 0.97\%  & 0.48\%  \\
\hline
\multicolumn{1}{c|}{\multirow{5}{*}{SGL}}      & Base     & 0.10230 & 0.18323 & 0.23440 & 0.07742  & 0.10307 & 0.11717 & 0.07436  & 0.15988 & 0.22108 & 0.06099 & 0.08926 & 0.10700 & 0.09254  & 0.18650 & 0.24951 & 0.07451 & 0.10527 & 0.12307 \\
\multicolumn{1}{c|}{}                          & RLMGen   & 0.10114 & \underline{0.19059} & \underline{0.24730} & 0.07778  & \underline{0.10602} & \underline{0.12143} & \underline{0.07786}  & \underline{0.16926} & \underline{0.23168} & \underline{0.06391} & \underline{0.09366} & \underline{0.11203} & 0.09414  & 0.18895 & 0.25423 & 0.07513 & 0.10615 & 0.12444 \\
\multicolumn{1}{c|}{}                          & RLMCon   & \underline{0.10296} & 0.18809 & 0.24367 & \underline{0.07859}  & 0.10540 & 0.12048 & 0.07625  & 0.16653 & 0.23008 & 0.06234 & 0.09129 & 0.11048 & \underline{0.09511}  & \underline{0.19005} & \underline{0.25513} & \underline{0.07601} & \underline{0.10703} & \underline{0.12526} \\
\multicolumn{1}{c|}{}                          & CLLMR    & \textbf{0.10941} & \textbf{0.19097 }& \textbf{0.24976} & \textbf{0.08083 } & \textbf{0.11468} & \textbf{0.12931} & \textbf{0.08018}  & \textbf{0.16992} & \textbf{0.23825} & \textbf{0.06521} & \textbf{0.09477} & \textbf{0.11438} & \textbf{0.09597}  & \textbf{0.19123} & \textbf{0.25584} & \textbf{0.07712} & \textbf{0.10792} & \textbf{0.12576} \\
\multicolumn{1}{c|}{}                          & \textbf{Imprv.}   & 6.26\%  & 0.20\%  & 0.99\%  & 2.85\%   & 8.17\%  & 6.49\%  & 0.87\%   & 0.18\%  & 1.78\%  & 0.74\%  & 0.41\%  & 1.15\%  & 0.90\%   & 0.62\%  & 0.28\%  & 1.46\%  & 0.83\%  & 0.40\%  \\
\hline
\multicolumn{1}{c|}{\multirow{5}{*}{DirectAU}} & Base     & 0.08714 & 0.17306 & 0.22761 & 0.06682  & 0.09376 & 0.10855 & 0.06905  & 0.15162 & 0.21356 & 0.05700 & 0.08419 & 0.10204 & 0.08619  & 0.17451 & 0.23589 & 0.06879 & 0.09750 & 0.11479 \\
\multicolumn{1}{c|}{}                          & RLMGen   & 0.09312 & 0.17760 & 0.23520 & 0.07064  & 0.09716 & 0.11254 & \underline{0.07786}  & \underline{0.16726} & \underline{0.23168} & \underline{0.06398} & \underline{0.09406} & \underline{0.11230} & \underline{0.09085}  & \underline{0.18339} & \underline{0.24694} & \underline{0.07252} & \underline{0.10281} & \underline{0.12073} \\
\multicolumn{1}{c|}{}                          & RLMCon   & \underline{0.09348} & \underline{0.18304} & \underline{0.24383} & \underline{0.07377}  & \underline{0.10232} & \underline{0.11814} & 0.07625  & 0.16653 & 0.23008 & 0.06234 & 0.09129 & 0.11048 & 0.08958  & 0.18163 & 0.24616 & 0.07209 & 0.10210 & 0.12020 \\
\multicolumn{1}{c|}{}                          & CLLMR    & \textbf{0.09693} & \textbf{0.18550}  & \textbf{0.24837} & \textbf{0.08321}  & \textbf{0.10814} & \textbf{0.12158} & \textbf{0.08026}  & \textbf{0.17005} & \textbf{0.23804} & \textbf{0.06540} & \textbf{0.09492} & \textbf{0.11438} & \textbf{0.09363}  & \textbf{0.18947} & \textbf{0.25574} & \textbf{0.07462} & \textbf{0.10617} & \textbf{0.12472} \\
\multicolumn{1}{c|}{}                          & \textbf{Imprv.}   & 3.69\%  & 1.34\%  & 1.86\%  & 12.80\%  & 5.69\%  & 2.91\%  & 3.08\%   & 1.67\%  & 2.75\%  & 2.22\%  & 0.91\%  & 1.85\%  & 3.06\%   & 3.32\%  & 3.56\%  & 2.90\%  & 3.27\%  & 3.30\%  \\
\hline
\multicolumn{1}{c|}{\multirow{5}{*}{SimGCL}}   & Base     & 0.09820 & 0.18676 & 0.24645 & 0.07405  & 0.10210 & 0.11824 & 0.07374  & 0.16295 & 0.23027 & 0.06208 & 0.09211 & 0.11071 & 0.09225  & 0.18610 & 0.25082 & 0.07418 & 0.10482 & 0.12294 \\
\multicolumn{1}{c|}{}                          & RLMGen   & 0.09866 & 0.19046 & 0.25150 & 0.07513  & 0.10400 & 0.12037 & 0.07554  & 0.16562 & 0.23040 & 0.06233 & 0.09200 & 0.11074 & \underline{0.09293}  & 0.18724 & \underline{0.25268} & 0.07515 & 0.10597 & 0.12427 \\
\multicolumn{1}{c|}{}                          & RLMCon   & \underline{0.09986} & \underline{0.19461} & \underline{0.25423} & \underline{0.07721}  & \underline{0.10682} & \underline{0.12277} & \underline{0.07795}  & \underline{0.16829} & \underline{0.23367} & \underline{0.06386} & \underline{0.09360} & \underline{0.11232} & 0.09270  & \underline{0.18822} & 0.25244 & \underline{0.07542} & \underline{0.10651} & \underline{0.12455} \\
\multicolumn{1}{c|}{}                          & CLLMR    & \textbf{0.10053} & \textbf{0.19497} & \textbf{0.25707} & \textbf{0.07918}  & \textbf{0.10996} & \textbf{0.12547} & \textbf{0.08185}  & \textbf{0.17445} & \textbf{0.24046} & \textbf{0.06695} & \textbf{0.09742} & \textbf{0.11658} &\textbf{ 0.09607} & \textbf{0.19196} & \textbf{0.25735 }& \textbf{0.07715} &\textbf{ 0.10847} & \textbf{0.12687} \\
\multicolumn{1}{c|}{}                          & \textbf{Imprv.}   & 0.67\%  & 0.18\%  & 1.12\%  & 2.55\%   & 2.94\%  & 2.20\%  & 5.00\%   & 3.66\%  & 2.91\%  & 4.84\%  & 4.08\%  & 3.79\%  & 9.33\%   & 1.99\%  & 1.85\%  & 2.29\%  & 1.84\%  & 1.86\%  \\
\hline
\multicolumn{1}{c|}{\multirow{5}{*}{LightGCL}} & Base     & 0.08908 & 0.17553 & 0.22902 & 0.06676  & 0.09383 & 0.10825 & 0.07341  & 0.15771 & 0.21868 & 0.06081 & 0.08974 & 0.10930 & 0.08225  & 0.17115 & 0.23351 & 0.06643 & 0.09539 & 0.11294 \\
\multicolumn{1}{c|}{}                          & RLMGen   & 0.08803 & 0.17214 & 0.22974 & 0.06658  & 0.09322 & 0.10870 & 0.07665  & 0.16462 & 0.23019 & 0.06104 & 0.09113 & 0.10992 & 0.08574  & 0.17379 & 0.23636 & 0.06843 & 0.09730 & 0.11409 \\
\multicolumn{1}{c|}{}                          & RLMCon   & \underline{0.09303} & \underline{0.18224} & \underline{0.24032} & \underline{0.07017}  & \underline{0.09821} & \underline{0.11381} & \underline{0.07709}  & \underline{0.16814} & \underline{0.23283} & \underline{0.06244} & \underline{0.09216} & \underline{0.11063} & \underline{0.08995}  & \underline{0.18292} & \underline{0.24593} & \underline{0.07165} & \underline{0.10208} & \underline{0.12005} \\
\multicolumn{1}{c|}{}                          & CLLMR    & \textbf{0.09868} & \textbf{0.18634} & \textbf{0.24663} & \textbf{0.07404}  & \textbf{0.10159} & \textbf{0.11786} & \textbf{0.07893}  & \textbf{0.17299} & \textbf{0.23824} & \textbf{0.06326} &\textbf{ 0.09238} & \textbf{0.11137} & \textbf{0.09091 } & \textbf{0.18547} & \textbf{0.24982} & \textbf{0.07815} & \textbf{0.10756} & \textbf{0.12194} \\
\multicolumn{1}{c|}{}                          & \textbf{Imprv.}   & 6.07\%  & 2.25\%  & 2.63\%  & 5.52\%   & 3.44\%  & 3.56\%  & 2.39\%   & 2.88\%  & 2.32\%  & 1.31\%  & 0.24\%  & 0.67\%  & 1.07\%   & 1.39\%  & 1.58\%  & 9.07\%  & 5.37\%  & 1.57\% \\
\hline
\end{tabular}}
\label{compresion}
\end{table*}

As shown in Table~\ref{dataset}, three commonly used public datasets are used in our evaluation. The Amazon dataset~\cite{he2016ups} includes user ratings and reviews related to books sold on the Amazon platform. The Yelp dataset~\cite{wang2019neural} collects a large amount of textual data related to various businesses. Additionlly, the Steam dataset~\cite{ren2024representation} contains textual feedback from users about games on the Steam platform. In the Amazon and Yelp datasets, we used a similar data pre-processing approach as in previous studies~\cite{ren2024representation, yu2022graph}, excluding interactions with ratings below 3. We then divided each dataset into training, validation, and testing subsets and maintained a 3:1:1 ratio. 

Regarding the side information of the datasets, we followed a methodology consistent with previous studies~\cite{ren2024representation} by synthesizing comprehensive profiles of users and items using the ChatGPT-3.5-turbo model. The specific instructions are shown in Table~\ref{instruction}. All side information was divided into two parts, `summarisation' and `reasoning', and the LLM responses were limited to 200 words.

\subsubsection{Evaluation Metrics}
To ensure a thorough and comprehensive evaluation of our proposed CLLMR method, we employed a ranking scheme across all experiments, consistent with methodologies from previous studies~\cite{he2020lightgcn, ren2024representation}. We selected two well-established ranking-based metrics to quantify the effectiveness of our approach: Recall@N and NDCG@N. By using these metrics, we aim to demonstrate the ability of our method to deliver high-quality results that align with user expectations and preferences. For clarity on the effectiveness of the improvement, we will calculate the magnitude of the improvement using the formula $\frac{b-a}{a} \times 100\%$, where $a$ is the best value and $b$ is the runner-up value.

\subsubsection{Implementation Details}
In our experiments, we standardised the dimensionality of the representations for all models to 32, ensuring a consistent basis for comparison. To optimise each model's performance, we meticulously tuned the hyperparameters through an exhaustive grid search. During the training phase, we uniformly employed the Adam optimiser across all models, chosen for its efficiency and adaptability in handling diverse data types. We set a fixed batch size of 4,096, balancing computational efficiency with the need for sufficient data representation in each training iteration. Additionally, we adopted a learning rate of $1 \times 10^{-3}$, a commonly used value for initializing the learning process in various machine learning tasks.

To further prevent overfitting and enhance model generalisability, we implemented an early stopping technique. This method monitors performance on a validation set and halts training when performance plateaus or begins to degrade.

\subsubsection{Baselines}
In this paper, we use six state-of-the-art collaborative filtering recommendation methods as backbone, while comparing two of the state-of-the-art LLM-based recommender systems.

\paragraph{Recommendation methods}
\begin{itemize} 
\item LRGCCF~\cite{chen2020revisiting}: This approach reevaluates the role of nonlinear operations in graph neural networks for recommendations and mitigates oversmoothing through residual connections. 
\item LightGCN~\cite{he2020lightgcn}: It introduces a streamlined approach to recommender systems by simplifying the neural architecture in graph message passing, resulting in a more efficient model that retains high performance. 
\item SGL~\cite{wu2021self}: This method employs node and edge dropout techniques as data augmentation, fostering a diversity of perspectives that enrich the contrastive learning process. 
\item SimGCL~\cite{yu2023xsimgcl}: SimGCL focuses on enhancing data representation through a noise mechanism that preserves the structural integrity of the underlying model, fostering a more robust and precise recommendation process. \item DirectAU~\cite{wang2022towards}: This method evaluates representation quality in collaborative filtering by focusing on alignment and uniformity. It ensures that representations are well-aligned with user-item relationships while exhibiting a uniform distribution on the hypersphere. 
\item LightGCL~\cite{cailightgcl}: LightGCL leverages singular value decomposition for contrast augmentation, enabling unconstrained structural refinement through global collaboration modeling. 
\end{itemize}

\paragraph{LLM-based recommender systems}
\begin{itemize} 
\item RLMGen~\cite{ren2024representation}: Aligning the side representation with the collaborative representation using contrastive learning effectively reduces feature noise. 
\item RLMCon~\cite{ren2024representation}: Feature reconstruction using a masked autoencoder aligns the side representation with the collaborative representation, effectively reducing feature noise. 
\end{itemize}

\begin{figure}[ht]
  \centering
    \subfloat[LRGCCF]{\includegraphics[width=0.35\textwidth]{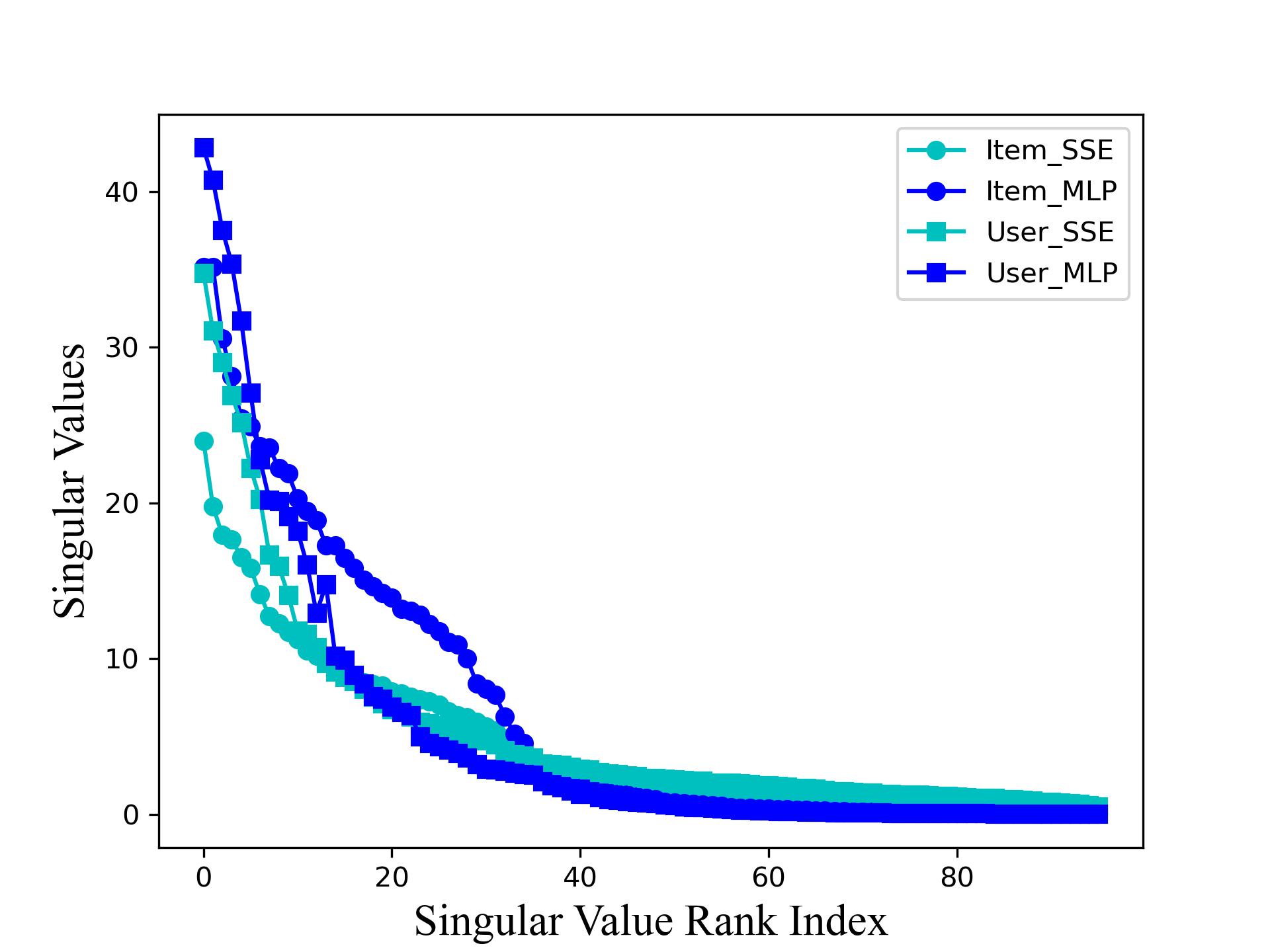}}
\subfloat[LightGCN]{\includegraphics[width=0.35\textwidth]{amazon.jpeg}}
  \subfloat[SGL]{\includegraphics[width=0.35\textwidth]{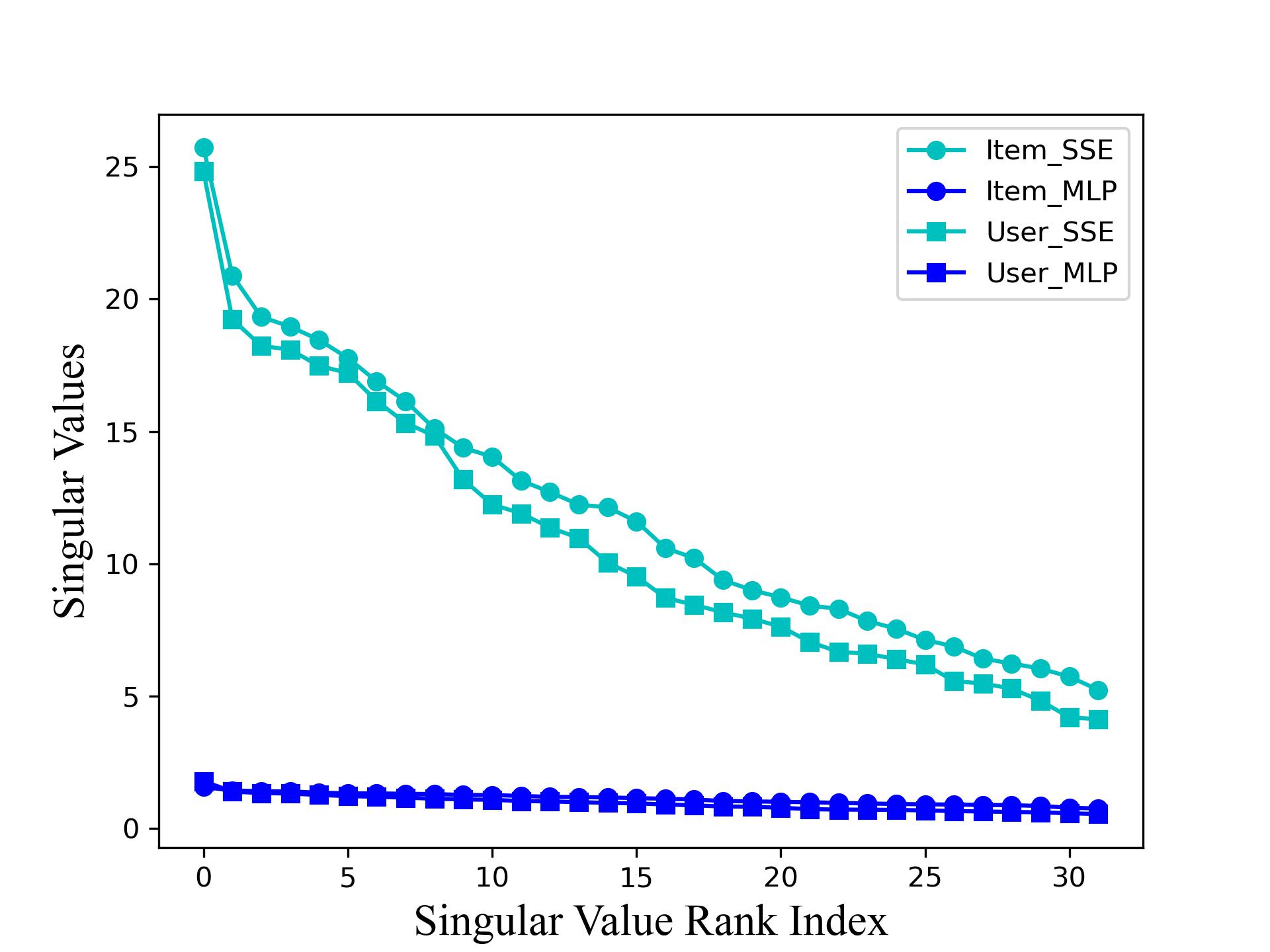}}
  \\
  \subfloat[DirectAU]{\includegraphics[width=0.35\textwidth]{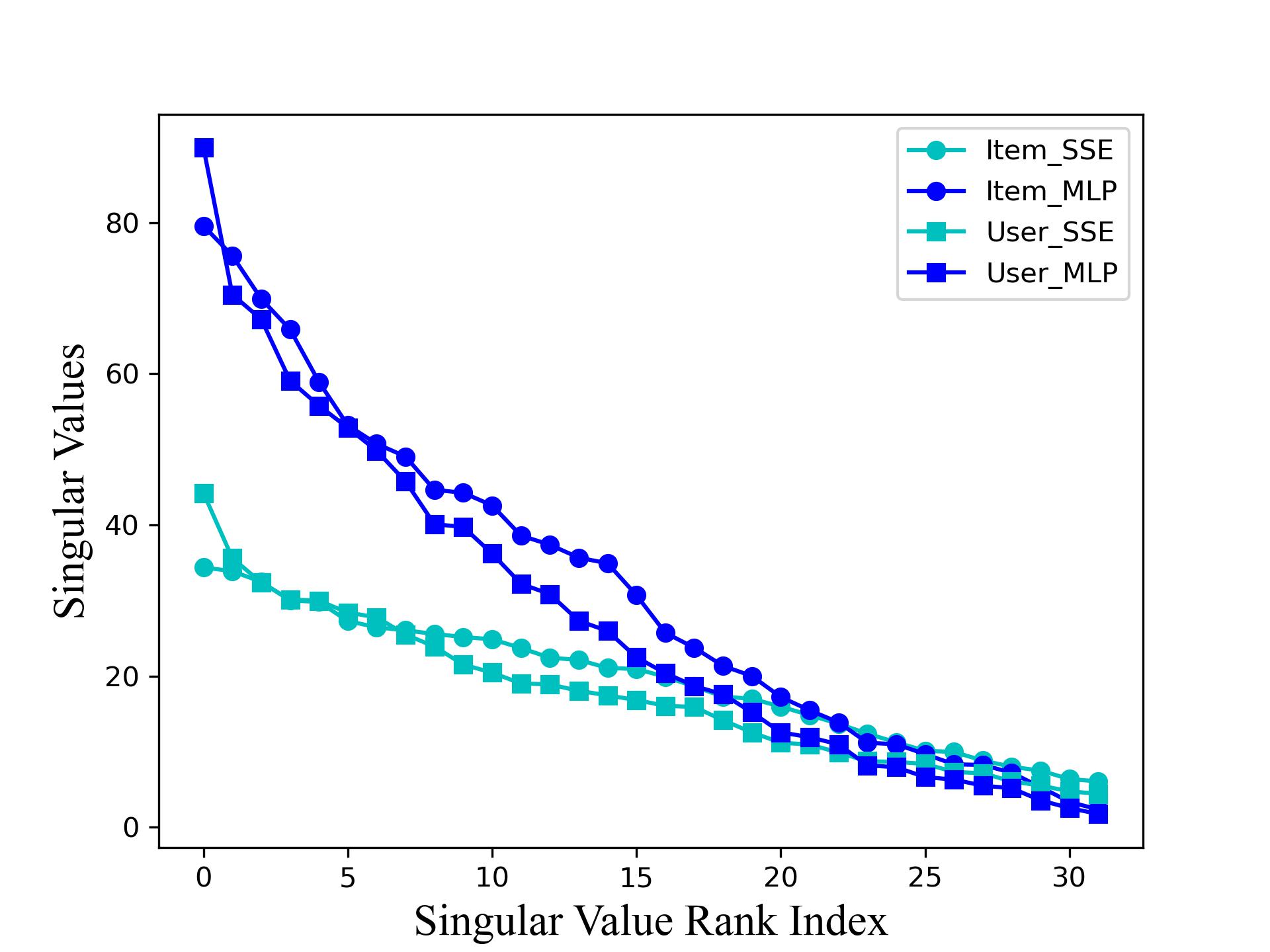}}
\subfloat[LightGCL]{\includegraphics[width=0.35\textwidth]{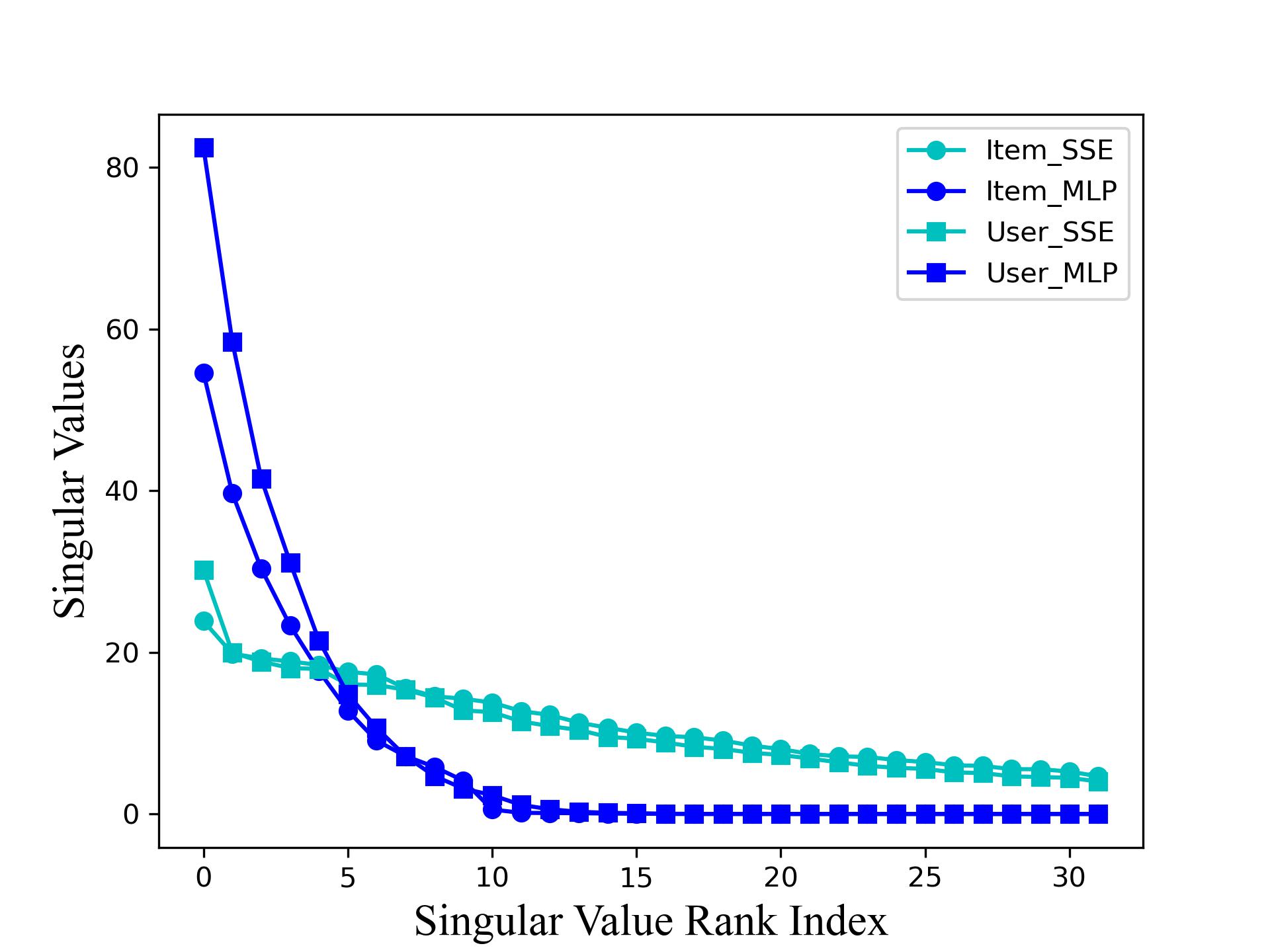}}
  \subfloat[SimGCL]{\includegraphics[width=0.35\textwidth]{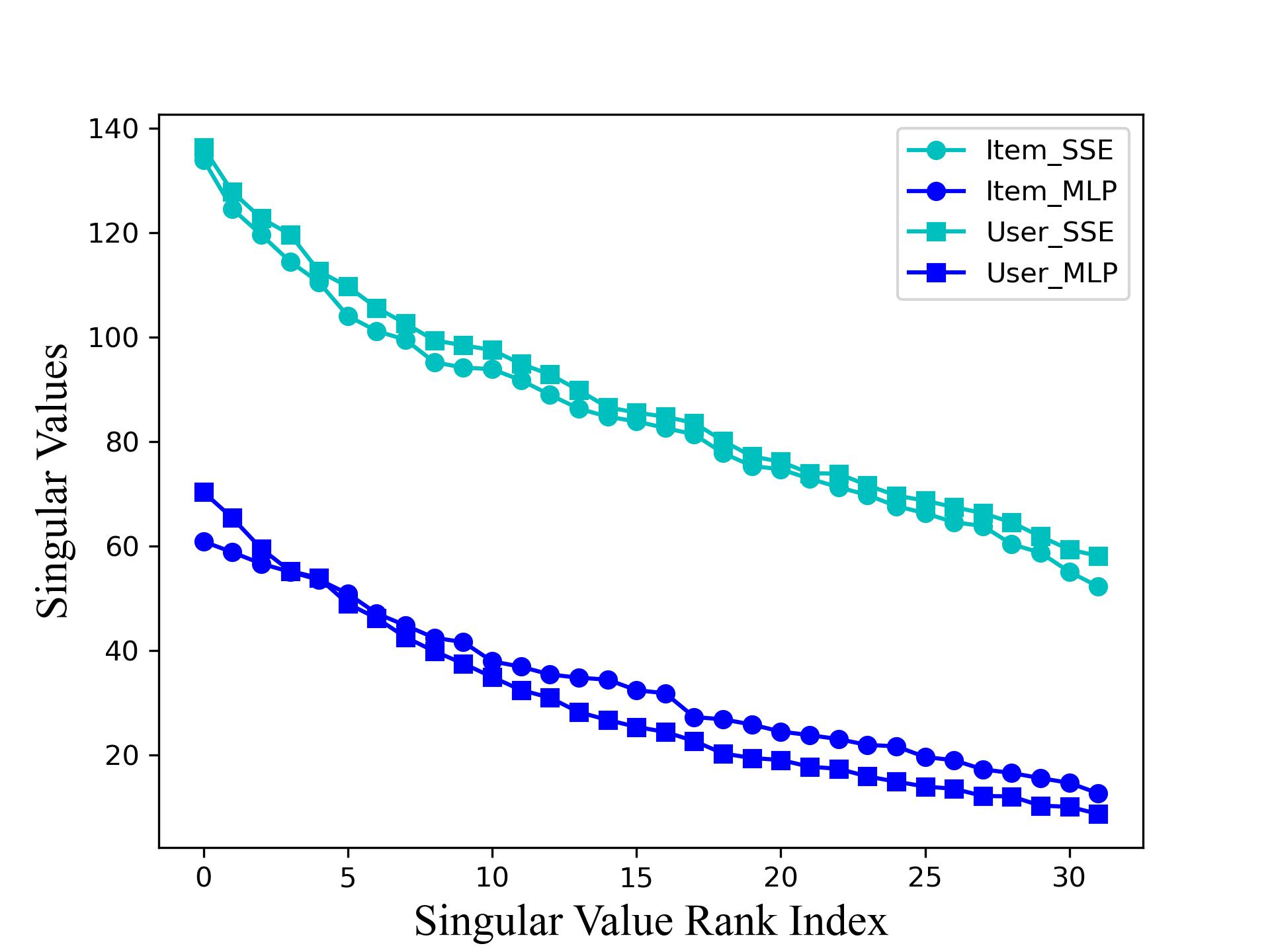}}
 \\
  \caption{Singular values of different recommendation methods on the Amazon dataset.}
  \label{svdamazon}
\end{figure}
\subsection{Performance Comparison}
To address \textbf{RQ1}, we integrated CLLMR into six state-of-the-art collaborative filtering models. The experimental results, averaged over five random initialisations, are presented in Table~\ref{compresion}. These results reveal several noteworthy observations, which we summarise as follows:

\begin{itemize} 
\item Overall, we consistently observe that integrating the side information generated by the LLM with the backbone recommender improves performance compared to the original versions. The LLM demonstrates strong emergence and generalizsation capabilities, storing extensive general world knowledge alongside linguistic understanding and expressive abilities. By generating side information for items enriched with real-world knowledge, the LLM provides not only basic item details but also deeper meanings and cultural context. This richer representation enhances the system's ability to interpret textual information, improving personalisation and recommendation accuracy by offering a more comprehensive view of the user.

\item For backbone methods based on contrastive learning, aligning the side information from the LLM with collaborative information through contrastive learning can improve the effectiveness. Contrastive learning aligns the side representation with the collaborative representation while preserving the relationships between sample pairs. Unlike reconstruction, which attempts to recover the original inputs without directly optimizing the distinction between positive and negative pairs, contrastive learning integrates both representations by maintaining their original pre-training objectives. This avoids the risk of losing key relationships, as contrastive learning retains the relative distances and structures between representations, resulting in better alignment.

\item CLLMR employs a spectrum-based side information encoder that deeply explores the intrinsic structure of historical user interaction data, constructing a rich and discriminative feature space. This encoding mechanism ensures feature diversity and significantly enhances the model's ability to distinguish between varying user preferences, effectively preventing feature collapse, where model outputs converge to a single solution. Additionally, CLLMR incorporates a counterfactual inference strategy to fine-tune the inherent bias in the LLM. By simulating decision-making in ``what-if'' scenarios, the model gains deeper insight into the user's true preferences, mitigating cultural biases embedded in the LLM. This reasoning approach enables the recommender system to capture personalised user needs more accurately. 
\end{itemize}

\begin{figure*}[t]
  \centering
    \subfloat[LRGCCF]{\includegraphics[width=0.35\textwidth]{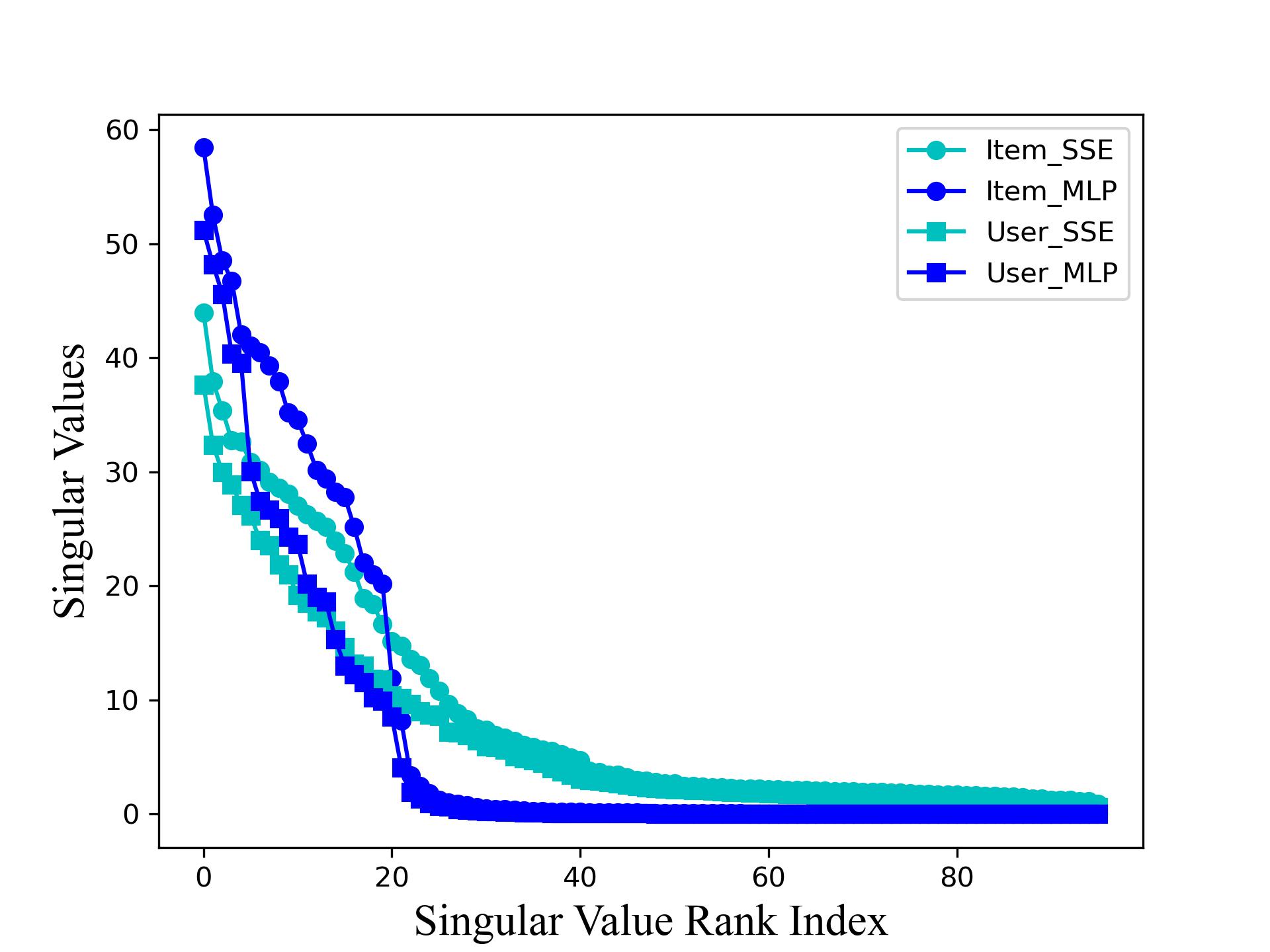}}
\subfloat[LightGCN]{\includegraphics[width=0.35\textwidth]{yelp.jpeg}}
  \subfloat[SGL]{\includegraphics[width=0.35\textwidth]{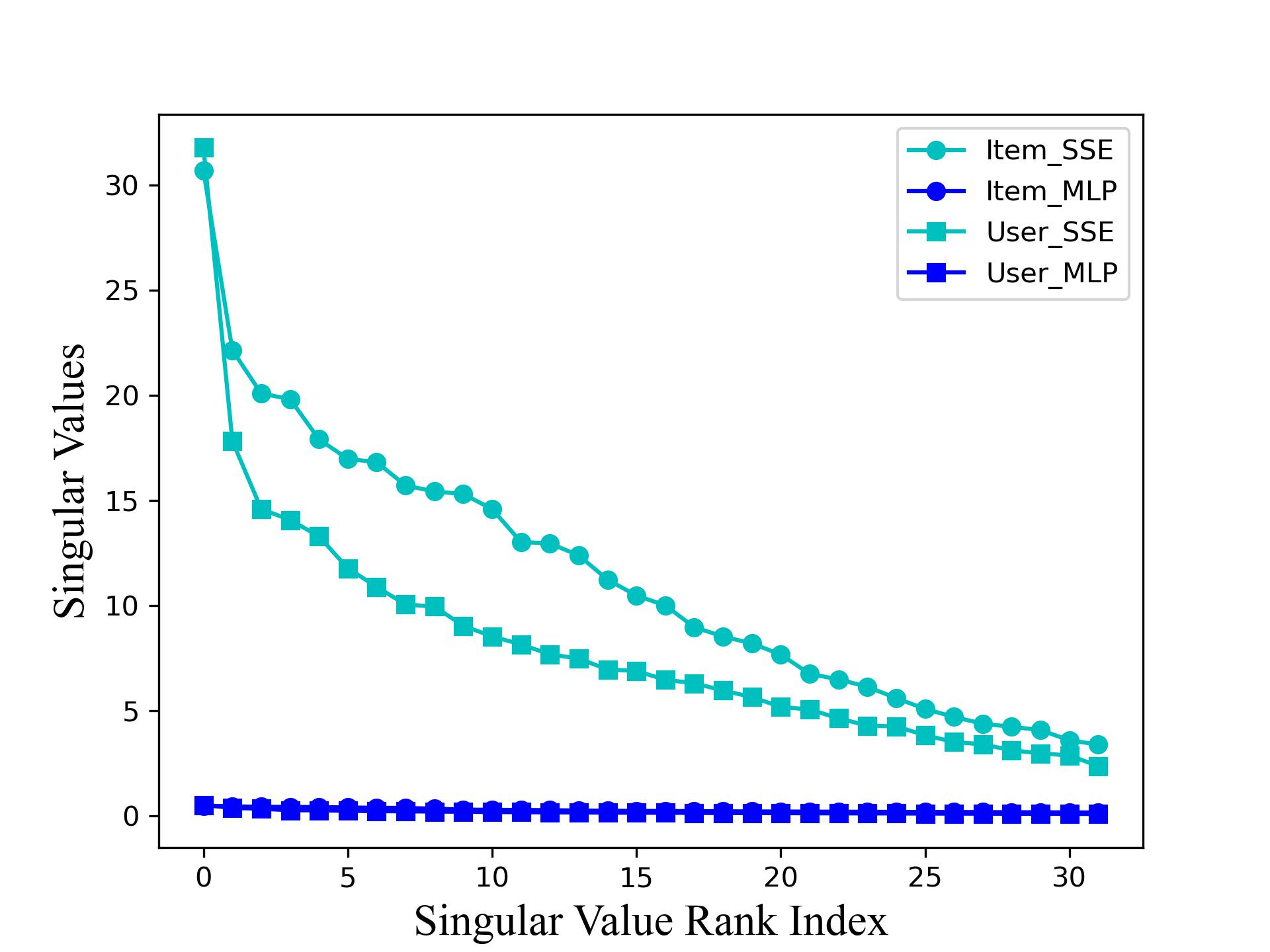}}
  \\
  \subfloat[DirectAU]{\includegraphics[width=0.35\textwidth]{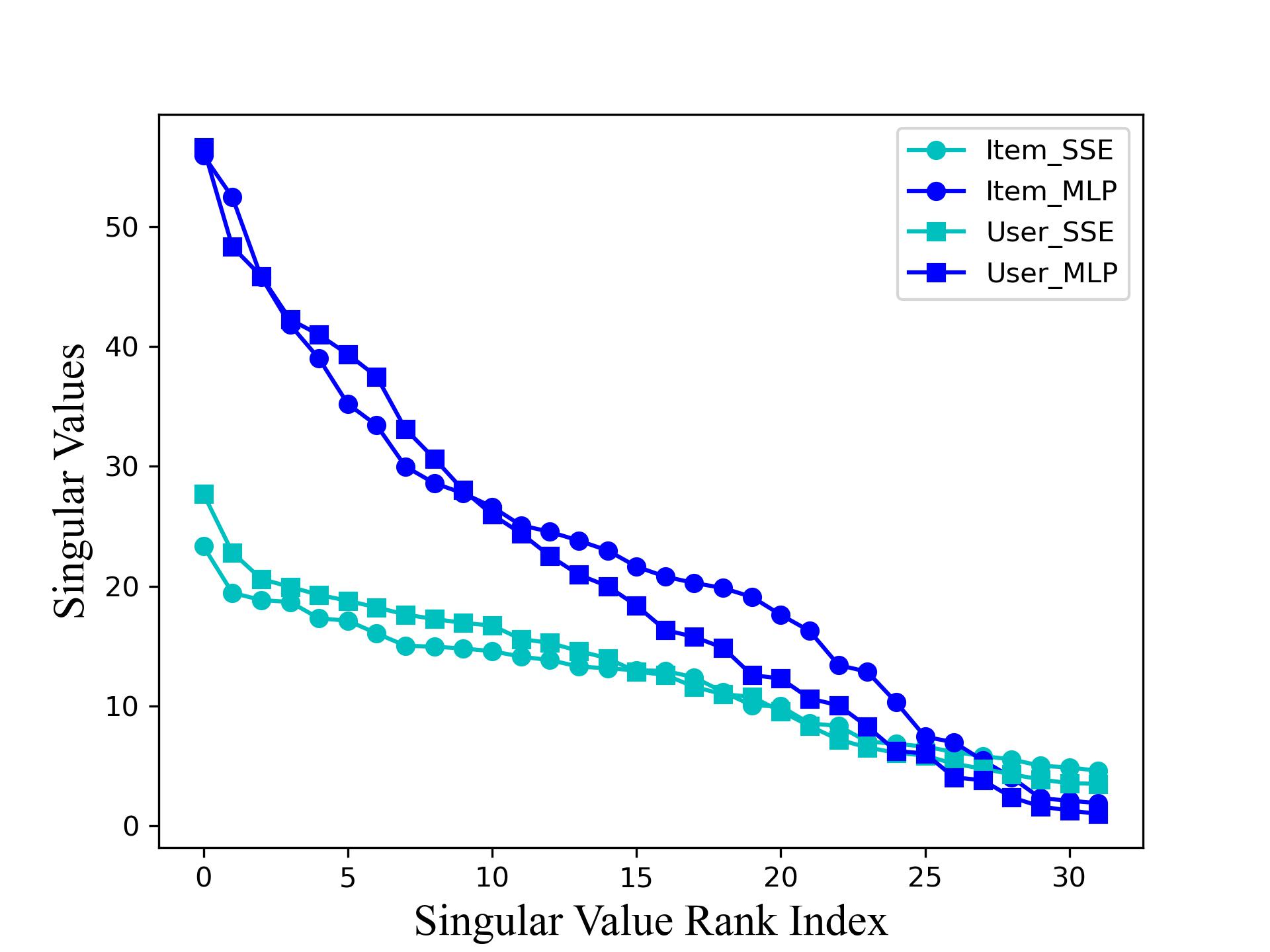}}
\subfloat[LightGCL]{\includegraphics[width=0.35\textwidth]{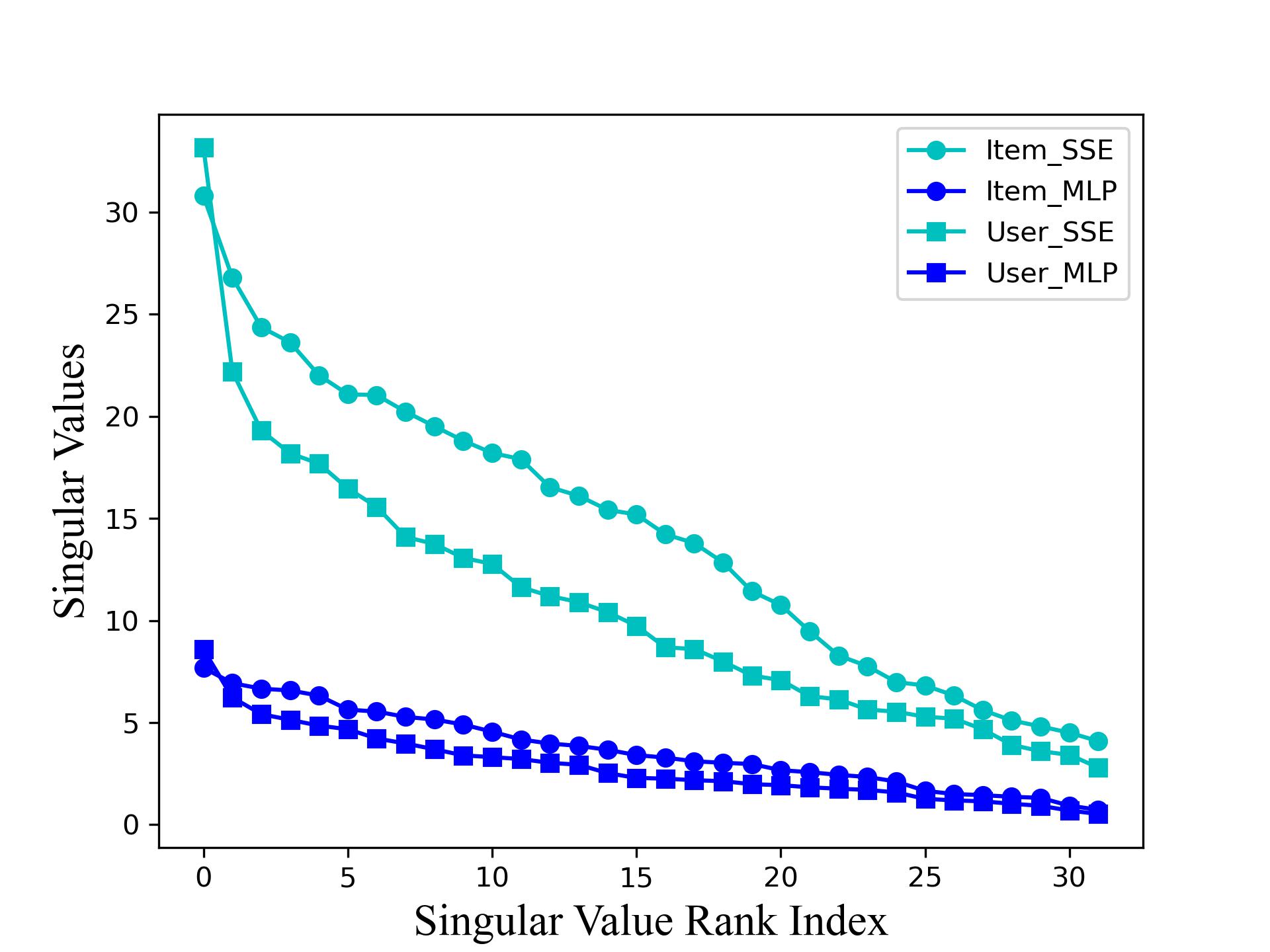}}
  \subfloat[SimGCL]{\includegraphics[width=0.35\textwidth]{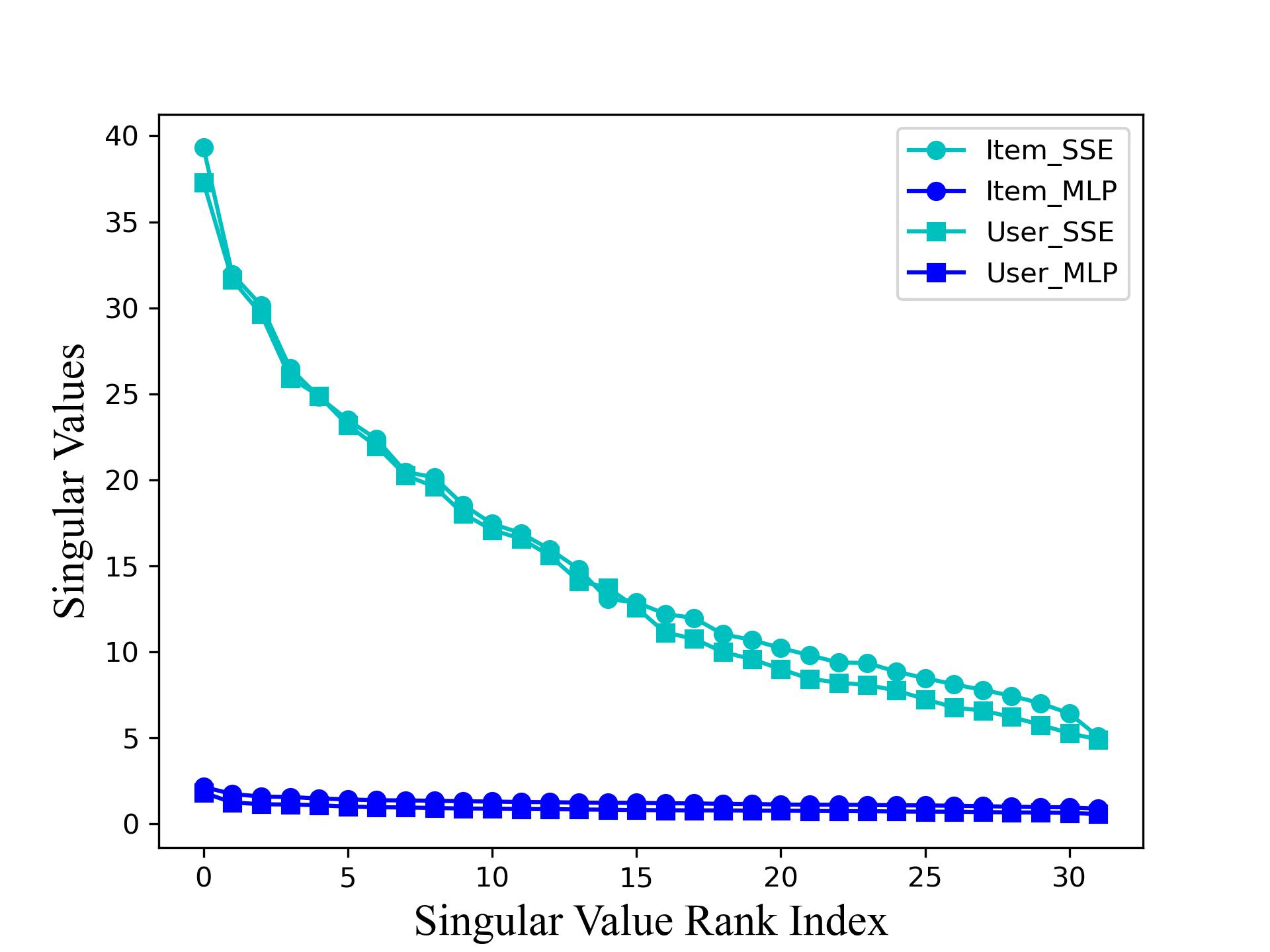}}
 \\
  \caption{Singular values of different recommendation methods on the Yelp dataset.}
  \label{svdyelp}
\end{figure*}

\begin{figure*}[ht]
  \centering
    \subfloat[LRGCCF]{\includegraphics[width=0.35\textwidth]{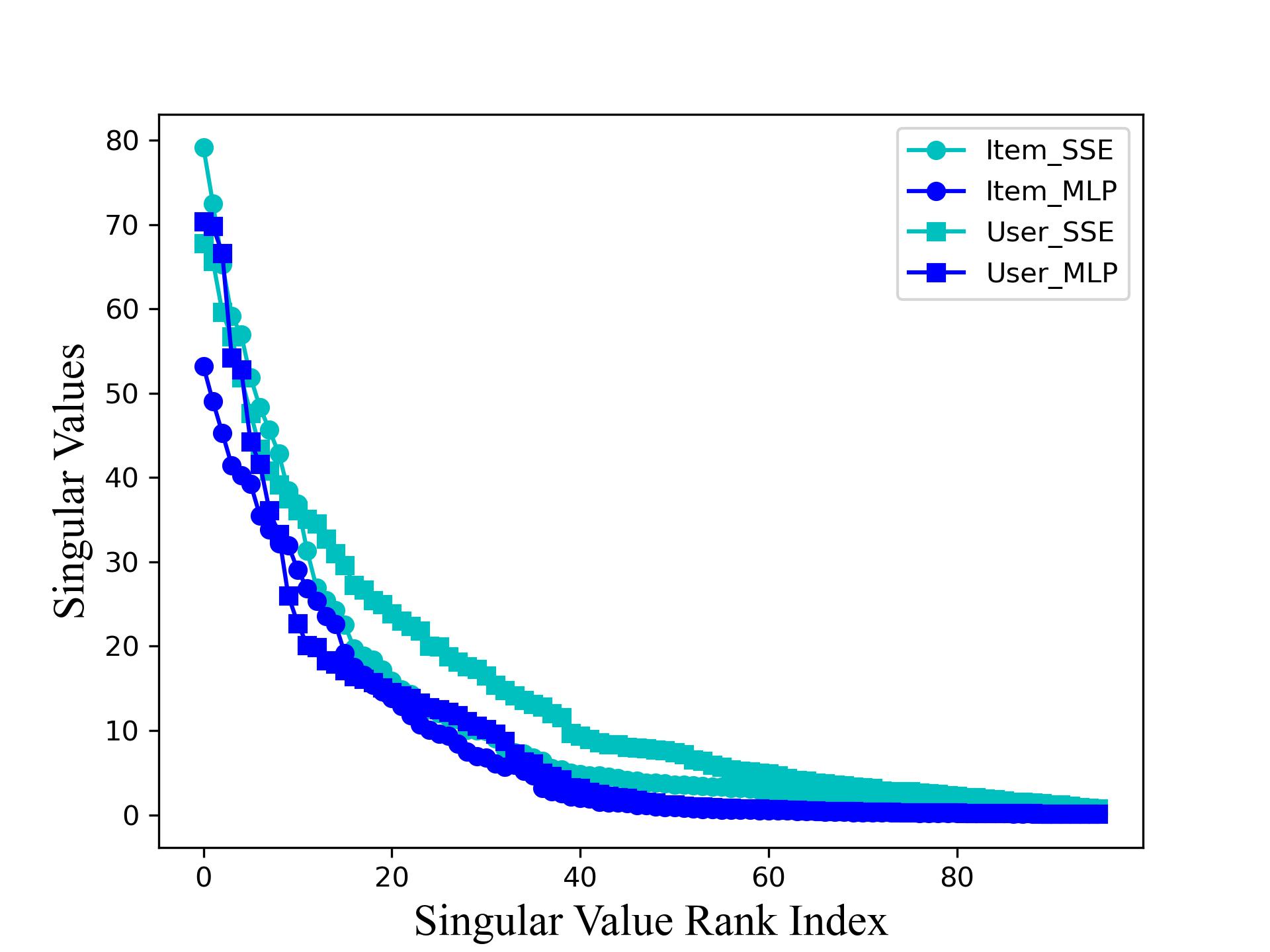}}
\subfloat[LightGCN]{\includegraphics[width=0.35\textwidth]{steam.jpeg}}
  \subfloat[SGL]{\includegraphics[width=0.35\textwidth]{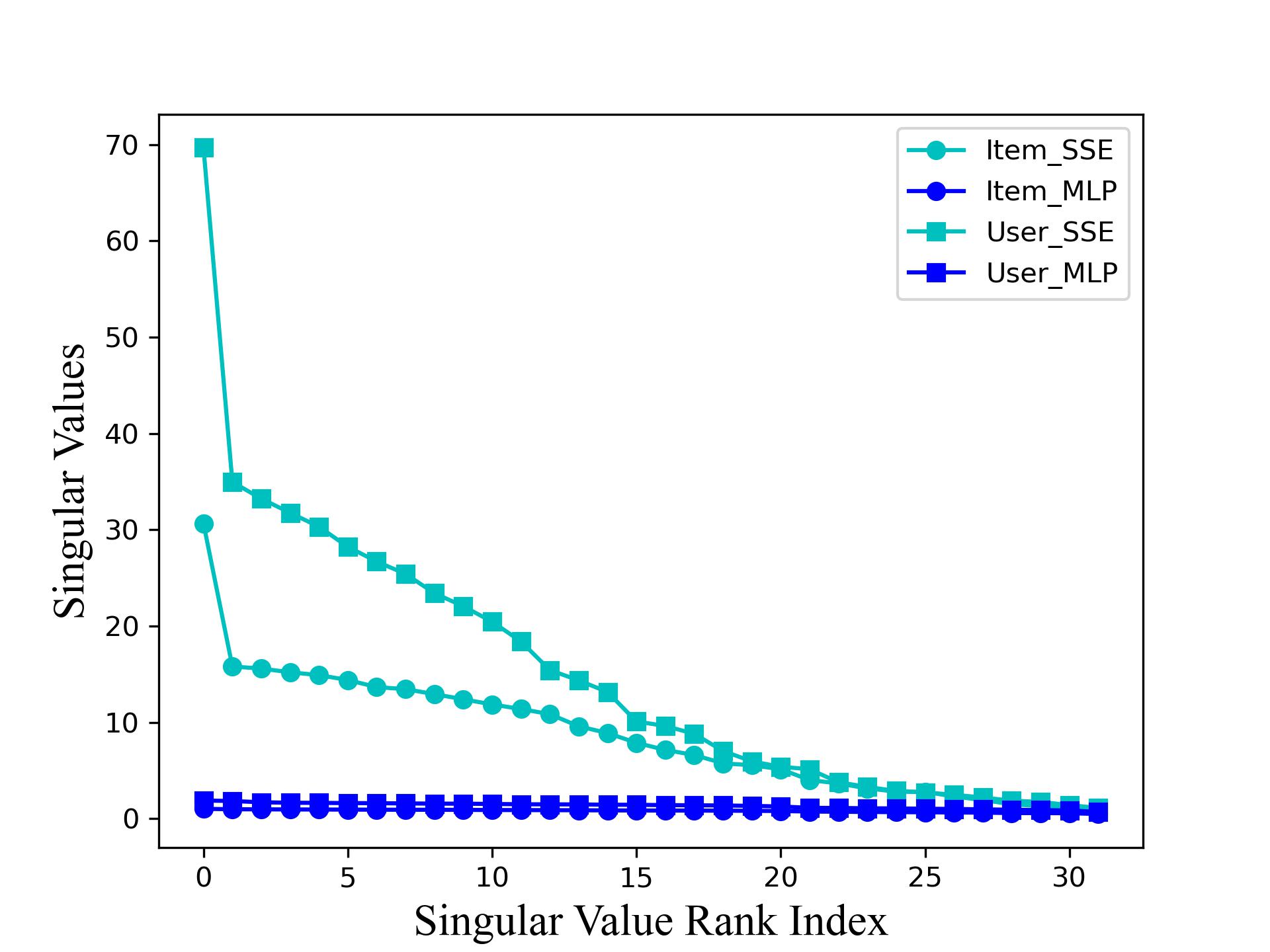}}
  \\
  \subfloat[DirectAU]{\includegraphics[width=0.35\textwidth]{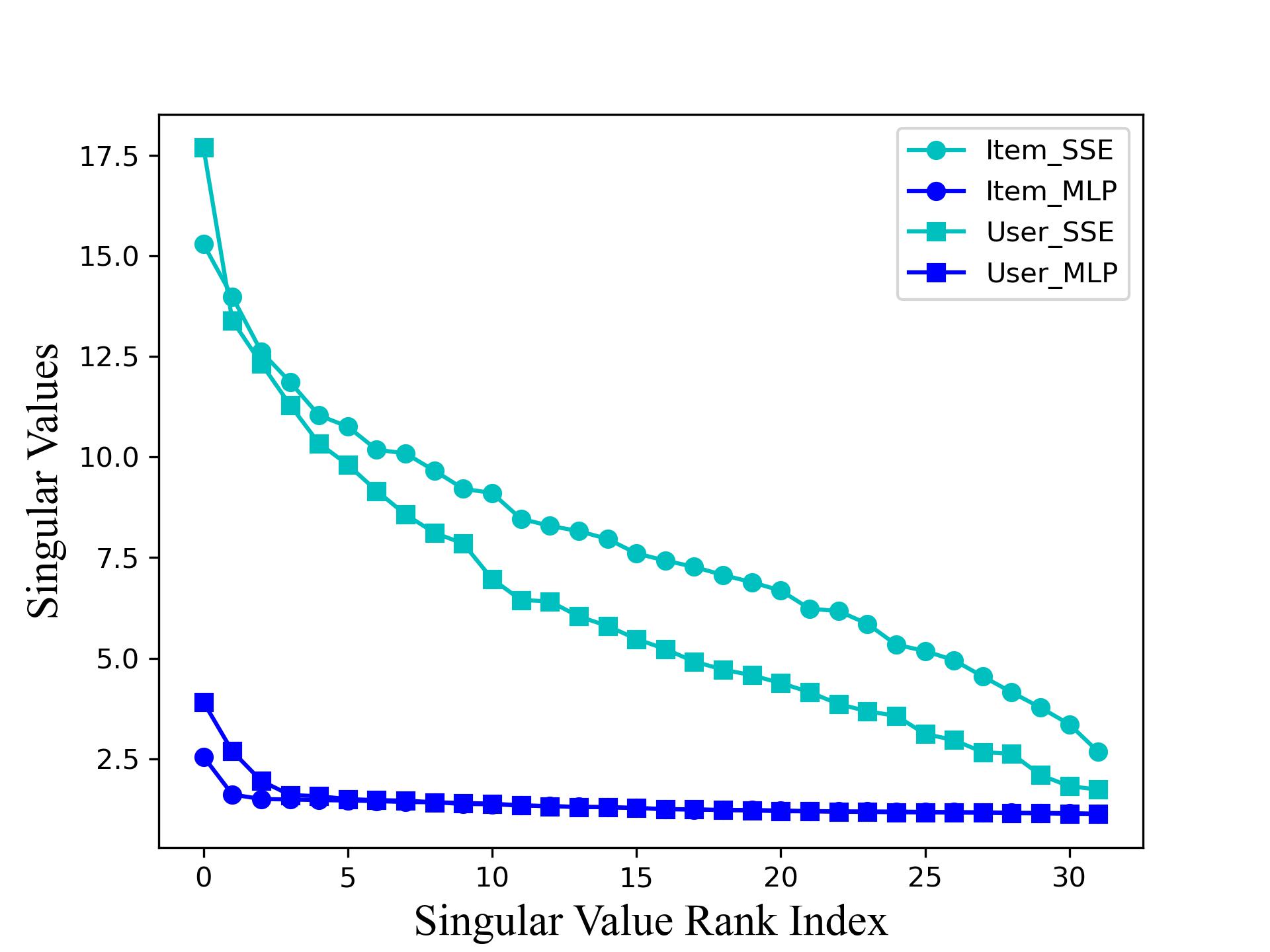}}
\subfloat[LightGCL]{\includegraphics[width=0.35\textwidth]{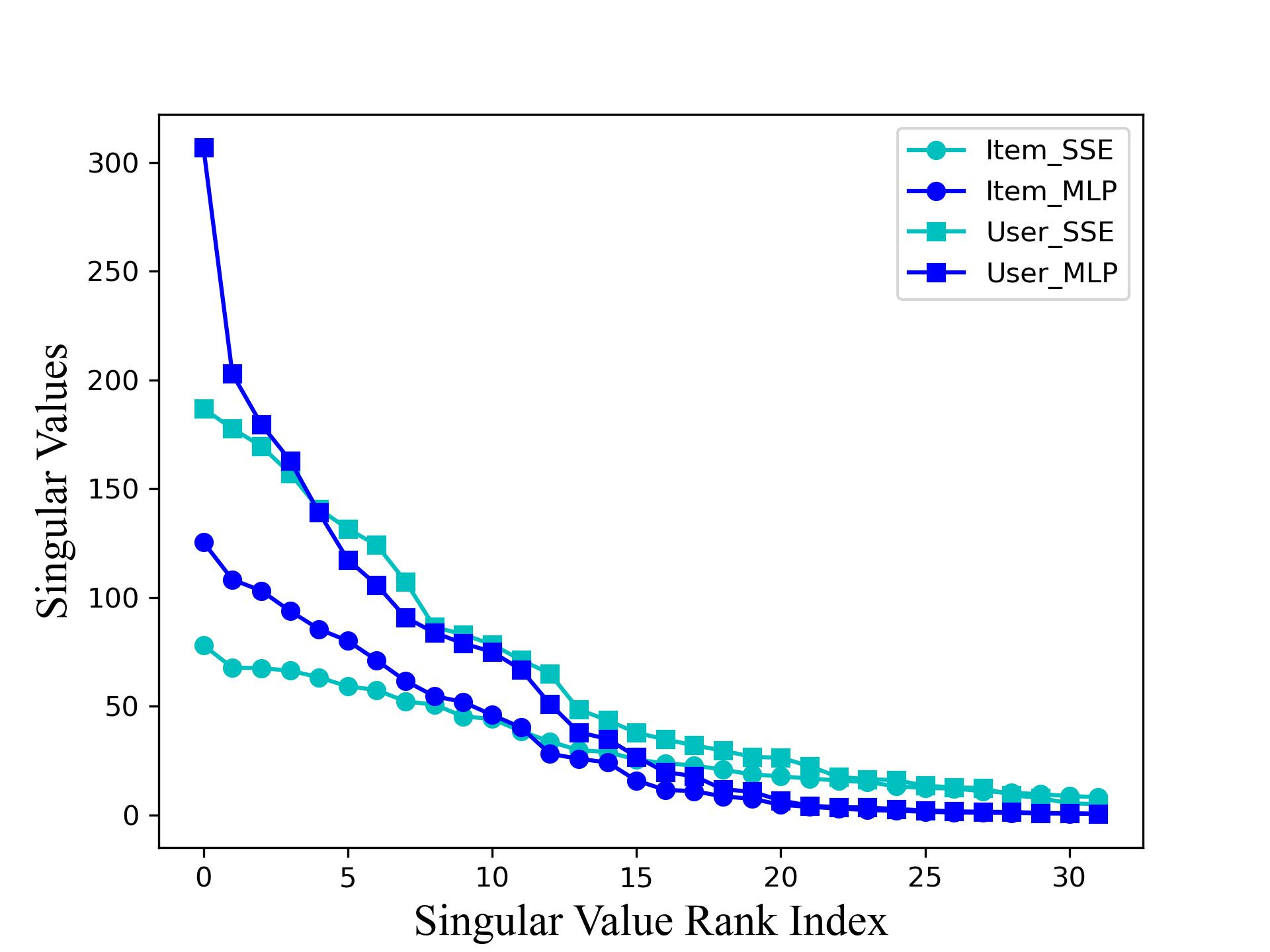}}
  \subfloat[SimGCL]{\includegraphics[width=0.35\textwidth]{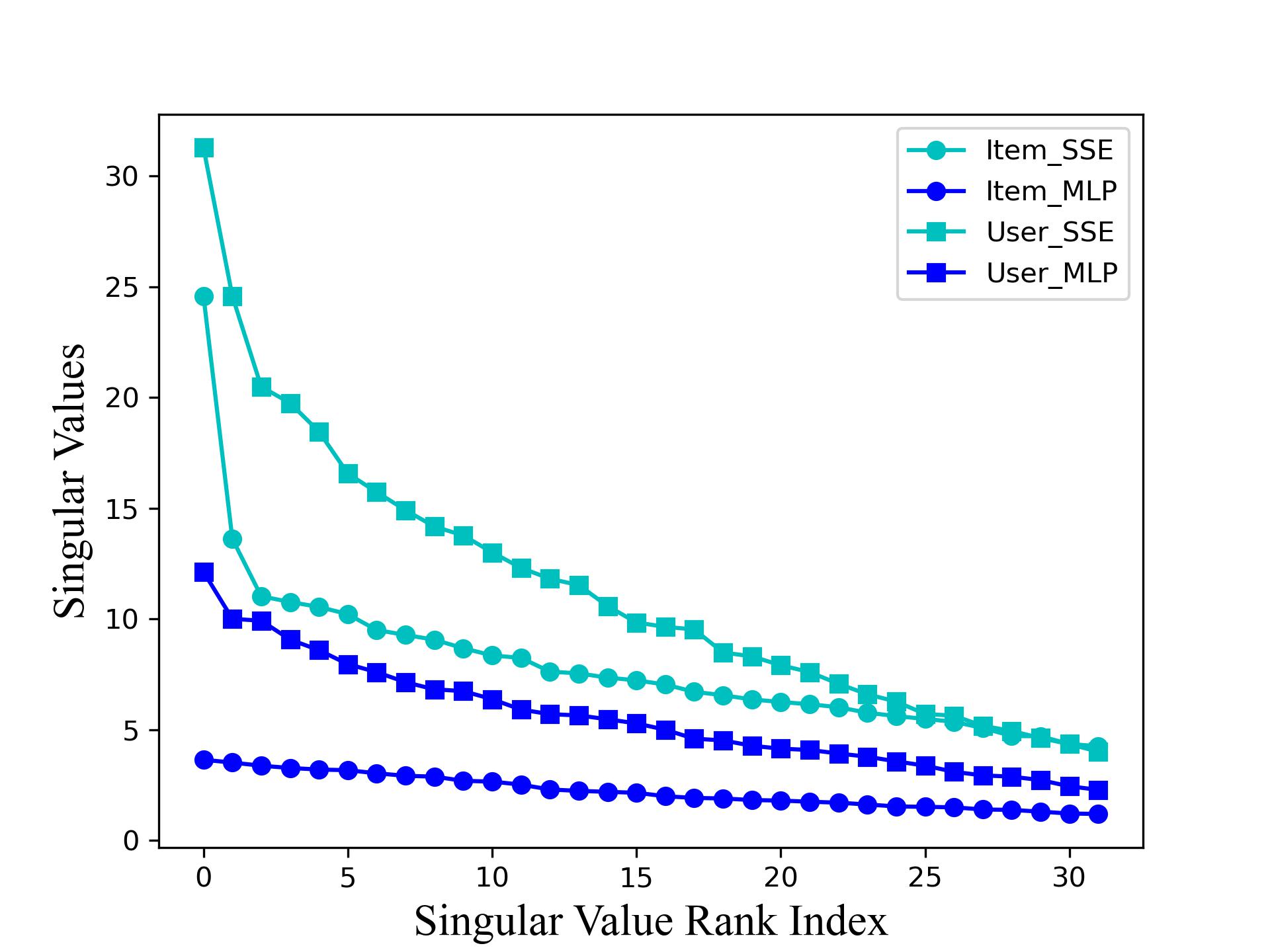}}
 \\
  \caption{Singular values of different recommendation methods on the Steam dataset.}
  \label{svdsteam}
\end{figure*}

\subsection{Ablation Study}
In this section, we construct ablation experiments to answer \textbf{RQ2}, \textbf{RQ3}, and \textbf{RQ4}.

\subsubsection{Necessity of each module of CLLMR}

To answer RQ2, we utilise MLP instead of SSE and perform a singular value decomposition of the user/item side representations. The singular value case resulting from MLP and SSE is shown in Figure.~\ref{svdamazon}, Figure.~\ref{svdyelp}, and Figure.~\ref{svdsteam}. From the results we have the following conclusions:

\begin{itemize}
\item SSE effectively avoids feature collapse relative to MLP. A healthy feature representation has multiple higher singular values, indicating that the features are distributed across multiple dimensions rather than being concentrated in just a few. SSE does not form a low-rank embedding matrix, meaning that the singular values of the matrix do not decay rapidly, thus maintaining the utilization of high-dimensional space. By introducing the spectrum as an observed variable under a condition, SSE achieves representational identifiability and disentanglement through reparameterization, allowing each dimension to capture an independent element of the data as much as possible, which helps prevent dimensional collapse.

\item LRGCCF mitigates the over-smoothing problem through residual linking, where the representation consists of multiple representations spliced together, resulting in higher dimensionality while making the singularities smoother. LRGCCF focuses on keeping the representations at a lower level during training, which may make the model less able to generalize to new, unseen data, thus reducing the difference in singularity. LightGCN is trained with fewer parameters than LRGCCF, effectively avoiding overfitting, which leads to more stable changes in singular values.

\item DirectAU, SGL, LightGCL and SimGCL, recommendation methods based on contrastive learning, effectively improve the learning ability of the model, but it is worth noting that their singular values vary greatly. This suggests that the model may have learned some overly complex features in the training data, potentially capturing noise rather than true data distribution features, leading to poor performance on new data. In contrast, SSE makes the singular values relatively smoother, effectively enhancing its generalization ability.
\end{itemize}

\begin{table*}[ht]
\LARGE
\caption{The experimental results of ablation studies on three datasets. In the “Variants” column, ``w/o Conf.'' indicates removing the component of counterfactual inference, ``w/o SSE'' indicates removing the component of the spectrum-based side information encoder, and ``De.’' indicates the magnitude due to the removal of a component.}
\renewcommand{\arraystretch}{1.5}
\resizebox{1.0\linewidth}{!}{
\begin{tabular}{cc|cccccc|cccccc|cccccc}
\hline
\multicolumn{2}{c}{Dataset}                                & \multicolumn{6}{|c}{Amazon}                                 & \multicolumn{6}{|c}{Yelp}                                   & \multicolumn{6}{|c}{Steam}                                  \\
\hline
\multicolumn{1}{c|}{Backbone}                  & Variants  & R@10    & R@30    & R@50    & N@10     & N@30    & N@50    & R@10    & R@30    & R@50    & N@10     & N@30    & N@50    & R@10     & R@30    & R@50    & N@10    & N@30    & N@50    \\
\hline
\multicolumn{1}{c|}{\multirow{5}{*}{LRGCCF}}     & CLLMR     & 0.09252 & 0.18098 & 0.23817 & 0.06906  & 0.09685 & 0.11233 & 0.07189 & 0.15758 & 0.22073 & 0.05896  & 0.07417 & 0.10513 & 0.08814  & 0.17912 & 0.24279 & 0.07214 & 0.10882 & 0.12093 \\
\multicolumn{1}{c|}{}                          & w/o Conf. & 0.09058 & 0.17904 & 0.23703 & 0.06812  & 0.09587 & 0.11139 & 0.06997 & 0.15457 & 0.21593 & 0.05812  & 0.08578 & 0.10315 & 0.08639  & 0.17782 & 0.24125 & 0.06960 & 0.09929 & 0.11699 \\
\multicolumn{1}{c|}{}  & Conf. De.  & -2.10\%  & -1.07\%  & -0.48\%  & -1.36\%   & -1.01\%  & -0.84\%  & -2.67\%  & -1.91\%  & -2.17\%  & -1.42\%   & 15.65\%  & -1.88\%  & -1.99\%   & -0.73\%  & -0.63\%  & -3.52\%  & -8.76\%  & -3.26\%   \\
\multicolumn{1}{c|}{}                          & w/o SSE   & 0.08840  & 0.17695 & 0.23391 & 0.06675  & 0.09462 & 0.11001 & 0.07048 & 0.15570 & 0.21672 & 0.05831  & 0.08643 & 0.10394 & 0.08760  & 0.17805 & 0.24099 & 0.07077 & 0.10093 & 0.11743 \\
\multicolumn{1}{c|}{}   & SSE De.   & -4.45\%  & -2.23\%  & -1.79\%  & -3.34\%   & -2.30\%  & -2.07\%  & -1.96\%  & -1.19\%  & -1.82\%  & -1.10\%   & 16.53\%  & -1.13\%  & -0.61\%   & -0.60\%  & -0.74\%  & -1.90\%  & -7.25\%  & -2.89\%    \\
\hline
\multicolumn{1}{c|}{\multirow{5}{*}{LightGCN}} & CLLMR  & 0.09813 & 0.18611 & 0.24551 & 0.07357 & 0.10128 & 0.11723 & 0.07738 & 0.16453 & 0.23119 & 0.06331  & 0.09205 & 0.11119 & 0.09179  & 0.18582 & 0.25055 & 0.07396 & 0.10397 & 0.12134 \\
\multicolumn{1}{c|}{}                          & w/o Conf. & 0.09500 & 0.18481 & 0.24374 & 0.07138  & 0.09981 & 0.11567 & 0.07327 & 0.15976 & 0.22355 & 0.06019  & 0.08865 & 0.10700 & 0.08868  & 0.18155 & 0.24532 & 0.07127 & 0.10136 & 0.11920 \\
\multicolumn{1}{c|}{}  & Conf. De.  & -3.19\%  & -0.70\%  & -0.72\%  & -2.98\%   & -1.45\%  & -1.33\%  & -5.31\%  & -2.90\%  & -3.30\%  & -4.93\%   & -3.69\%  & -3.77\%  & -3.39\%   & -2.30\%  & -2.09\%  & -3.64\%  & -2.51\%  & -1.76\% \\
\multicolumn{1}{c|}{}                          & w/o SSE   & 0.09503 & 0.18469 & 0.24349 & 0.07241  & 0.10053 & 0.11691 & 0.07415 & 0.16427 & 0.22667 & 0.06043  & 0.09011 & 0.10826 & 0.08873  & 0.18056 & 0.24371 & 0.07124 & 0.10127 & 0.11913 \\
\multicolumn{1}{c|}{} & SSE De.   & -3.16\%  & -0.76\%  & -0.82\%  & -1.58\%   & -0.74\%  & -0.27\%  & -4.17\%  & -0.16\%  & -1.96\%  & -4.55\%   & -2.11\%  & -2.64\%  & -3.33\%   & -2.83\%  & -2.73\%  & -3.68\%  & -2.60\%  & -1.82\%  \\
\hline
\multicolumn{1}{c|}{\multirow{5}{*}{SGL}}      & CLLMR     & 0.10941 & 0.19097 & 0.24976 & 0.08083  & 0.11468 & 0.12931 & 0.08018  & 0.16992 & 0.23825 & 0.06521 & 0.09477 & 0.11438 & 0.09597  & 0.19123 & 0.25584 & 0.07712 & 0.10792 & 0.12576 \\
\multicolumn{1}{c|}{}                          & w/o Conf. & 0.09535 & 0.18697 & 0.24537 & 0.07245  & 0.10131 & 0.11709 & 0.07766 & 0.16838 & 0.23535 & 0.06326  & 0.09372 & 0.11265 & 0.09357  & 0.18879 & 0.25348 & 0.07406 & 0.10534 & 0.12348 \\
\multicolumn{1}{c|}{}  & Conf. De.  & -12.85\% & -2.09\%  & -1.76\%  & -10.37\%  & -11.66\% & -9.45\%  & -3.24\%  & -0.98\%  & -1.13\%  & -3.27\%   & -1.26\%  & -1.51\%  & -2.50\%   & -1.28\%  & -0.92\%  & -3.97\%  & -2.39\%  & -1.81\%   \\
\multicolumn{1}{c|}{}                          & w/o SSE   & 0.10188 & 0.19031 & 0.24314 & 0.078300 & 0.10598 & 0.12036 & 0.07688 & 0.16523 & 0.23106 & 0.06269  & 0.09172 & 0.11064 & 0.09452  & 0.18965 & 0.25304 & 0.07665 & 0.10777 & 0.12557 \\
\multicolumn{1}{c|}{}     & SSE De.   & -6.88\%  & -0.35\%  & -2.65\%  & -3.13\%   & -7.59\%  & -6.92\%  & -4.21\%  & -2.83\%  & -2.93\%  & -4.14\%   & -3.37\%  & -3.27\%  & -1.51\%   & -0.83\%  & -1.09\%  & -0.61\%  & -0.14\%  & -0.15\%   \\
\hline
\multicolumn{1}{c|}{\multirow{5}{*}{DirectAU}} & CLLMR     & 0.09693 & 0.18550 & 0.24837 & 0.08321  & 0.10814 & 0.12158 & 0.07854 & 0.16846 & 0.23581 & 0.06438  & 0.09439 & 0.11334 & 0.09363  & 0.18947 & 0.25574 & 0.07462 & 0.10617 & 0.12472 \\
\multicolumn{1}{c|}{}                          & w/o Conf. & 0.09260 & 0.18738 & 0.24741 & 0.07025  & 0.09982 & 0.11608 & 0.06716 & 0.14476 & 0.20394 & 0.05508  & 0.08035 & 0.09745 & 0.08173  & 0.16732 & 0.22448 & 0.06457 & 0.09259 & 0.10868 \\
\multicolumn{1}{c|}{}  & Conf. De.  & -4.47\%  & -0.55\%  & -0.39\%  & -15.58\%  & -7.69\%  & -4.52\%  & -14.49\% & -14.07\% & -13.52\% & -14.45\%  & -14.87\% & -14.02\% & -12.71\%  & -11.69\% & -12.22\% & -13.47\% & -12.79\% & -12.86\%  \\
\multicolumn{1}{c|}{}                          & w/o SSE   & 0.09169 & 0.17980 & 0.23839 & 0.06980  & 0.09751 & 0.11343 & 0.07295 & 0.16056 & 0.22430 & 0.06020  & 0.08912 & 0.10943 & 0.08930  & 0.18367 & 0.24740 & 0.07176 & 0.10262 & 0.12045 \\
\multicolumn{1}{c|}{}  & SSE De.   & -5.41\%  & -3.07\%  & -4.02\%  & -16.12\%  & -9.83\%  & -6.70\%  & -7.12\%  & -4.69\%  & -4.88\%  & -6.49\%   & -5.58\%  & -3.45\%  & -4.62\%   & -3.06\%  & -3.26\%  & -3.83\%  & -3.34\%  & -3.42\%    \\
\hline
\multicolumn{1}{c|}{\multirow{5}{*}{SimGCL}}   & CLLMR     & 0.10053 & 0.19497 & 0.25707 & 0.07918  & 0.10996 & 0.12547 & 0.08185 & 0.17445 & 0.24046 & 0.06695  & 0.09742 & 0.11658 & 0.096076 & 0.19196 & 0.25735 & 0.07715 & 0.10847 & 0.12687 \\
\multicolumn{1}{c|}{}                          & w/o Conf. & 0.09939 & 0.19425 & 0.25458 & 0.07589  & 0.10570 & 0.12213 & 0.08119 & 0.17311 & 0.23908 & 0.06593  & 0.09615 & 0.11529 & 0.09555  & 0.19126 & 0.25515 & 0.07678 & 0.10832 & 0.12532 \\
\multicolumn{1}{c|}{}   & Conf. De.  & -1.13\%  & -0.37\%  & -0.97\%  & -4.16\%   & -3.87\%  & -2.66\%  & -0.81\%  & -0.77\%  & -0.57\%  & -1.52\%   & -1.30\%  & -1.11\%  & -0.55\%   & -0.36\%  & -0.85\%  & -0.48\%  & -0.14\%  & -1.22\%  \\
\multicolumn{1}{c|}{}                          & w/o SSE   & 0.09951 & 0.19188 & 0.25201 & 0.07481  & 0.10408 & 0.12032 & 0.07767 & 0.16749 & 0.23381 & 0.06324  & 0.09280 & 0.11184 & 0.09471  & 0.19004 & 0.25504 & 0.07645 & 0.10754 & 0.12578 \\
\multicolumn{1}{c|}{}                          & SSE De.   & -1.01\%  & -1.58\%  & -1.97\%  & -5.52\%   & -5.35\%  & -4.10\%  & -5.11\%  & -3.99\%  & -2.77\%  & -5.54\%   & -4.74\%  & -4.07\%  & -1.42\%   & -1.00\%  & -0.90\%  & -0.91\%  & -0.86\%  & -0.86\%    \\
\hline
\multicolumn{1}{c|}{\multirow{5}{*}{LightGCL}} & CLLMR     & 0.09868 & 0.18634 & 0.24663 & 0.07404  & 0.10159 & 0.11786 & 0.07893 & 0.17299 & 0.23824 & 0.06326  & 0.09238 & 0.11137 & 0.09091  & 0.18547 & 0.24982 & 0.07815 & 0.10756 & 0.12194 \\
\multicolumn{1}{c|}{}                          & w/o Conf. & 0.09526 & 0.18282 & 0.24256 & 0.07260  & 0.10018 & 0.11629 & 0.07691 & 0.16574 & 0.22967 & 0.06270  & 0.09201 & 0.11050 & 0.08875  & 0.18316 & 0.24495 & 0.07195 & 0.10277 & 0.12011 \\
\multicolumn{1}{c|}{}                           & Conf. De.  & -3.47\%  & -1.89\%  & -1.65\%  & -1.94\%   & -1.39\%  & -1.33\%  & -2.56\%  & -4.19\%  & -3.60\%  & -0.89\%   & -0.40\%  & -0.78\%  & -2.38\%   & -1.25\%  & -1.95\%  & -7.93\%  & -4.45\%  & -1.50\%   \\
\multicolumn{1}{c|}{}                          & w/o SSE   & 0.08662 & 0.17124 & 0.22883 & 0.06562  & 0.09201 & 0.10747 & 0.06870 & 0.15220 & 0.21241 & 0.055195 & 0.08271 & 0.10008 & 0.08644  & 0.17653 & 0.23953 & 0.06874 & 0.09825 & 0.11590\\
\multicolumn{1}{c|}{}      & SSE De.   & -12.22\% & -8.10\%  & -7.22\%  & -11.37\%  & -9.43\%  & -8.82\%  & -12.96\% & -12.02\% & -10.84\% & -12.75\%  & -10.47\% & -10.14\% & -4.92\%   & -4.82\%  & -4.12\%  & -12.04\% & -8.66\%  & -4.95\%   \\
\hline
\end{tabular}}
\label{rq2}
\end{table*}

To answer \textbf{RQ3}, we conduct an ablation study on CLLMR, specifically removing SSE and counterfactual inference respectively. The experimental results, shown in Table~\ref{rq2}, demonstrate the following phenomena:

\begin{itemize}
\item When counterfactual inference is removed (``w/o Conf.''), the performance decreases compared to CLLMR, but it remains higher than directly using contrastive learning for alignment. This demonstrates the effectiveness of counterfactual inference. By utilizing spectrum information as auxiliary input, we learn a side representation that aligns with the real-world historical interaction distribution, reflecting the global nature of the data and providing a richer and more comprehensive representation. In contrast, MLP, as a basic neural network structure, learns feature representations primarily through local non-linear transformations and is unable to effectively capture this global structure. Additionally, the introduction of spectrum information helps avoid the phenomenon of feature collapse, where the representation vectors learned by the model become identical or highly similar, impairing the model's ability to distinguish between different inputs. Spectrum information serves as auxiliary information to ensure the identifiability of the representations, improving encoding quality and maintaining data differentiation.

\item When only counterfactual inference is performed (``w/o SSE''), the results are not only lower than CLLMR, but also lower than directly using contrastive learning for alignment. Counterfactual inference considers different outcomes under varying conditions, requiring the model to have a deep understanding of the intrinsic structure and patterns of the data. SSE introduces structural information into the representation, providing global properties of the data and effectively preventing dimensional collapse. If dimensional collapse occurs in the embedding representation, the model may struggle to differentiate between subtle feature differences, directly affecting the accuracy of counterfactual inference. In recommender systems, if two different users or items are represented too similarly in the embedding space, counterfactual inference may fail to correctly assess the specific impact of changing a certain condition on the recommendation result. Understanding user behavior is crucial for counterfactual inference, as it allows the model to predict how users might behave differently in various contexts. Once SSE is replaced with MLP, merely performing counterfactual inference may lead to negative results due to the dimensional collapse of side information.
\end{itemize}

\begin{figure*}[ht]
  \centering
  \subfloat[LRGCCF]{\includegraphics[width=0.325\textwidth]{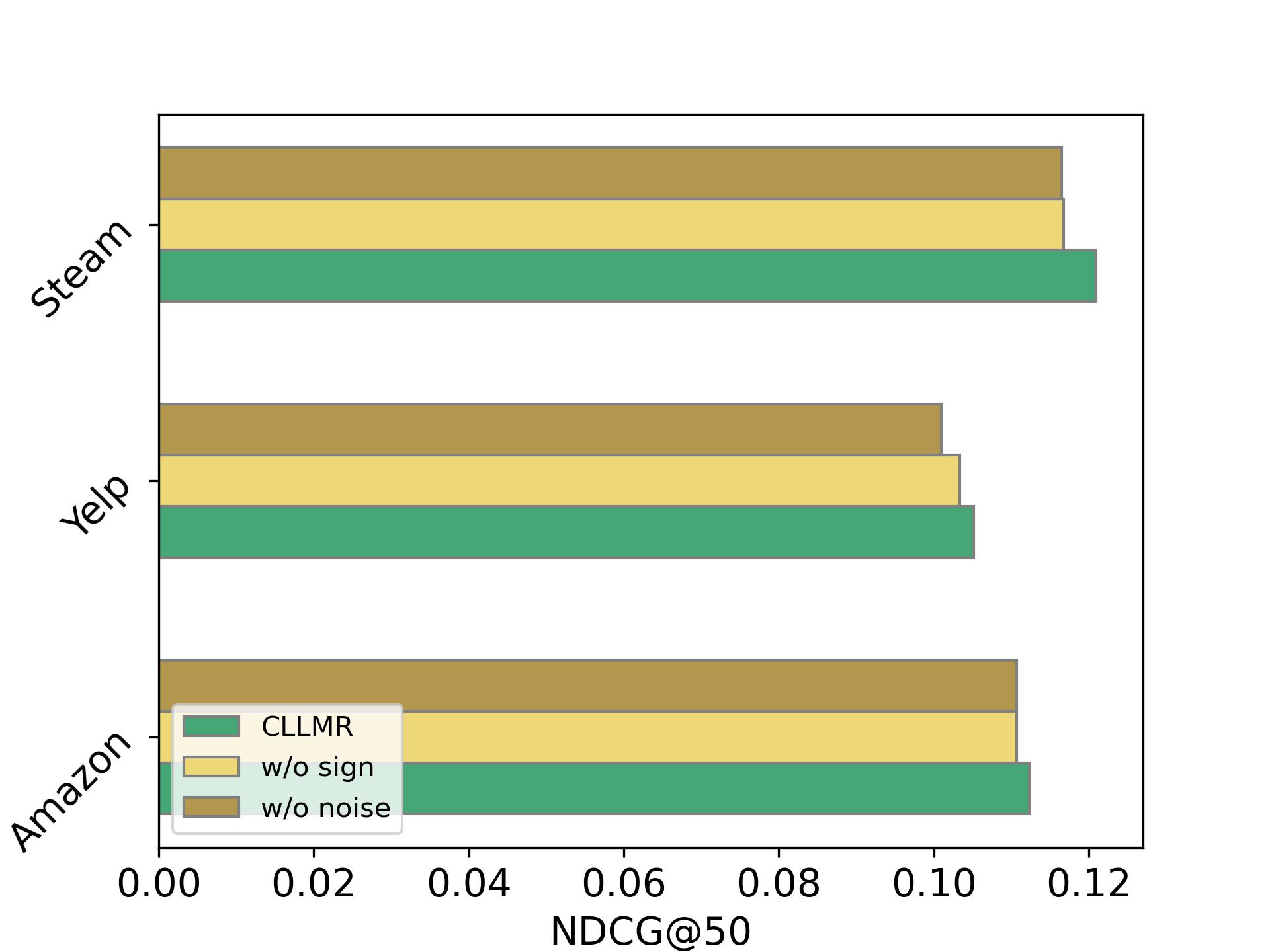}\captionsetup{font={tiny}}}
\subfloat[LightGCN]{\includegraphics[width=0.325\textwidth]{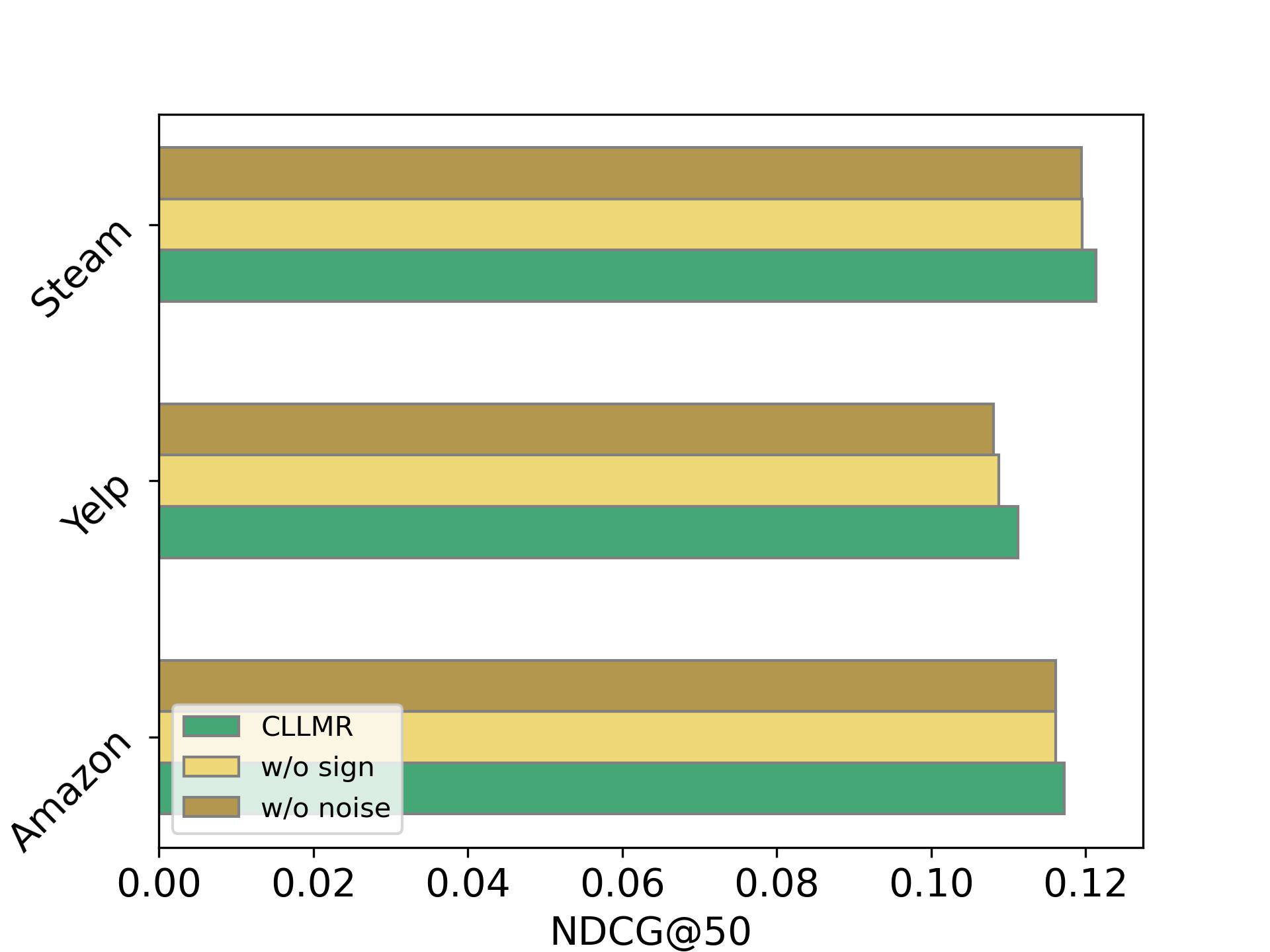}\captionsetup{font={tiny}}}
\subfloat[SGL]{\includegraphics[width=0.325\textwidth]{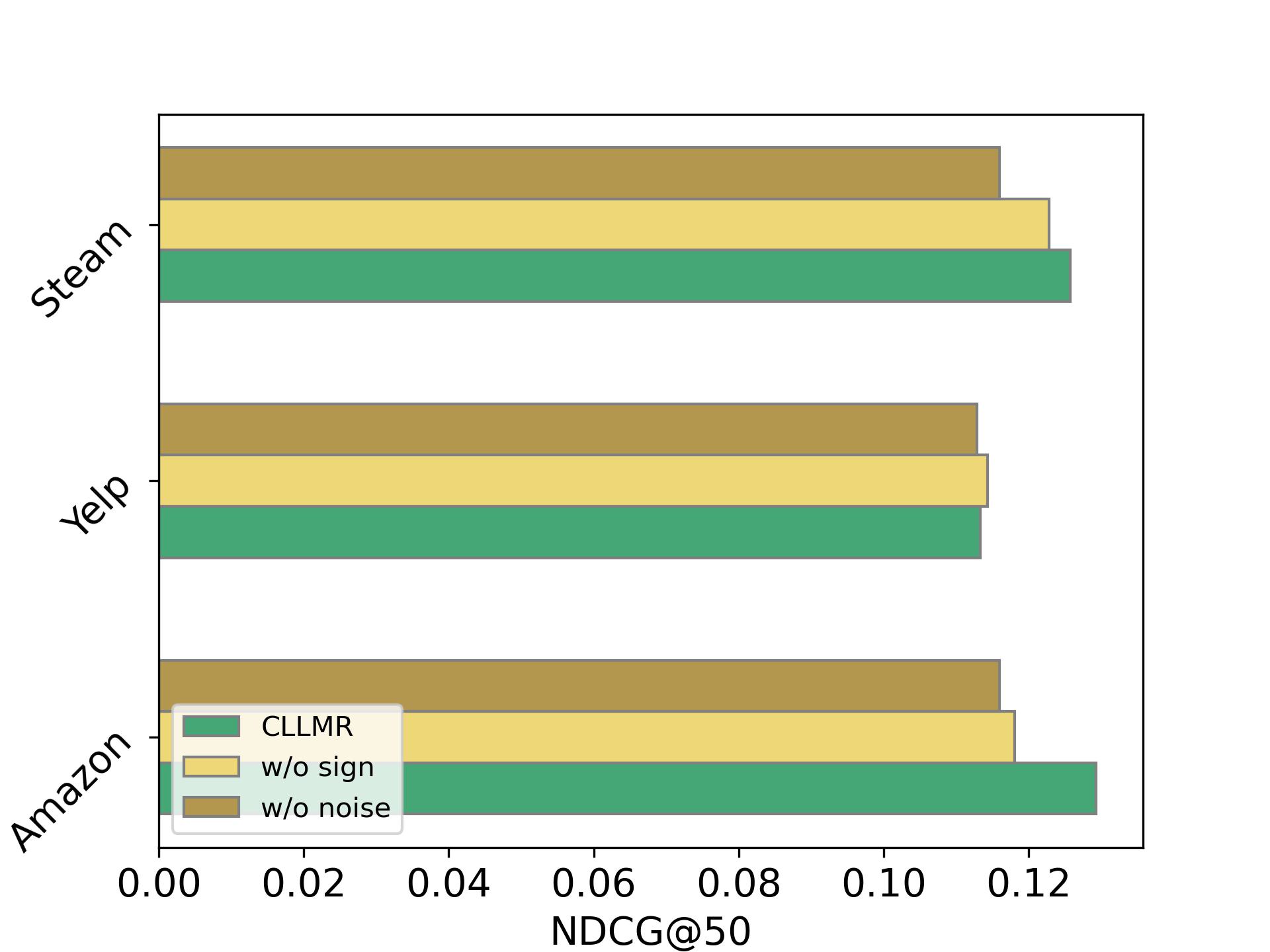}}
\\
  \subfloat[DirectAU]{\includegraphics[width=0.325\textwidth]{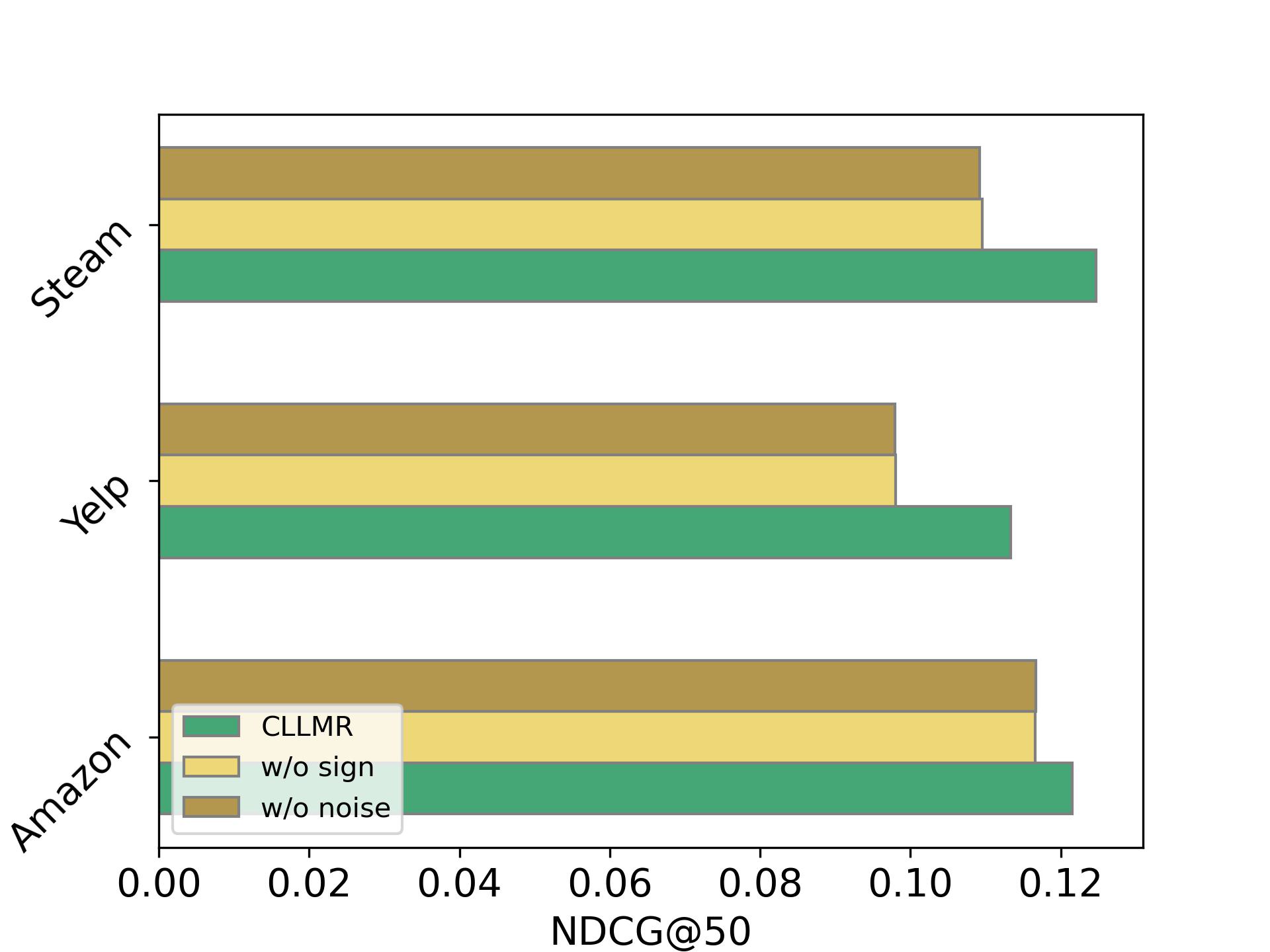}}
\subfloat[SimGCL]{\includegraphics[width=0.325\textwidth]{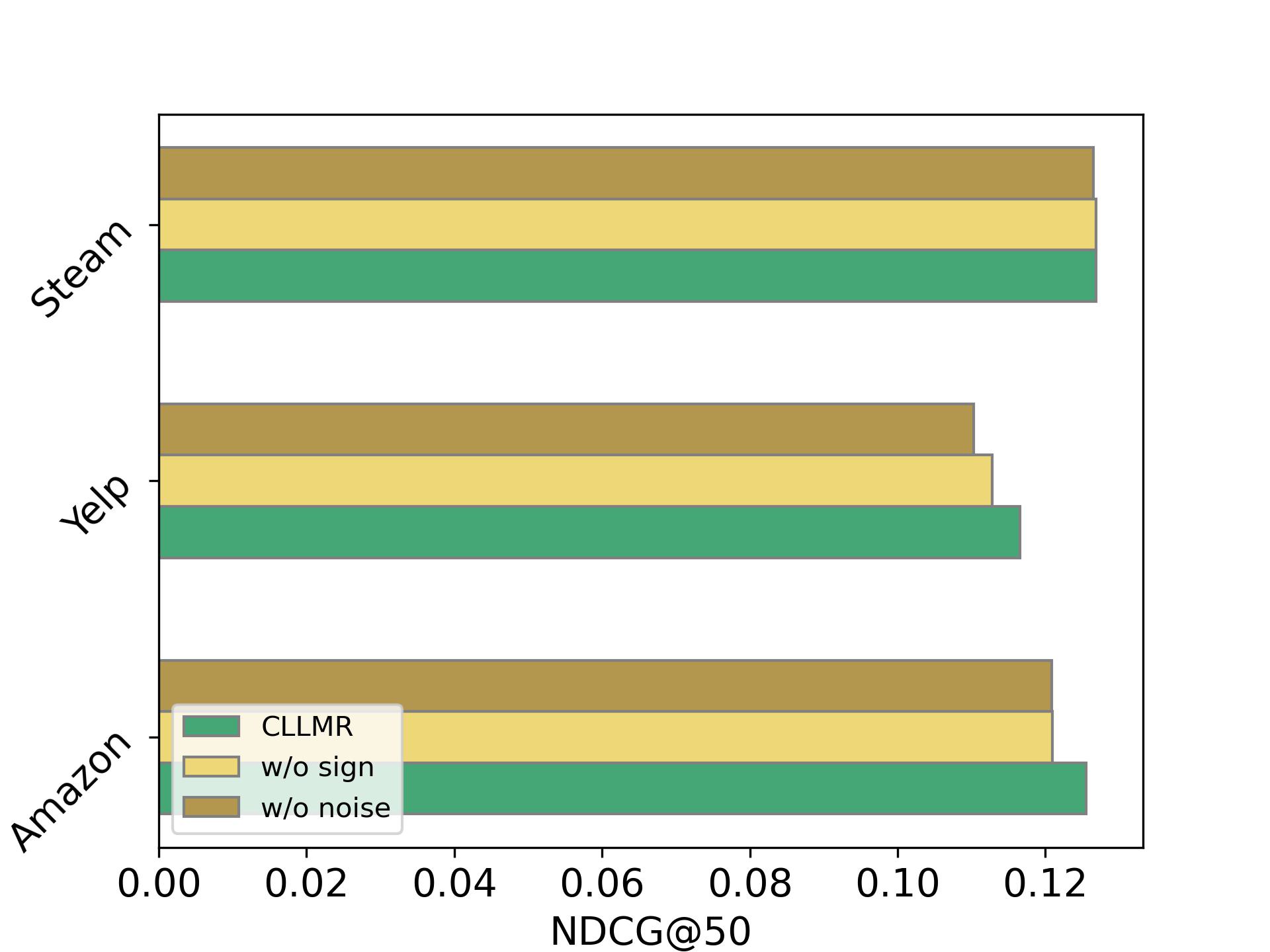}}
\subfloat[LightGCL]{\includegraphics[width=0.325\textwidth]{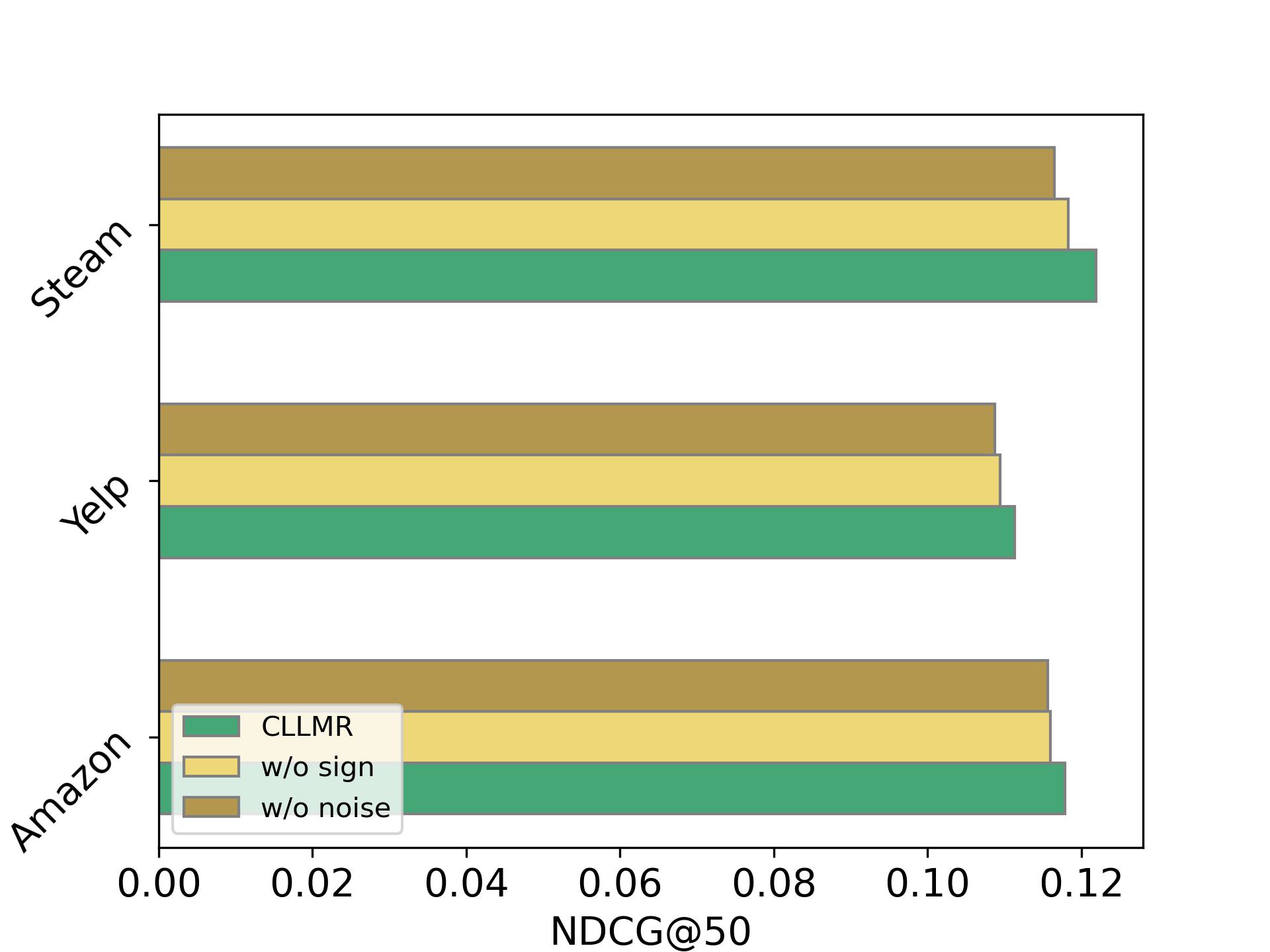}}
  \caption{Necessity of each module of SSE}
  \label{rq3}
\end{figure*}

\begin{figure*}[ht]
  \centering
  \subfloat[LRGCCF]{\includegraphics[width=0.325\textwidth]{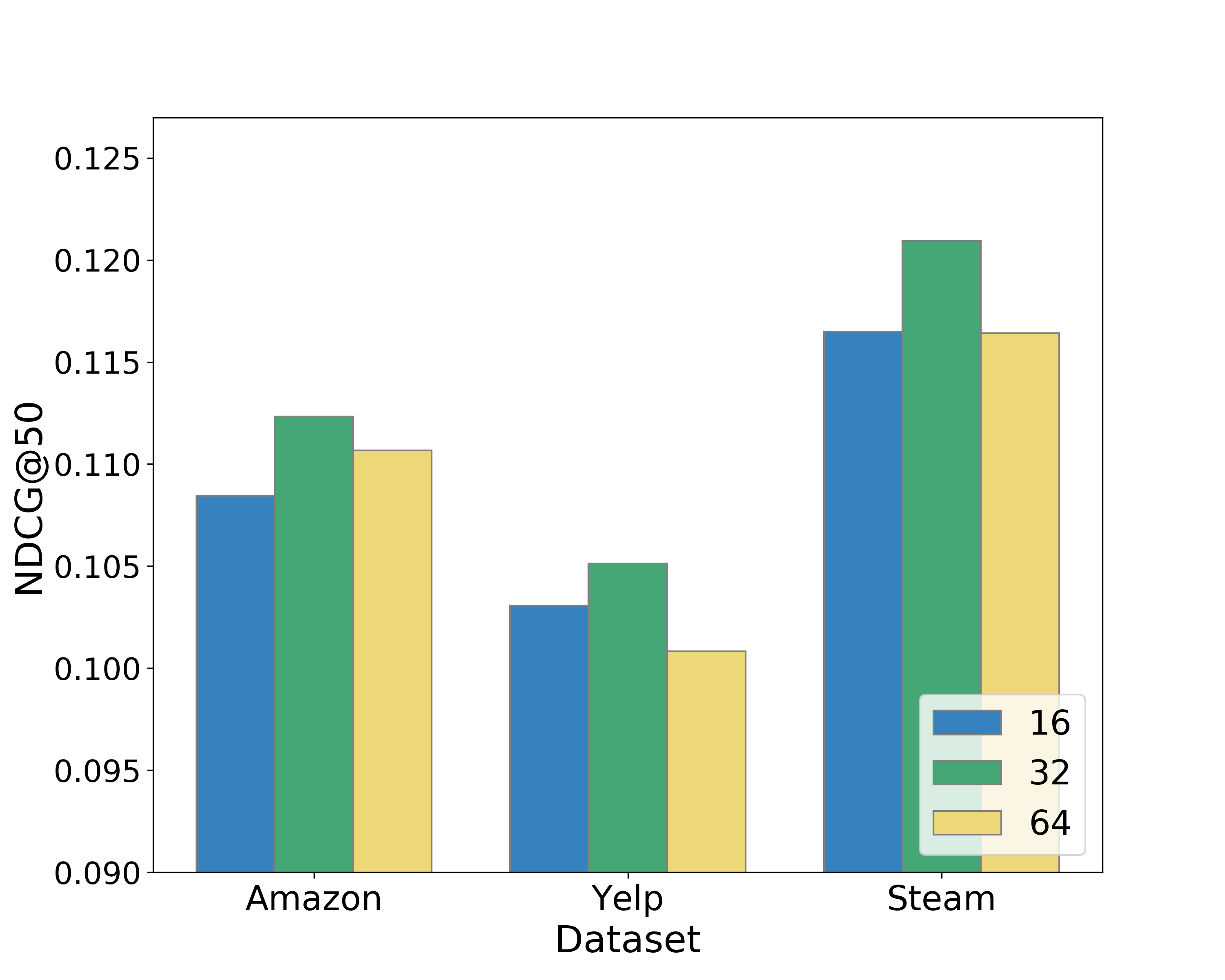}}
\subfloat[LightGCN]{\includegraphics[width=0.325\textwidth]{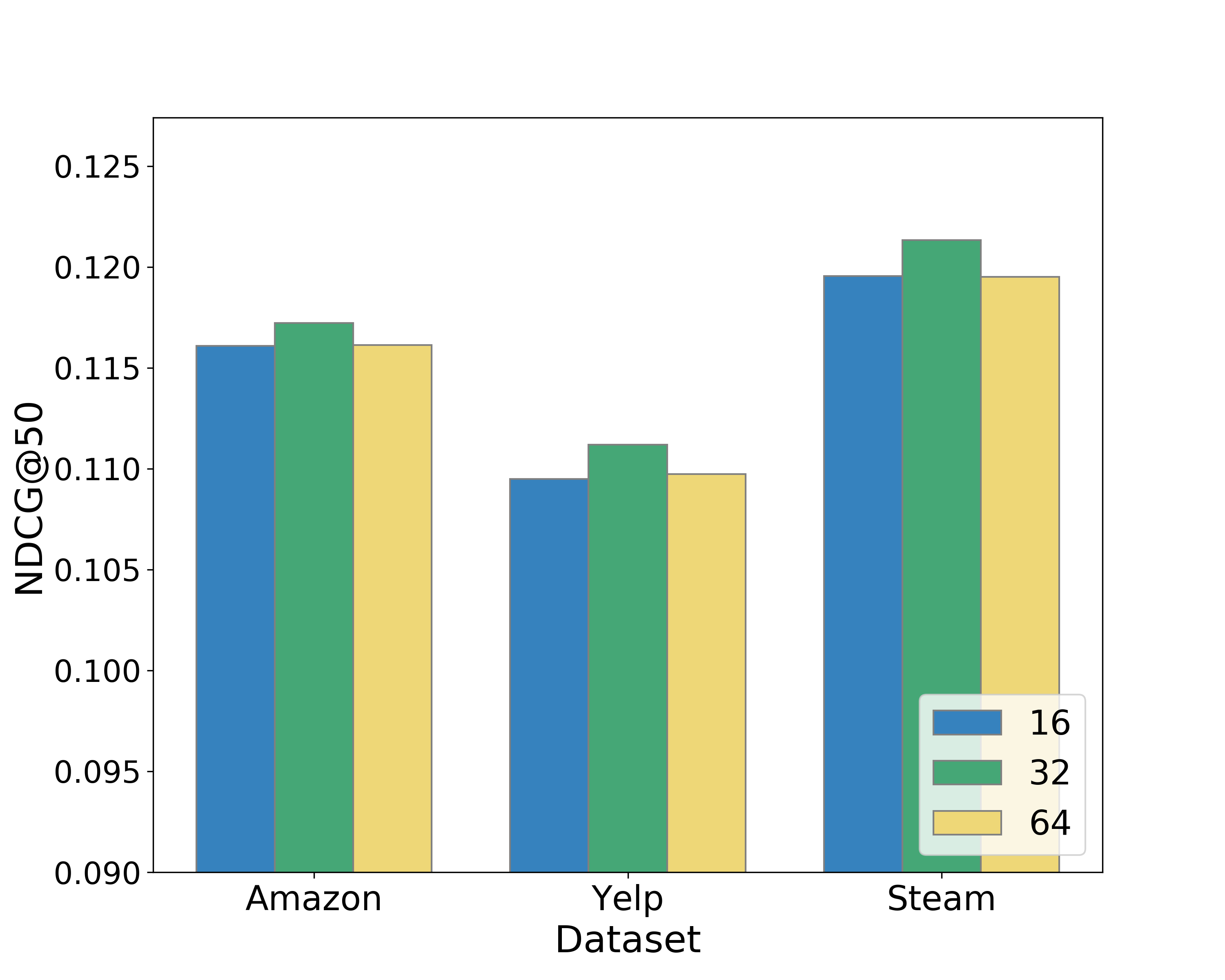}}
\subfloat[SGL]{\includegraphics[width=0.325\textwidth]{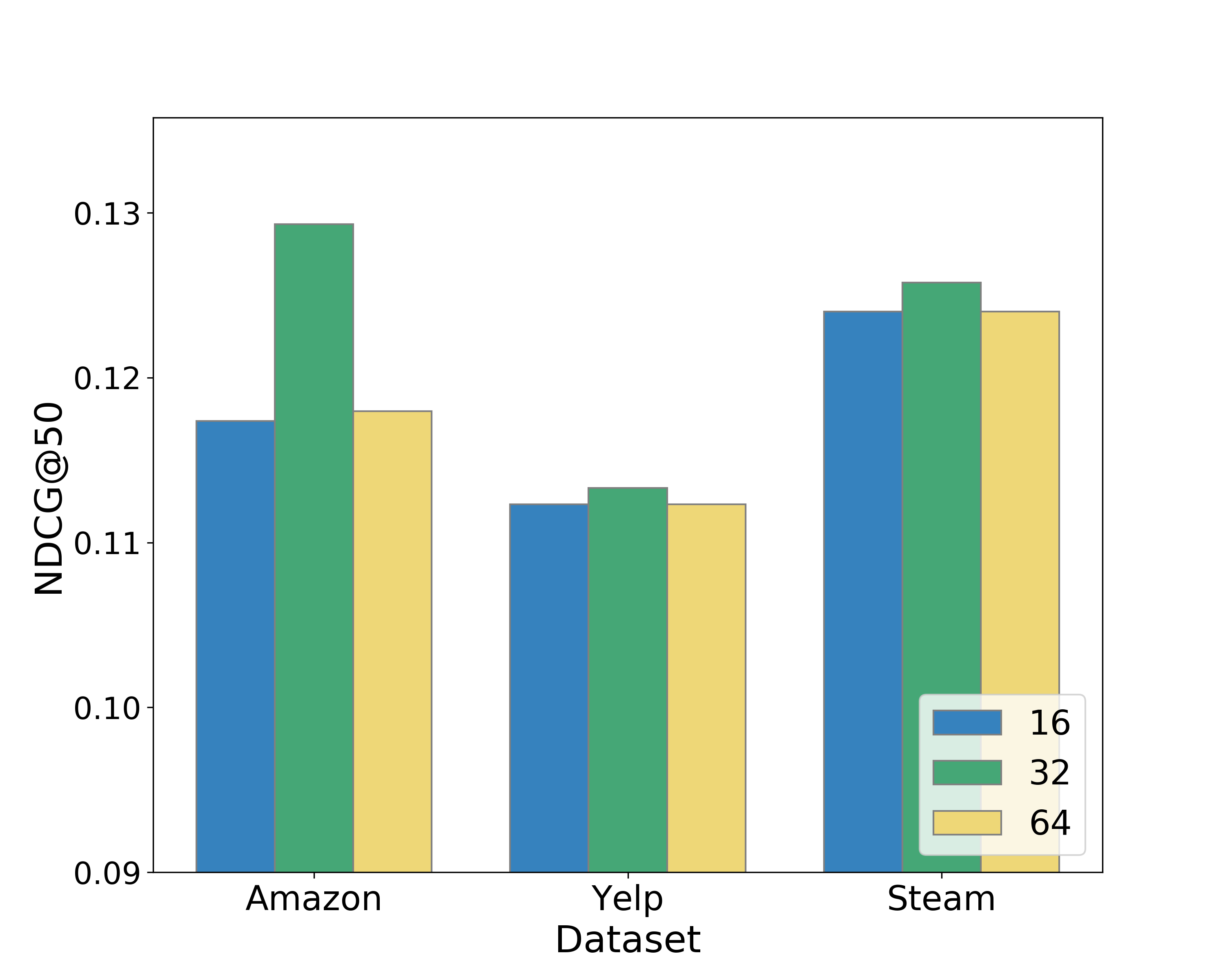}}
\\
  \subfloat[DirectAU]{\includegraphics[width=0.325\textwidth]{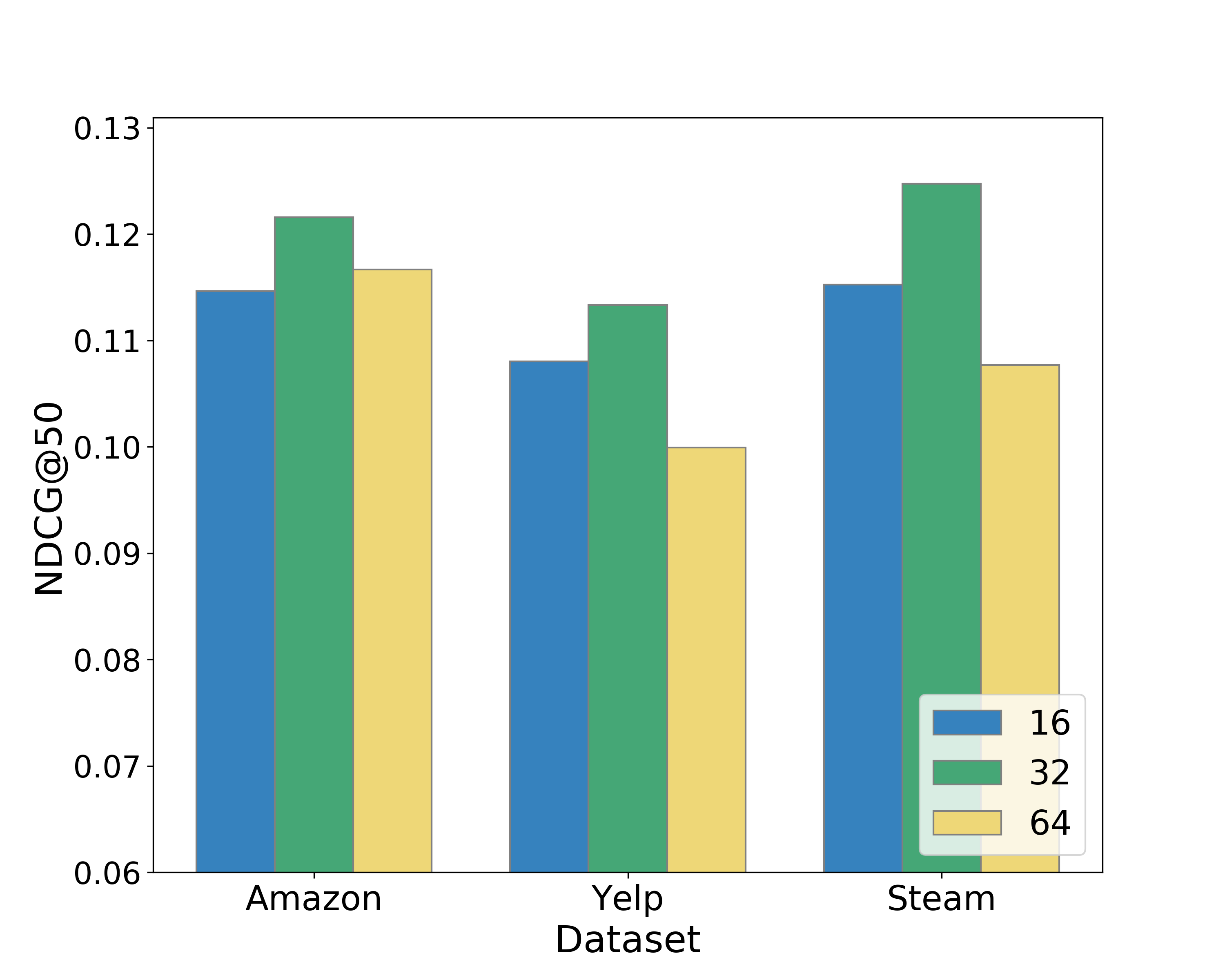}}
\subfloat[SimGCL]{\includegraphics[width=0.325\textwidth]{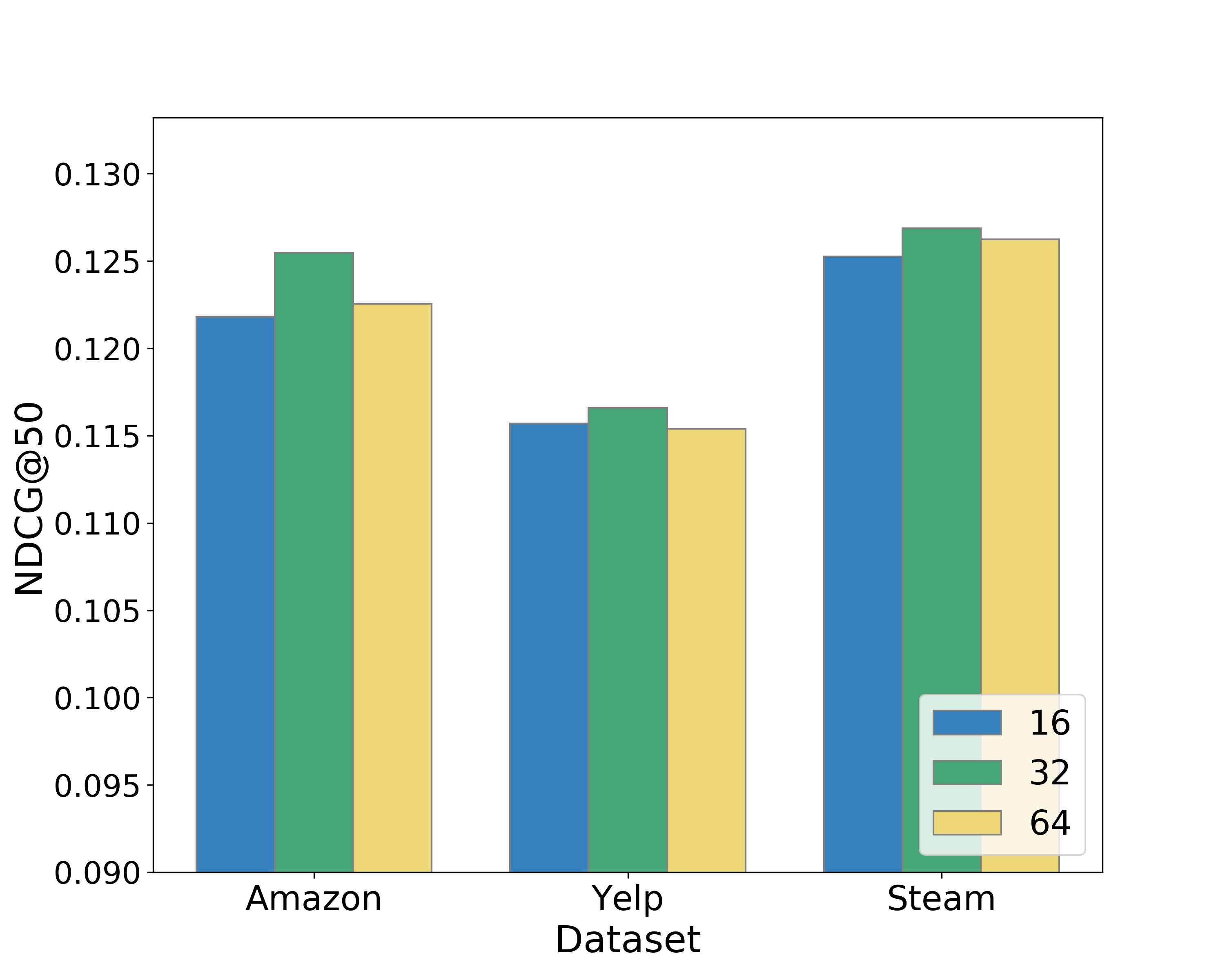}}
\subfloat[LightGCL]{\includegraphics[width=0.325\textwidth]{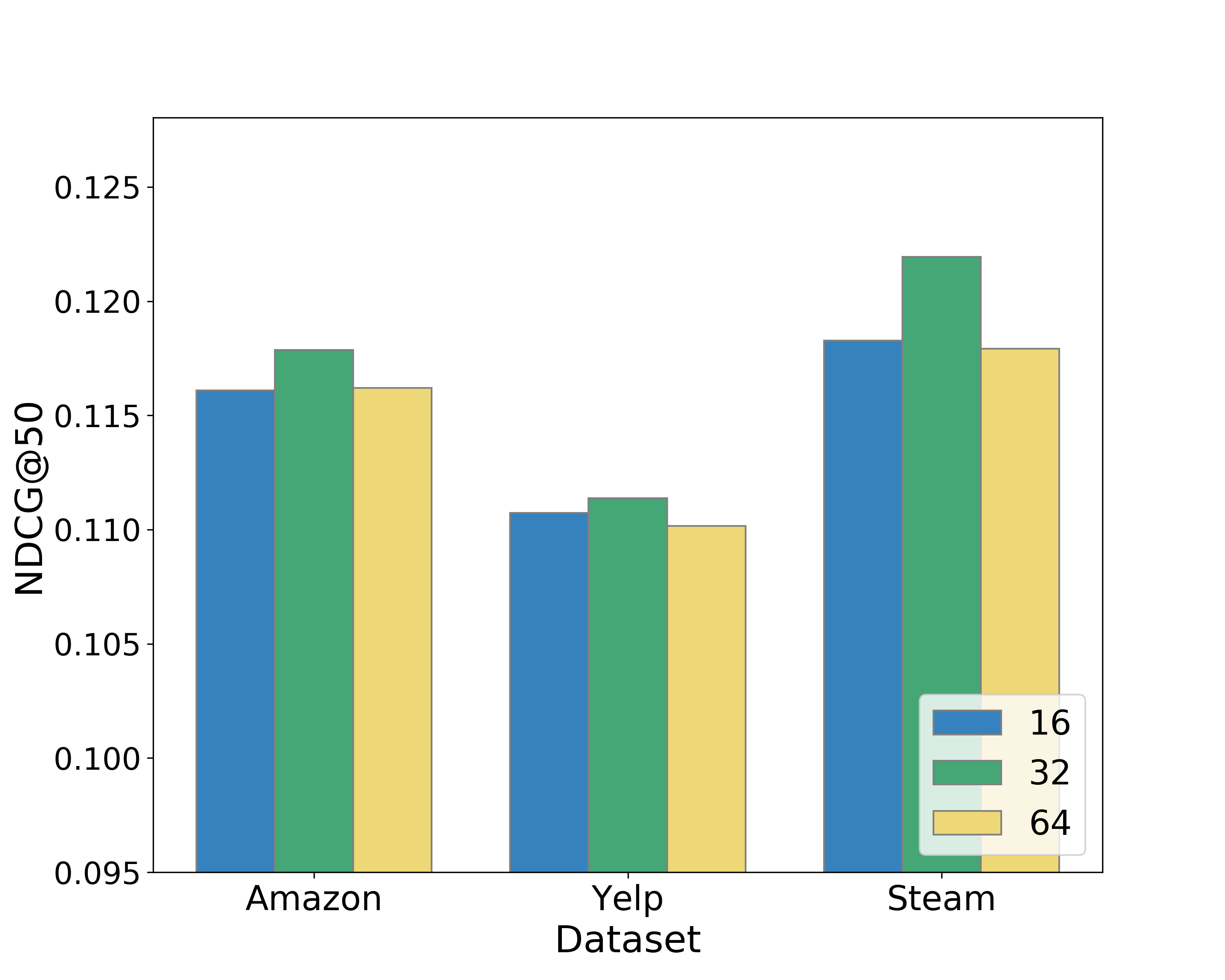}}
  \caption{Hyperparameter analysis for the rank of the spectrum in SSE.}
  \label{rank}
\end{figure*}

\subsubsection{Necessity of each module of SSE}

To answer \textbf{RQ4}, we successively removed the noise sign (``w/o sign'') and random noise (``w/o noise'') to assess their necessity. The experimental results are displayed in Figure~\ref{rq3}. It can be observed that the removal of each module reduces the model's performance. In the real world, user behavior is often accompanied by uncertainty and randomness. By adding random noise to the spectrum when learning the side information representation, the model can better adapt to this uncertainty, thereby improving its performance in real-world environments.

On the other hand, when we add random noise to the node representation in the recommender system, the use of the sign function ensures that the addition of noise does not change the orientation of the original node embedding in the feature space. This is because the sign function returns the sign of each element of the original vector, allowing the added noise to maintain the same direction as the original vector. As a result, the node's orientation in the feature space remains consistent even after the noise is added. This is crucial for the model's generalization ability, as it ensures that the model remains stable despite input changes, enabling better generalization to unseen data. In practice, this approach enhances the model's robustness to noise, particularly when the recommender system is faced with complex and dynamic user behavior and item characteristics. This enhanced generalization capability allows the recommender system to make reasonable recommendations when confronted with new users or new items, ultimately improving overall system performance.

\subsection{Hyperparameter Analysis}

In this section, we do the hyperparameter analyses of the rank of the spectrum and  the noise size in SSE, respectively.

\paragraph{The rank of the spectrum in SSE}
In this subsection, we conduct a hyperparameter analysis for the rank of the spectrum in SSE to answer \textbf{RQ5}, with the experimental results shown in Figure~\ref{rank}. We set the rank to 16, 32, and 64, respectively, and it is important to note that the hidden layer dimension is set to 32 for all the experiments in other sections. The results show that our proposed method achieves the best performance when the rank is set to 32.

The rank of the spectrum obtained through singular value decomposition (SVD) of the graphical data affects how much of the graph's structural information is retained. A higher rank captures more details, while a lower rank reduces noise. Specifically, setting the rank to match the hidden layer dimension ensures that all relevant graph structures are represented. The hidden layer learns the representation of nodes based on the input graph, and if the rank is lower than the hidden dimension, part of the hidden layer's capacity is left unused because it cannot access all the structural information from the input. Conversely, if the rank is higher, it may retain noise that exceeds the effective modeling capacity of the hidden layer, potentially affecting performance. Using the same rank as the hidden layer dimension creates a one-to-one mapping between the graph structure and the learned node representations, allowing the hidden layer to fully utilize the spectral inputs without overfitting.

\begin{figure*}[ht]
  \centering
  \subfloat[LRGCCF]{\includegraphics[width=0.325\textwidth]{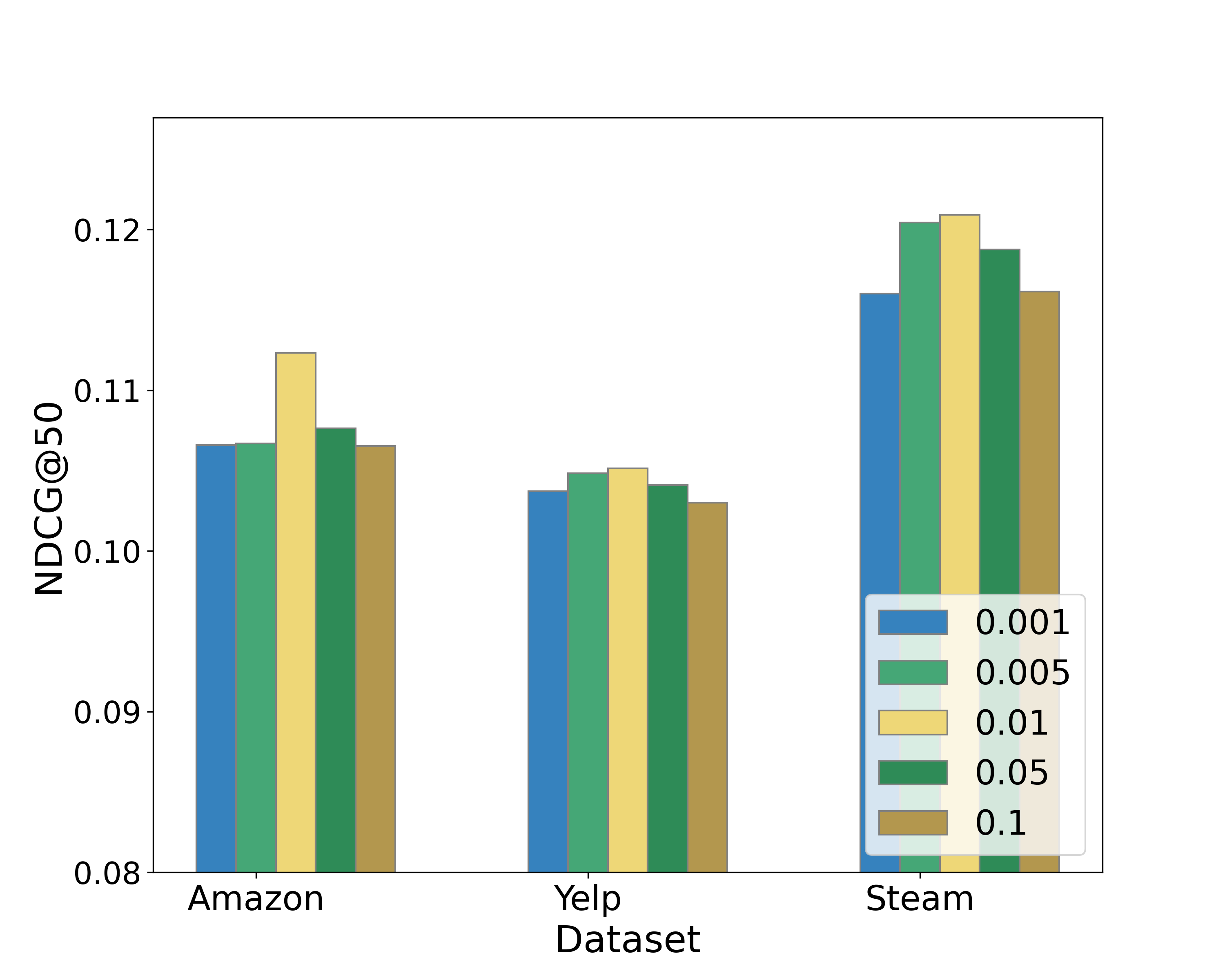}}
\subfloat[LightGCN]{\includegraphics[width=0.325\textwidth]{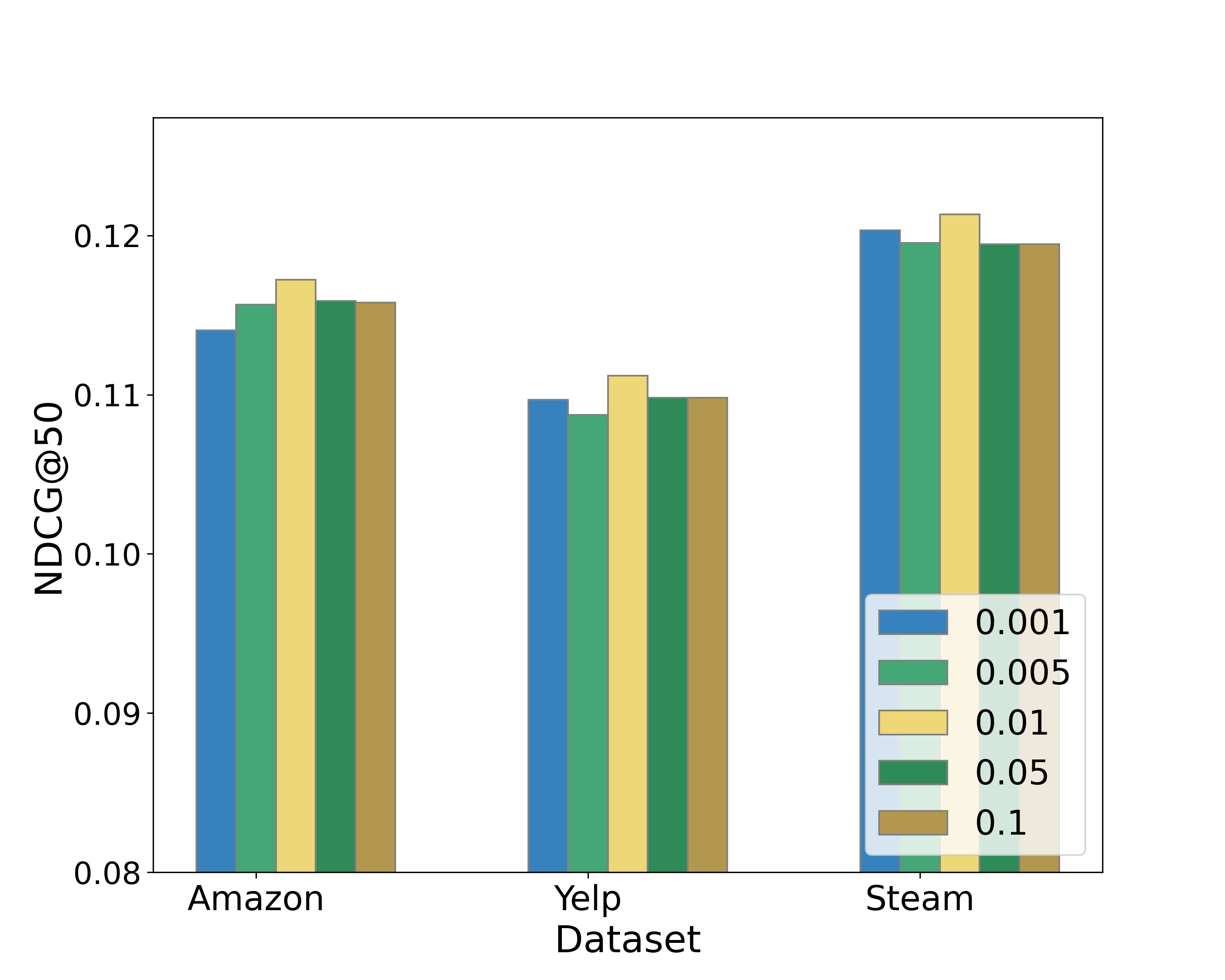}}
\subfloat[SGL]{\includegraphics[width=0.325\textwidth]{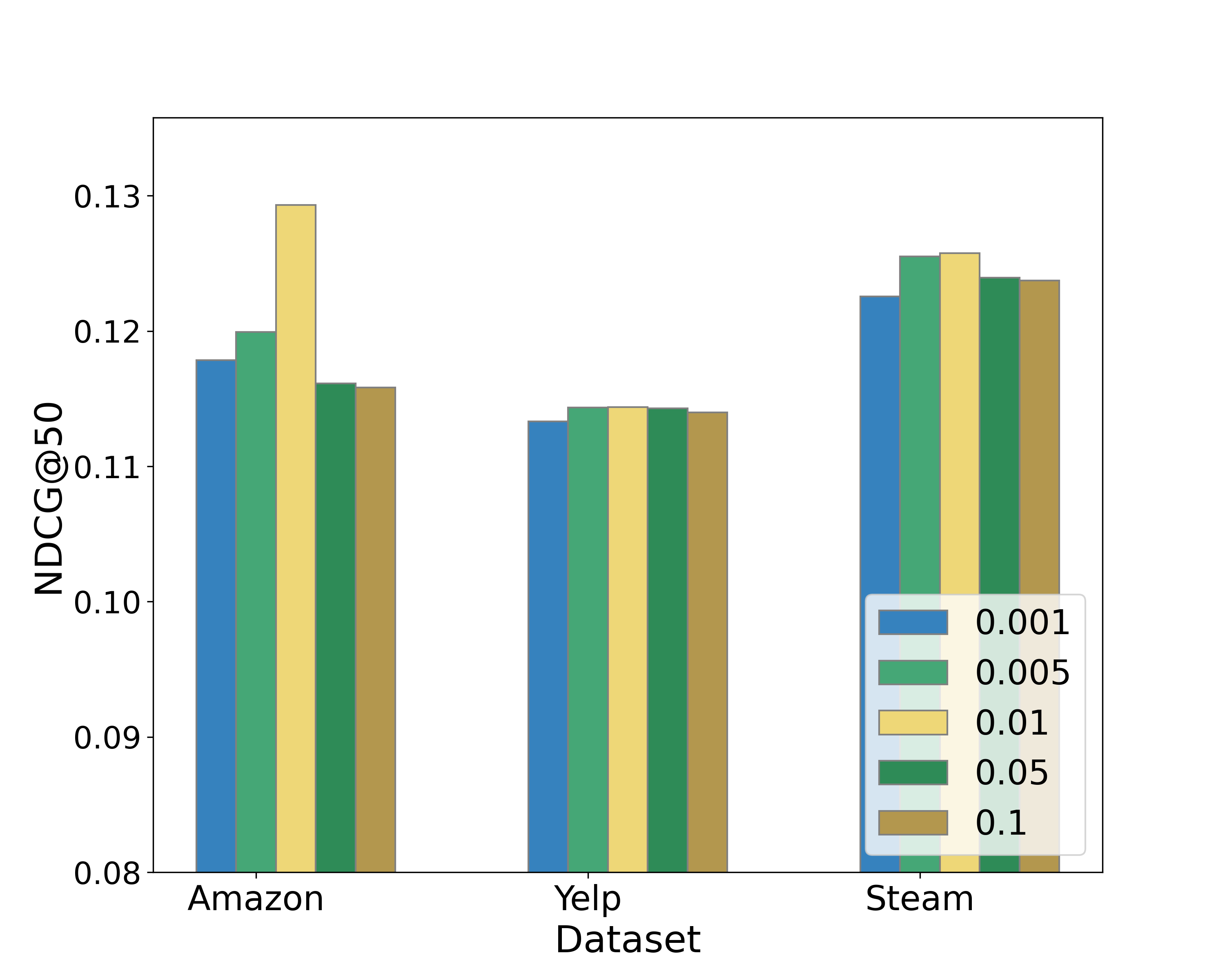}}
\\
  \subfloat[DirectAU]{\includegraphics[width=0.325\textwidth]{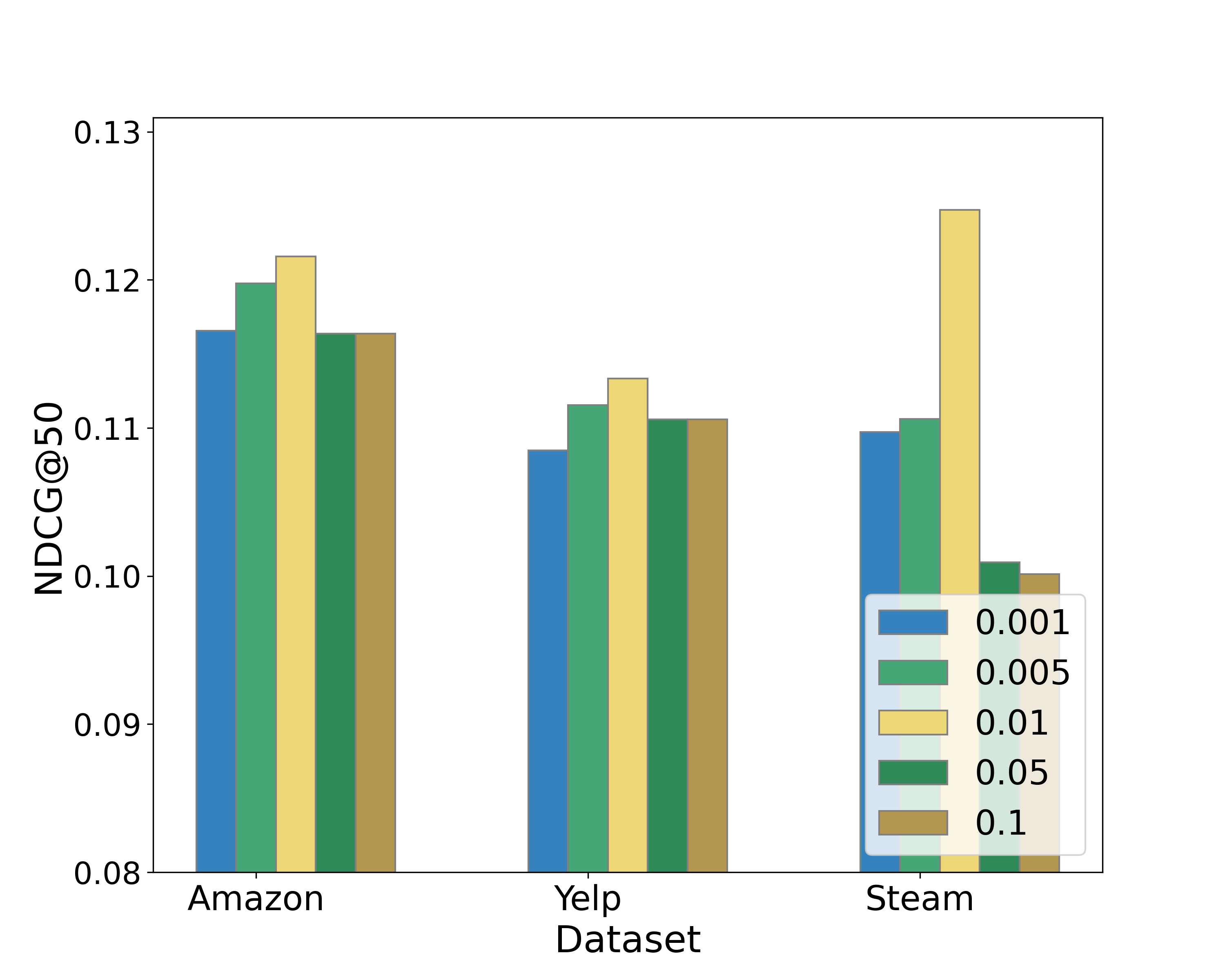}}
\subfloat[SimGCL]{\includegraphics[width=0.325\textwidth]{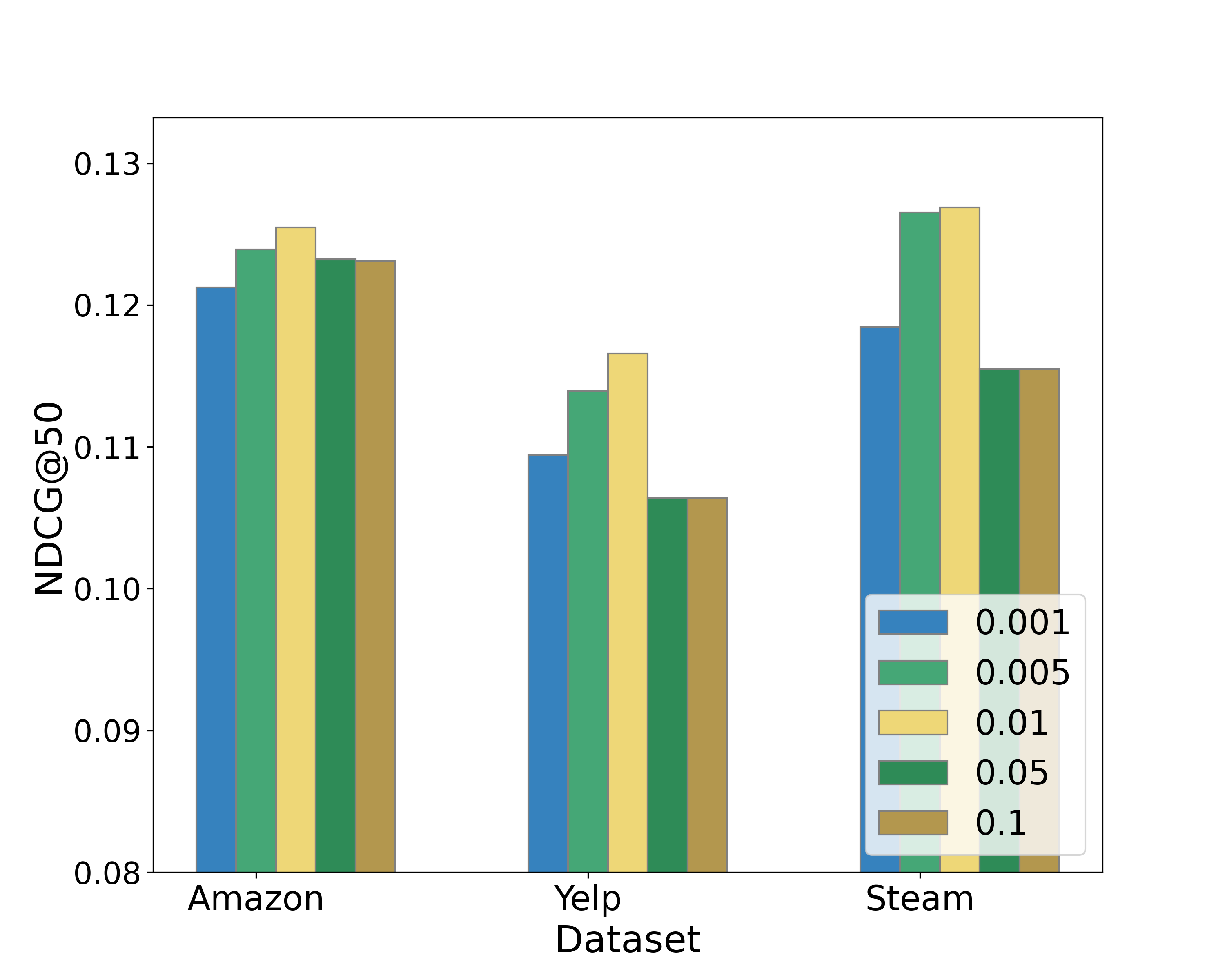}}
\subfloat[LightGCL]{\includegraphics[width=0.325\textwidth]{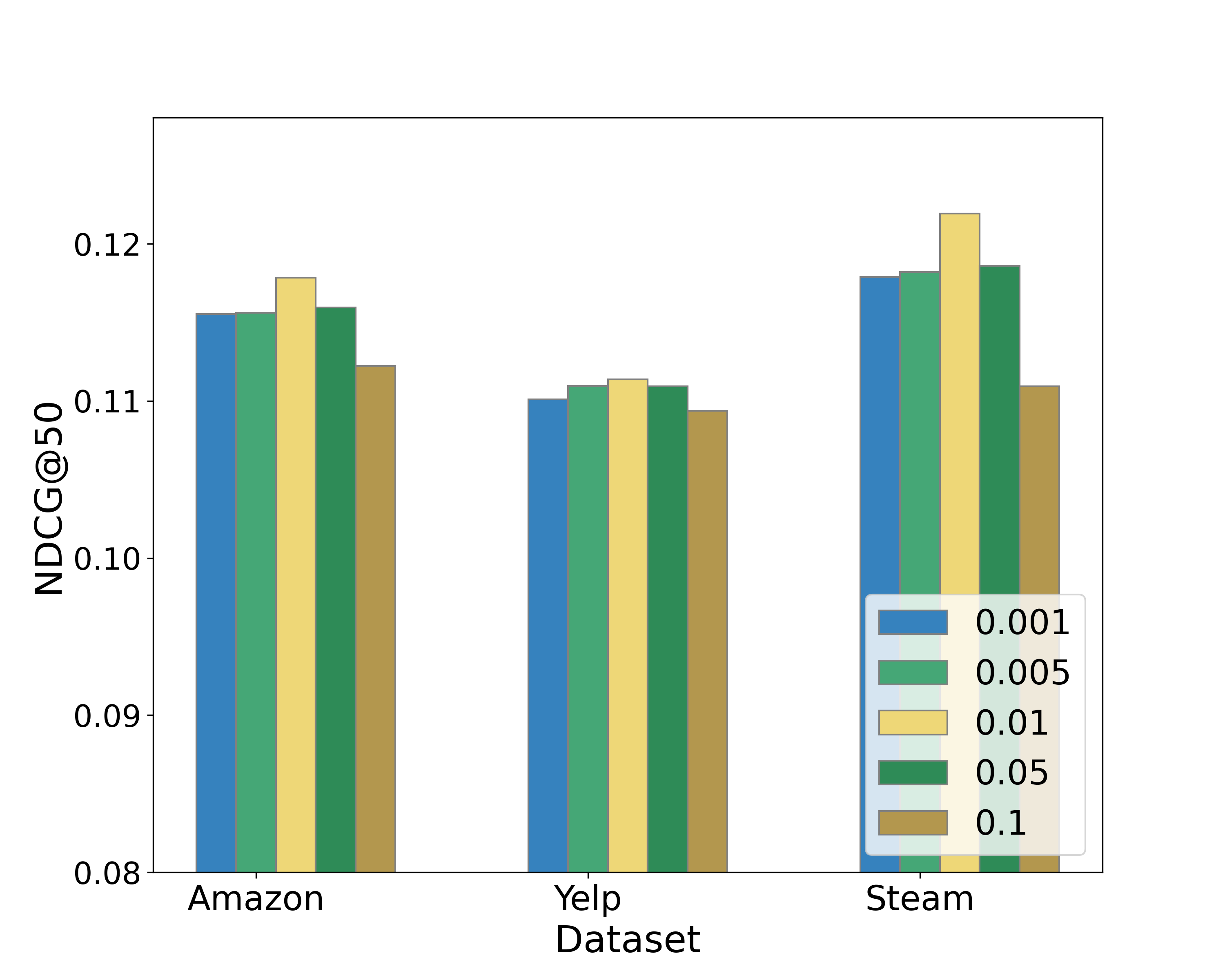}}
  \caption{Hyperparameter analysis for the noise size in SSE.}
  \label{noise}
\end{figure*}

\paragraph{The noise size in SSE}
To better answer \textbf{RQ4}, we conduct a hyperparameter analysis of the noise size in SSE, with the experimental results shown in Figure~\ref{noise}. We evaluate the rank settings to 0.001, 0.005, 0.01, 0.05 and 0.1, respectively. It is important to note that the hidden layer dimension is fixed at 0.01 for all the experiments in other sections. The results show that our proposed method achieves the best performance when the rank is set to 0.01.

The experimental results reveal that the performance of the model gradually improves when the noise size is increased from 0.001 to 0.01. This proves that adding noise to the spectral information enhances the generalisation ability of the model by enabling it to learn more diverse data during training. However, when the noise increases beyond 0.1, the model's effectiveness appears to decrease, or in some scenarios, remains unchanged. This is because as the noise increases, the information in the spectrum itself is affected or even overwritten, which makes the model performance converge. At the same time, we can observe that when the noise obtains a smaller value, the model performs better than when the noise is larger, which once again verifies that a small amount of noise helps to improve the generalisation ability of the model.

\section{Conclusion}
\label{sec:cons}
In this paper, we have proposed CLLMR, a counterfactual inference framework designed to effectively eliminate the inherent propensity bias of LLMs and enhance recommendation accuracy. To address the issue of dimensional collapse, we have also introduced a spectrum-based encoder that extracts the side information generated by LLMs. This is crucial because when dimensional collapse occurs, counterfactual inference becomes ineffective, as the model is unable to distinguish subtle differences between features. This validates the necessity of each module within the CLLMR framework. Additionally, we have incorporated contrastive learning as an alignment technique to synchronize collaborative representations, based on historical interactions, with LLM-derived side representations. This approach effectively combines the strengths of traditional recommendation methods with LLM capabilities and has been rigorously evaluated on three real-world datasets. Experiments have demonstrated that CLLMR effectively mitigates propensity bias, avoids dimensional collapse and achieves the state-of-the-art results.

\section*{ACKNOWLEDGMENTS}
This work has been supported in part by the National Natural Science Foundation of China under Grant 71774159, the China Postdoctoral Science Foundation under Grant 2021T140707, the Jiangsu Postdoctoral Science Foundation under grant number 2021K565C, the Science and Technology Foundation of Xuzhou under Grant KC22047, Australian Research Council under Grant DP230101122, the Graduate Innovation Program of China University of Mining and Technology 2024WLKXJ183, the Fundamental Research Funds for the Central Universities 2024-10949, the Postgraduate Research \& Practice Innovation Program of Jiangsu Province KYCX24\_2781, Research Fund of Guangxi Key Lab of Multi-source Information Mining \& Security (MIMS24-13), and Guangxi Key Laboratory of Big Data in Finance and Economics (Grant No. FEDOP2022A03).

\bibliographystyle{ACM-Reference-Format}
\bibliography{sample}










\end{document}